\documentclass[]{aa}  

\usepackage{graphicx}
\usepackage{txfonts}
\usepackage{natbib}
\bibpunct{(}{)}{;}{a}{}{,} 

\begin{document}

   \title{The stellar metallicity gradients of Local Group dwarf galaxies}

   \subtitle{}

   \author{
        S.~Taibi\inst{1,2,3}
        \and
        G.~Battaglia\inst{2,3}
        \and 
        R.~Leaman\inst{4}
        \and 
        A.~Brooks\inst{5,6}
        \and
        C.~Riggs\inst{5}
        \and
        F.~Munshi\inst{7,5,8}
        \and
        Y.~Revaz\inst{9}
        \and
        P.~Jablonka\inst{9,10}
        }

   \institute{
            Leibniz-Institut für Astrophysik Potsdam (AIP), An der Sternwarte 16, D-14482 Potsdam, Germany\\
            \email{staibi@aip.de}
         \and
             Instituto de Astrofísica de Canarias, Calle Vía Láctea s/n, E-38206 La Laguna, Tenerife, Spain
        \and
            Universidad de La Laguna, Avda. Astrofísico Fco. Sánchez, E-38205 La Laguna, Tenerife, Spain
        \and 
            Department of Astrophysics, University of Vienna, Türkenschanzstrasse 17, 1180 Vienna, Austria
        \and
            Department of Physics \& Astronomy, Rutgers, The State University of New Jersey, 136 Frelinghuysen Road, Piscataway, NJ 08854, USA
        \and
            Center for Computational Astrophysics, Flatiron Institute, 162 Fifth Avenue, New York, NY 10010, USA
        \and
            Homer L. Dodge Department of Physics \& Astronomy, University of Oklahoma, 440 W. Brooks St., Norman, OK 73019, USA
        \and
            Department of Physics \& Astronomy, Vanderbilt University, PMB 401807, Nashville, TN 37206, USA
        \and
            Institute of Physics, Laboratory of Astrophysics, École Polytechnique Fédérale de Lausanne (EPFL), 1290 Sauverny, Switzerland
        \and
            GEPI, CNRS UMR 8111, Observatoire de Paris, PSL University, 92125 Meudon, Cedex, France
             }

   \date{Received; accepted}

 
  \abstract
   {}
   {We explore correlations between the strength of metallicity gradients in Local Group dwarf galaxies and their stellar mass, star formation history timescales and environment.}
   {We perform a homogeneous analysis of literature spectroscopic data of red giant stars and determine radial metallicity profiles for 30 Local Group dwarf galaxies. This is the largest compilation to date of this type.}
   {The dwarf galaxies in our sample show a variety of metallicity profiles, most of them decreasing with radius and some having rather steep profiles. The derived metallicity gradients as a function of the half-light radius, $\nabla_{\rm [Fe/H]} (R/R_e)$, show no statistical differences when compared with the galaxies' morphological type, nor with their distance from the Milky Way or M31. No correlations are found with either stellar mass or star formation timescales. In particular, we do not find the linear relationship between $\nabla_{\rm [Fe/H]} (R/R_e)$ and the galaxies' median age $t_{50}$, as instead shown in the literature for a set of simulated systems. On the other hand, the presence of high angular momentum in some of our galaxies does not seem to have an impact on the gradient strengths. The strongest gradients in our sample are observed in systems that are likely to have experienced a past merger event. Excluding such merger candidates, the analysed dwarf galaxies show mild gradients (i.e., $\sim -0.1$~dex\,$R_e^{-1}$) with little scatter between them, regardless of their stellar mass, dynamical state, and their star formation history. These results are in good agreement with different sets of simulations presented in the literature and analysed using the same method as the observed dwarf galaxies.}
   {The interplay between the multitude of factors that could drive the formation of metallicity gradients likely combine in complex ways to produce in general comparable mild $\nabla_{\rm [Fe/H]} (R/R_e)$ values, regardless of stellar mass and star formation history. The strongest driver of steep gradients seems to be dwarf-dwarf merger events in a system's past.}


   \maketitle
%

\section{Introduction}

In the Local Group (LG), a morphology-density relation \citep[e.g.,][]{VanDenBergh1994} is observed: gas-poor, passively evolving dwarf galaxies (also known as dwarf spheroidals, or dSph) are preferentially satellites of the Milky Way (MW) or M31, while gas-rich systems (dwarf irregulars, dIrr, and transition types, dTr) are typically found in isolation. Although exceptions are found in both groups (e.g., the Magellanic Clouds; a few isolated dSphs, like Cetus and Tucana), the question to what extent has the environment influenced the emergence of the different morphological types, and the possible evolutionary link between them, remains open. 

The analysis of the star formation histories (SFH) of LG dwarf galaxies from deep colour-magnitude diagrams has revealed that their current morphology does not necessarily reflect their past evolution. For instance, some satellite dSphs have an SFH that differs from that of an isolated gas-rich dwarf only in the last Gyr, probably due to a late fall in the potential of their host galaxy \citep{Gallart2015}. This suggests that the evolutionary path of LG dwarf galaxies may have been imprinted by the initial conditions under which they formed. In addition, LG dwarf galaxies do not show significant differences among themselves in terms of scaling relations: they follow the same trend in stellar mass and size \citep{Tolstoy2009}, as well as in stellar mass and average metallicity \citep{Kirby2013}. The length of their SFH also appears to be broadly correlated with the stellar mass, with higher-mass systems having formed stars for a longer period compared to the lower-mass ones, which may show a wider range of observed SFHs at a given stellar mass (see e.g., \citealp{Tolstoy2009,Weisz2015}, but also \citealp{Cole2014}).

To shed light on the different formation scenarios and the main mechanisms involved in the evolution of dwarf galaxies, detailed studies of their internal properties are needed.
The stellar metal content of galaxies can be in this respect a powerful observational parameter to obtain information on the mechanisms involved in their evolution. 
In particular, the inspection of their radial metallicity profiles can provide important clues on the interplay between internal dynamics and star formation processes, as well as on possible external perturbations. 

Numerical hydrodynamic simulations have recently reached the necessary resolution to study the physical processes of baryonic matter within simulated dwarf galaxies. They have shown that for an isolated dwarf galaxy, factors such as its star formation history, mass, angular momentum, and specific accretion events, can all contribute to producing a radial metallicity gradient \citep[e.g.,][]{Marcolini2008,Schroyen2013,Benitez-Llambay2016,Revaz+Jablonka2018,Mercado2021}. However, it remains unclear if metallicity gradients form mainly from star formation being more prolonged in the central regions than in the outskirts of the galaxy \citep{Schroyen2013,Revaz+Jablonka2018}, or by recurrent stellar feedback events pushing the old and metal-poor stars outwards with time (\citealp{Mercado2021}; see also \citealp{Pontzen+Governato2014}).

For satellite galaxies, instead, it is possible that tidal and ram-pressure stripping could alter the radial metallicity profiles, although their impact could be rather complex to reconstruct. In fact, depending on infall time, orbital history and gas content, a satellite can have its star formation stopped or centrally reignited, as well as have its stellar orbits modified. All these factors could both strengthen or weaken the metallicity gradient \citep[based on][]{Mayer2007,Sales2010,Hausammann2019,Fillingham2019arXiv,Miyoshi+Chiba2020,DiCintio2021}.
It is therefore reasonable to expect that satellite galaxies should exhibit a larger spread of metallicity gradients than isolated dwarfs.

Stellar metallicity gradients appear to be a common characteristic in LG dwarf galaxies. They have been observed both in MW and M31 satellites \citep[e.g.,][]{Harbeck2001,Tolstoy2004,Battaglia2006,Battaglia2011,Gullieuszik2009,Kirby2011,Ho2015,Spencer2017,Pace2020}, as well as in some systems found in isolation \citep{Kirby2013Erratum,Kirby2013,Kirby2017,Leaman2013,Kacharov2017,Taibi2018,Taibi2020,Hermosa2020}. However, only few studies have carried out a systematic characterisation of metallicity gradients in LG dwarf galaxies, or conducted theoretical studies on the physical processes involved in the formation of such gradients. The reason is mainly historical, as only in the last two decades it has been possible to carry out extensive observing campaigns for many of the dwarf galaxies inside and outside the MW virial radius, which have provided sizeable and spatially extended spectroscopic data-sets (see e.g., the sources considered in this work and listed in Table~\ref{tab:sample}).

A first attempt in studying the radial metallicity properties of dwarf galaxies was conducted by \citet{Kirby2011} who analysed a sample of 8 MW satellite dSphs and found that only half of them showed significantly negative slopes. However, their sample was too limited in angular extent to draw firm conclusions about the presence of radial metallicity gradients. Another notable past analysis of the metallicity gradients in LG dwarf galaxies was conducted by \citet{Leaman2013}. Using the same stellar tracers (i.e red giant branch stars), they compared the metallicity profile of the isolated WLM dIrr to that of 8 MW-satellites (6 dSphs plus the Magellanic Clouds), finding a dichotomy between the dSphs, showing decreasing trends, and the dIrrs, having instead flat profiles. They attributed this difference to the presence of internal stellar rotation in the dIrrs, which could form a centrifugal barrier preventing gas to efficiently move towards their centre and thus to build a radially decreasing metallicity profile as observed in the dSphs \citep[see e.g.,][]{Schroyen2011,Schroyen2013}. Their results were reinforced by the  shallow metallicity profiles observed in M31's dEs, which also showed signs of internal rotation \citep[see e.g.,][]{Geha2006,Geha2010}. However, \citet{Leaman2013} noted that systems with high angular momentum also tend to have a larger mass and prolonged SFHs, so understanding which process is causal is difficult. Hence, large samples covering a range of parameters of interest are clearly needed to tease out the interdependence of attributes between dwarf galaxies.

The presence of a strong metallicity gradient in both the Phoenix dTr and the Andromeda~II dSph, unique examples of LG dwarf galaxies showing prolate stellar rotation \citep{Ho2015,Kacharov2017}, has been instead attributed to a possible gas-accretion or dwarf-dwarf merger event \citep[see also][]{Amorisco2014,delPino2017,Cardona-Barrero2021}. An early merger between two dwarf galaxies can have the effect of dispersing the old, metal-poor stellar populations, while a late accretion adds gas reigniting star formation in the galaxy centre, eventually leading to the formation of a young, metal-rich population and thus a negative metallicity gradient \citep[see][]{Benitez-Llambay2016}.

In this study, we present results of an homogeneous analysis performed on publicly available spectroscopic catalogues of red giant stars in 30 LG dwarf galaxies. Our aim is to explore the spatial variation of their [Fe/H] measurements and to look for correlations with their physical properties, in order to possibly distinguish between different mechanisms leading to the formation of metallicity gradients. To date, this is the largest compilation of this type. 
The article is structured as follows. In Sect.~\ref{sec:sample} we present the sample of dwarf galaxies we analysed, together with their associated spectroscopic data-sets of stellar metallicities. In Sect.~\ref{sec:method} we describe the method we followed for obtaining the radial variation of the stellar metallicities and to calculate the metallicity gradients. Section~\ref{sec:analysis} is dedicated to the analysis of the gradient strengths compared to different quantities: host distance (Sect.~\ref{subsec:met-grad-dist}), stellar mass/luminosity (Sect.~\ref{subsec:met-grad-lum}), star formation timescales (Sect.~\ref{subsec:met-grad-sfh}), and angular momentum (Sect.~\ref{subsec:met-grad-rotation}). In Sect.~\ref{subsec:met-grad-mergers} we further inspect the role of mergers in the formation of strong gradients. 
Section~\ref{sec:met-grad-simul} is dedicated to the detailed comparison with different simulation sets from the literature. 
Finally, Sect.~\ref{sec:end} is for the summary and our conclusions, while the appendices contain results of the consistency tests performed and supporting material for the main analysis.


\section{Sample and spectroscopic data-sets}
\label{sec:sample}

\begin{table*}
\caption{Sample of the Local Group systems analysed in this work, together with the number of probable member stars within each data-set, the corresponding morphological type (dwarf spheroidal - dSph; compact/dwarf elliptical - cE/dE; dwarf irregular or transition type - dIrr and dTr), the method used to obtain the metallicities (by direct fit of the available Fe~I absorption lines - SpF; by full spectral-fitting using synthetic spectra - F-SpF; applying the (semi-)empirical calibration between the equivalent width of the near infrared Ca~II triplet lines and [Fe/H] - labelled as CaT) in each of the cited works reported in the last column.\\ 
\textbf{Notes:} for Sextans and Sculptor, we adopt the membership criteria as in \citet{Cicuendez2018} and \citet{Iorio2019}, respectively.} 
\label{tab:sample}
\centering
\begin{tabular}{lcccc}
\hline
\hline
  \multicolumn{1}{c}{Galaxy} &
  \multicolumn{1}{c}{Morph. Type} &
  \multicolumn{1}{c}{N} &
  \multicolumn{1}{c}{Method} &
  \multicolumn{1}{c}{Reference} \\
\hline
  \multicolumn{5}{c}{Milky Way satellites} \\\hline
  SMC                      & dIrr & 3032 & CaT   & \citet{Dobbie2014} \\
  Fornax (Fnx)             & dSph & 822  & CaT   & \citet{Battaglia+Starkenburg2012} \\
  Leo~I                    & dSph & 813  & SpF   & \citet{Kirby2013} \\
  Sculptor (Scl)           & dSph & 539  & CaT   & \citet{Battaglia+Starkenburg2012} \\
  Leo~II                   & dSph & 258  & SpF   & \citet{Kirby2013} \\
  Carina (Car)             & dSph & 392  & CaT   & \citet{Koch2006} \\
  Sextans (Sxt)            & dSph & 162  & CaT   & \citet{Battaglia2011} \\
  Antlia~II (AntlII)       & dSph & 234  & CaT   & \citet{Ji2021} \\
  Ursa Minor (UMi)         & dSph & 718  & SpF   & \citet{Pace2020} \\
  Draco (Dra)              & dSph & 455  & F-SpF & \citet{Walker2015} \\
  Canes Venatici~I (CVenI) & dSph & 163  & SpF   & \citet{Kirby2013} \\
  Crater~II (CratII)       & dSph &  81  & CaT   & \citet{Ji2021} \\\hline
  \multicolumn{5}{c}{M31 satellites} \\\hline
  M32                      & cE   &  64  & CaT   & \citet{Ho2015} \\
  NGC~205 (N205)           & dE   & 224  & CaT   & \citet{Ho2015} \\
  NGC~185 (N185)           & dE   & 321  & CaT   & \citet{Ho2015} \\
  NGC~147 (N147)           & dE   & 230  & CaT   & \citet{Ho2015} \\
  Andromeda~VII (AndVII)   & dSph & 104  & CaT   & \citet{Ho2015} \\
  Andromeda~II (AndII)     & dSph & 300  & CaT   & \citet{Ho2015} \\
  Andromeda~V (AndV)       & dSph &  81  & SpF   & \citet{Kirby2020} \\\hline
  \multicolumn{5}{c}{Local Group Isolated} \\\hline
  NGC~6822 (N6822)         & dIrr & 279  & SpF   & \citet{Kirby2013} \\
  IC~1613                  & dIrr & 296  & CaT   & Taibi et al. (in prep.)\\
  WLM                      & dIrr & 126  & CaT   & \citet{Leaman2013} \\
  UGC~4879 (VV124)         & dTr  &  63  & SpF   & \citet{Kirby2013Erratum} \\
  Leo~A                    & dIrr & 113  & SpF   & \citet{Kirby2017} \\
  Pegasus-DIG (PegDIG)     & dIrr &  95  & SpF   & \citet{Kirby2013} \\
  Sagittarius-DIG (SagDIG) & dIrr &  43  & SpF   & \citet{Kirby2017} \\
  Cetus (Cet)              & dSph &  54  & CaT   & \citet{Taibi2018} \\
  Aquarius (Aqu)           & dIrr &  45  & CaT   & \citet{Hermosa2020} \\
  Phoenix (Phx)            & dTr  & 193  & CaT   & \citet{Kacharov2017} \\
  Tucana (Tuc)             & dSph &  54  & CaT   & \citet{Taibi2020} \\
\hline
\hline
\end{tabular}
\end{table*}

The Local Group dwarf galaxies considered in this work are listed in Table~\ref{tab:sample}, together with the number of probable member stars and sources of the samples of metallicity ([Fe/H]) measurements. We used publicly available catalogues, analysing systems with $\gtrsim 50$ individual member stars with [Fe/H] measurements. For the Milky Way satellites we could select all the known systems from the Small Magellanic Cloud (SMC) down in luminosity to Crater~II (excluding the tidally disrupted Sagittarius dSph), forming a set of 12 dwarf galaxies.
For the M31 satellites we included the six brightest systems (excluding IC~10, for which the needed data are not available) and with Andromeda~V. For the isolated systems we included all those known down to the luminosity of Tucana and out to the distance of VV~124 (excluding the Antlia-Sextans group), for a set of 11 dwarf galaxies. 

In all the literature works a membership classification was available, either as binary classification or a continuous probability (in which case we took targets with a probability $>0.95$). The only exception was the source for Draco; however, the authors provide broad membership selection criteria, which we adopted \citep[see Fig.~10 in][]{Walker2015}. Further, we excluded the spectroscopic probable members which have a  probability of membership $<0.05$ in \citet{Battaglia2022}. In that work, the probabilities of membership were based on a joint analysis of the spatial distribution, and location on the colour-magnitude and proper motion plane of sources detected in {\it Gaia}-eDR3 \citep{Gaia-eDR3_2021}. This step removes few stars per system (between 0 and 15, i.e., $\lesssim1\%$ of the starting samples), showing the goodness of the initial membership selection. 

The celestial coordinates, distances, morphological types, and structural parameters that we adopted were mostly those compiled by \citet{Battaglia2022}. For those systems not included there, we adopted the values from the references in Table~\ref{tab:data_ref}. For the choice of the half-light radii, as in \citet{Battaglia2022}, we gave preference to those studies tracing the galaxies' surface density profiles from resolved RGB (and AGB) stars, to make the radial metallicity distribution of systems with and without recent star formation directly comparable. The half-light radii used for our sample represent the semi-major axis of the ellipse (projected onto the sky) that encloses half of the light.

The type of stars for which [Fe/H] measurements are available across the galaxies are mainly RGB stars\footnote{The catalogues of Draco and Ursa Minor include also red horizontal branch stars. For all the galaxies it is likely that the samples also include a fraction of asymptotic giant branch stars.}. 
However, metallicities were obtained in different ways depending on the study under consideration, namely by direct fit of the available Fe~I absorption lines, by full spectral-fitting using synthetic spectra, or by applying the (semi-)empirical calibration between the equivalent width (EW) of the near infrared Ca~II triplet (CaT) lines and [Fe/H] (see Table~\ref{tab:sample}). 

The validity of the CaT method in its application to stars in complex stellar populations as those of dwarf galaxies has been widely addressed in the literature \citep[e.g.,][]{Cole2004,Pont2004,Battaglia2008, Starkenburg2010, Carrera2013}. Being an empirical method, different calibrations have been proposed and the selected data-sets are not homogeneous in this sense. Among those using the CaT method, 18 out of 20 used either the \citet{Starkenburg2010} or the \citet{Carrera2013} calibrations to obtain metallicities, while metallicity estimates were made on the \citet{Carretta+Gratton1997} scale for the remaining two systems.

Differences between calibration methods are generally small and mainly limited to systematic shifts in the range $-2.5<{\rm [Fe/H]}<-0.5$, where most of dwarf galaxies measurements are distributed \citep[see, e.g., discussion in][]{Battaglia2008}. Nevertheless, they can have an impact on the metallicity distributions, and radial trends can also potentially be affected \citep[see e.g.,][in particular their Figs.~2 and 3]{Leaman2013}. The \citet{Starkenburg2010} and \citet{Carrera2013} calibrations, which we recall are used in 90\% of systems with CaT-metallicites, provide comparable results in the range $-3.5<{\rm [Fe/H]}<-0.5$ with no significant shifts between them \citep[see e.g.,][]{Carrera2013,Kacharov2017}; therefore we expect a negligible impact on the radial [Fe/H] distributions.
For the remaining two systems (i.e., Carina and the SMC), we were able to recalculate their metallicities using the \citet{Starkenburg2010} and \citet{Carrera2013} calibrations, concluding that the recovered differences have no impact on our analysis, so we continue using the original values. We refer to Appendix~\ref{sec:apx0}, for further details.

Regarding metallicities obtained with different methodologies (CaT vs spectral fitting, or between different spectral fitting techniques), a general agreement was demonstrated in works focusing on specific galaxies \citep[see e.g.,][]{Ho2015, Walker2015, Hermosa2020, Pace2020}.

Since several other galaxies in our sample have been included in different works, but in some cases using different spectroscopic data-sets, which also implies [Fe/H] values derived with other methodologies, we conducted additional sanity checks. We verified that the galaxies' global metallicities that we calculated (i.e., their weighted average [Fe/H]) are in good agreement with those expected according to the \citet[][]{Kirby2013} stellar luminosity-metallicity relation (see Fig.~\ref{fig:lum_met} in Appendix~\ref{sec:apx0}). In addition, we do not recover significant offsets with respect to such relation for those galaxies with [Fe/H] measurements derived with the CaT calibration method (see further details in Appendix~\ref{sec:apx0}). Such result shows how possible variations in the global [Ca/Fe] ratio, to which the CaT-method could be sensitive to \citep[see e.g.,][]{Mucciarelli2012}, do not have a significant impact in the determination of the average metallicity properties.

\section{Radial variation of the stellar metallicity properties}
\label{sec:method}

\begin{table*}
\caption{Assumed half-light radii (Col.~2), distances (Col.~3), V-band luminosities (Col.~4), and star formation timescales (i.e., $t_{50}$ and $t_{90}$; Cols.~5-6) for the samples of galaxies analysed in this work (Col.~1), together with their calculated central [Fe/H] values (i.e., the initial value of the GPR curves; Col.~7) and [Fe/H] gradients (Cols.~8-9). 
Values in columns 3-4 obtained from the V-band magnitudes listed in the updated catalogue (January~2021) of \citet{McConnachie2012} and the distance modules listed in \citet{Battaglia2022} and the references of Table~\ref{tab:data_ref}; Cols.~5-6 from the compilation of \citet{Weisz2014}, except for values reported in \citet{Albers2019} and \citet{Bettinelli2018}.}
\label{tab:met_grad}
\centering
\begin{tabular}{ccccrrcrr}
\hline
\hline
  \multicolumn{1}{c}{Galaxy} &
  \multicolumn{1}{c}{$R_e$} &
  \multicolumn{1}{c}{$D$} &
  \multicolumn{1}{c}{$L_V$} &
  \multicolumn{1}{c}{log($t_{90}/yr$)} &
  \multicolumn{1}{c}{log($t_{50}/yr$)} &
  \multicolumn{1}{c}{${\rm [Fe/H]}_0$} &
  \multicolumn{1}{c}{$\nabla_{\rm [Fe/H]} (R/R_e)$} &
  \multicolumn{1}{c}{$\nabla_{\rm [Fe/H]} (R)$} \\
  \multicolumn{1}{c}{} &
  \multicolumn{1}{c}{[arcmin]} &
  \multicolumn{1}{c}{[kpc]} &
  \multicolumn{1}{c}{[$10^6 L_\odot$]} &
  \multicolumn{2}{c}{[dex]} &
  \multicolumn{1}{c}{[dex]} &
  \multicolumn{1}{c}{[dex $R_e^{-1}$] } &
  \multicolumn{1}{c}{[dex kpc$^{-1}$]} \\
\hline 
SMC     & 81.0  & $64   \pm 4 $ &  461   $\pm$ 99    &  --                       &  --                       & -0.94  &  $-0.101 \pm 0.006$ & $-0.067 \pm 0.004 $ \\
Fnx     & 18.5  & $139  \pm 3 $ &  18.5  $\pm$ 2.5   &  $9.38 ^{+0.14}_{-0.06}$  &  $9.87 ^{+0.05}_{-0.07}$  & -0.92  &  $-0.23  \pm 0.02 $ & $-0.31  \pm 0.03  $ \\
LeoI    & 3.53  & $269  \pm 12$ &  5.0   $\pm$ 1.4   &  $9.23 ^{+0.05}_{-0.05}$  &  $9.73 ^{+0.11}_{-0.07}$  & -1.32  &  $-0.09  \pm 0.01 $ & $-0.33  \pm 0.04  $ \\
Scl     & 12.4  & $84   \pm 2 $ &  1.74  $\pm$ 0.23  &  $10.03^{+0.05}_{-0.17}$  &  $10.08^{+0.02}_{-0.01}$  & -1.58  &  $-0.14  \pm 0.01 $ & $-0.45  \pm 0.04  $ \\
LeoII   & 2.46  & $217  \pm 11$ &  0.58  $\pm$ 0.06  &  $9.81 ^{+0.05}_{-0.04}$  &  $9.93 ^{+0.04}_{-0.06}$  & -1.53  &  $-0.21  \pm 0.03 $ & $-1.36  \pm 0.19  $ \\
Car     & 10.2  & $106  \pm 6 $ &  0.51  $\pm$ 0.06  &  $9.46 ^{+0.11}_{-0.12}$  &  $9.69 ^{+0.28}_{-0.01}$  & -1.69  &  $-0.04  \pm 0.03 $ & $-0.12  \pm 0.09  $ \\
Sxt     & 21.4  & $85   \pm 1 $ &  0.42  $\pm$ 0.20  &  $10.06^{+0.02}_{-0.01}$  &  $10.10^{+0.02}_{-0.01}$  & -2.02  &  $-0.32  \pm 0.05 $ & $-0.60  \pm 0.09  $ \\
AntlII  & 76.2  & $132  \pm 7 $ &  0.35  $\pm$ 0.06  &  --                       &  --                       & -1.75  &  $-0.15  \pm 0.07 $ & $-0.05  \pm 0.02  $ \\
UMi     & 18.2  & $76   \pm 4 $ &  0.35  $\pm$ 0.04  &  $9.96 ^{+0.07}_{-0.18}$  &  $9.99 ^{+0.10}_{-0.01}$  & -2.02  &  $-0.15  \pm 0.02 $ & $-0.37  \pm 0.06  $ \\
Dra     & 9.61  & $81   \pm 3 $ &  0.29  $\pm$ 0.02  &  $10.01^{+0.06}_{-0.11}$  &  $10.06^{+0.04}_{-0.01}$  & -1.86  &  $-0.07  \pm 0.01 $ & $-0.32  \pm 0.07  $ \\
CVnI    & 7.5   & $211  \pm 5 $ &  0.22  $\pm$ 0.04  &  $9.92 ^{+0.06}_{-0.11}$  &  $10.10^{+0.00}_{-0.10}$  & -1.91  &  $ 0.00  \pm 0.09 $ & $ 0.00  \pm 0.19  $ \\
CratII  & 31.2  & $117  \pm 4 $ &  0.16  $\pm$ 0.02  &  --                       &  --                       & -2.16  &  $ 0.00  \pm 0.09 $ & $ 0.00  \pm 0.09  $ \\
M32     & 0.47  & $805  \pm 78$ &  319   $\pm$ 68    &  $9.23 ^{+0.49}_{-0.01}$  &  $10.10^{+0.00}_{-0.00}$  & -1.19  &  $ 0.00  \pm 0.05 $ & $ 0.00  \pm 0.43  $ \\
N205    & 2.46  & $824  \pm 27$ &  334   $\pm$ 38    &  $9.34 ^{+0.47}_{-0.04}$  &  $9.94 ^{+0.16}_{-0.11}$  & -0.85  &  $ 0.00  \pm 0.02 $ & $ 0.00  \pm 0.03  $ \\
N185    & 2.94  & $619  \pm 20$ &  68.5  $\pm$ 8.0   &  $9.56 ^{+0.40}_{-0.05}$  &  $9.97 ^{+0.13}_{-0.02}$  & -0.44  &  $-0.20  \pm 0.03 $ & $-0.37  \pm 0.06  $ \\
N147    & 6.7   & $711  \pm 20$ &  68.5  $\pm$ 7.0   &  $9.43 ^{+0.53}_{-0.01}$  &  $9.79 ^{+0.31}_{-0.01}$  & -0.54  &  $ 0.00  \pm 0.07 $ & $ 0.00  \pm 0.05  $ \\
AndVII  & 3.5   & $762  \pm 35$ &  16.5  $\pm$ 5.0   &  $9.75 ^{+0.07}_{-0.05}$  &  $10.10^{+0.00}_{-0.00}$  & -1.26  &  $ 0.00  \pm 0.14 $ & $ 0.00  \pm 0.18  $ \\
AndII   & 6.2   & $631  \pm 15$ &  8.5   $\pm$ 1.6   &  $9.75 ^{+0.07}_{-0.05}$  &  $9.99 ^{+0.11}_{-0.16}$  & -0.77  &  $-0.39  \pm 0.07 $ & $-0.34  \pm 0.06  $ \\
AndV    & 1.4   & $741  \pm 22$ &  0.5   $\pm$ 0.1   &  $9.87 ^{+0.23}_{-0.31}$  &  $10.04^{+0.06}_{-0.06}$  & -1.79  &  $-0.10  \pm 0.05 $ & $-0.33  \pm 0.17  $ \\
N6822   & 11.95 & $470  \pm 37$ &  109   $\pm$ 26    &  $9.00 ^{+0.10}_{-0.01}$  &  $9.61 ^{+0.22}_{-0.17}$  & -0.89  &  $-0.46  \pm 0.13 $ & $-0.28  \pm 0.08  $ \\
IC1613  & 7.57  & $759  \pm 5 $ &  103   $\pm$ 10    &  $9.30 ^{+0.09}_{-0.02}$  &  $9.87 ^{+0.0 }_{-0.06}$  & -1.10  &  $-0.06  \pm 0.08 $ & $-0.04  \pm 0.05  $ \\
WLM     & 4.1   & $933  \pm 34$ &  43    $\pm$ 5     &  $9.02 ^{+0.06}_{-0.01}$  &  $9.71 ^{+0.03}_{-0.07}$  & -1.27  &  $ 0.00  \pm 0.05 $ & $ 0.00  \pm 0.04  $ \\
VV124   & 1.15  & $1324 \pm 79$ &  7.9   $\pm$ 1.7   &  --                       &  --                       & -1.25  &  $-0.12  \pm 0.04 $ & $-0.28  \pm 0.10  $ \\
LeoA    & 2.3   & $718  \pm 17$ &  4.8   $\pm$ 0.9   &  $8.93 ^{+0.06}_{-0.02}$  &  $9.61 ^{+0.01}_{-0.03}$  & -1.39  &  $-0.16  \pm 0.05 $ & $-0.32  \pm 0.10  $ \\
PegDIG  & 3.81  & $759  \pm 70$ &  4.5   $\pm$ 1.2   &  $9.23 ^{+0.82}_{-0.10}$  &  $10.10^{+0.00}_{-0.74}$  & -1.21  &  $-0.29  \pm 0.12 $ & $-0.34  \pm 0.14  $ \\
SagDIG  & 1.43  & $1067 \pm 88$ &  3.5   $\pm$ 0.9   &  $8.91 ^{+0.35}_{-0.30}$  &  $9.82 ^{+0.28}_{-0.33}$  & -1.84  &  $ 0.00  \pm 0.08 $ & $ 0.00  \pm 0.18  $ \\
Cet     & 3.2   & $755  \pm 24$ &  2.8   $\pm$ 0.5   &  $9.95 ^{+0.04}_{-0.05}$  &  $10.05^{+0.04}_{-0.03}$  & -1.55  &  $-0.10  \pm 0.05 $ & $-0.14  \pm 0.06  $ \\
Aqu     & 1.63  & $1072 \pm 40$ &  1.6   $\pm$ 0.2   &  $9.33 ^{+0.09}_{-0.02}$  &  $9.85 ^{+0.03}_{-0.02}$  & -1.53  &  $-0.08  \pm 0.08 $ & $-0.15  \pm 0.16  $ \\
Phx     & 2.3   & $409  \pm 23$ &  0.75  $\pm$ 0.30  &  $9.42 ^{+0.03}_{-0.02}$  &  $10.03^{+0.03}_{-0.08}$  & -0.98  &  $-0.35  \pm 0.05 $ & $-1.28  \pm 0.17  $ \\
Tuc     & 1.1   & $887  \pm 49$ &  0.6   $\pm$ 0.1   &  $9.88 ^{+0.04}_{-0.13}$  &  $10.11^{+0.00}_{-0.11}$  & -1.56  &  $-0.07  \pm 0.04 $ & $-0.24  \pm 0.16  $ \\
\hline
\hline
\end{tabular}
\end{table*}

Our main goal is to examine the radial variation of the metallicity ([Fe/H]) properties of the galaxies' stellar component and explore whether the strength of the stellar [Fe/H] gradients correlates with physical properties such as the galaxies' stellar mass, the global age of their stellar population, and their environment. Hereafter, [Fe/H] will be used interchangeably with ``metallicity'', and whenever we refer to a metallicity gradient, we refer to that of the stellar component, unless said otherwise.

In order to follow the radial metallicity variation of a galaxy, we looked for a tool that could find the general radial trend in the data, while being robust to outliers and without the need to choose a functional form a priori (e.g., a linear least-square fit). Other works that have performed a similar analysis \citep[e.g.,][]{Leaman2013} have used a running-boxcar average, but the smoothness of the resulting [Fe/H] profile depends in this case on the size of the boxcar.
We decided instead to apply the Gaussian process regression (GPR) method to our sample (\citealp[see e.g.,][for a previous application to a LG dwarf galaxy in this context]{Hermosa2020}; \citealp[but also][and references therein]{Williams2022}).

The GPR is a non-parametric Bayesian method that allows a data-set to be modelled without making any assumptions about the underlying functional form. In fact, the GPR models the correlation between data-points (in our case a radial correlation in the metallicity distribution) by assuming a covariance kernel. The outcome is a smoothed posterior probability distribution of [Fe/H] values as a function of radius, which allows the calculation of associated confidence intervals. 
We used the \textsc{python} package \texttt{GaussianProcessRegressor} in \texttt{scikit-learn} \citep{scikit-learn}, implementing a Gaussian kernel together with a noise component to take into account the intrinsic scatter of the data. We checked that the choice of initial parameters for the GPR analysis does not lead to local maxima in the log-likelihood function. 

Since the stellar component of LG dwarf galaxies is not spherical, we traced the radial behaviour as a function of projected semi-major axis (SMA) radius (also referred to as ``elliptical'' radius $R$). Qualitatively this choice appears justified by the shape of the 2D metallicity maps of dwarf galaxies in environments other than the LG \citep[e.g.,][]{Rys2013, Sybilska2017} obtained with integral-field-unit spectroscopy, whose spatial coverage is homogeneous and not affected by slit/fibre allocation to the target stars. It is also supported by the behaviour of the 2D median metallicity maps of NIHAO simulated galaxies of stellar masses $< 10^9 M_\odot$ (Cardona-Barrero et al., Subm.). 

\begin{figure*}
    \centering
    \includegraphics[width=\textwidth]{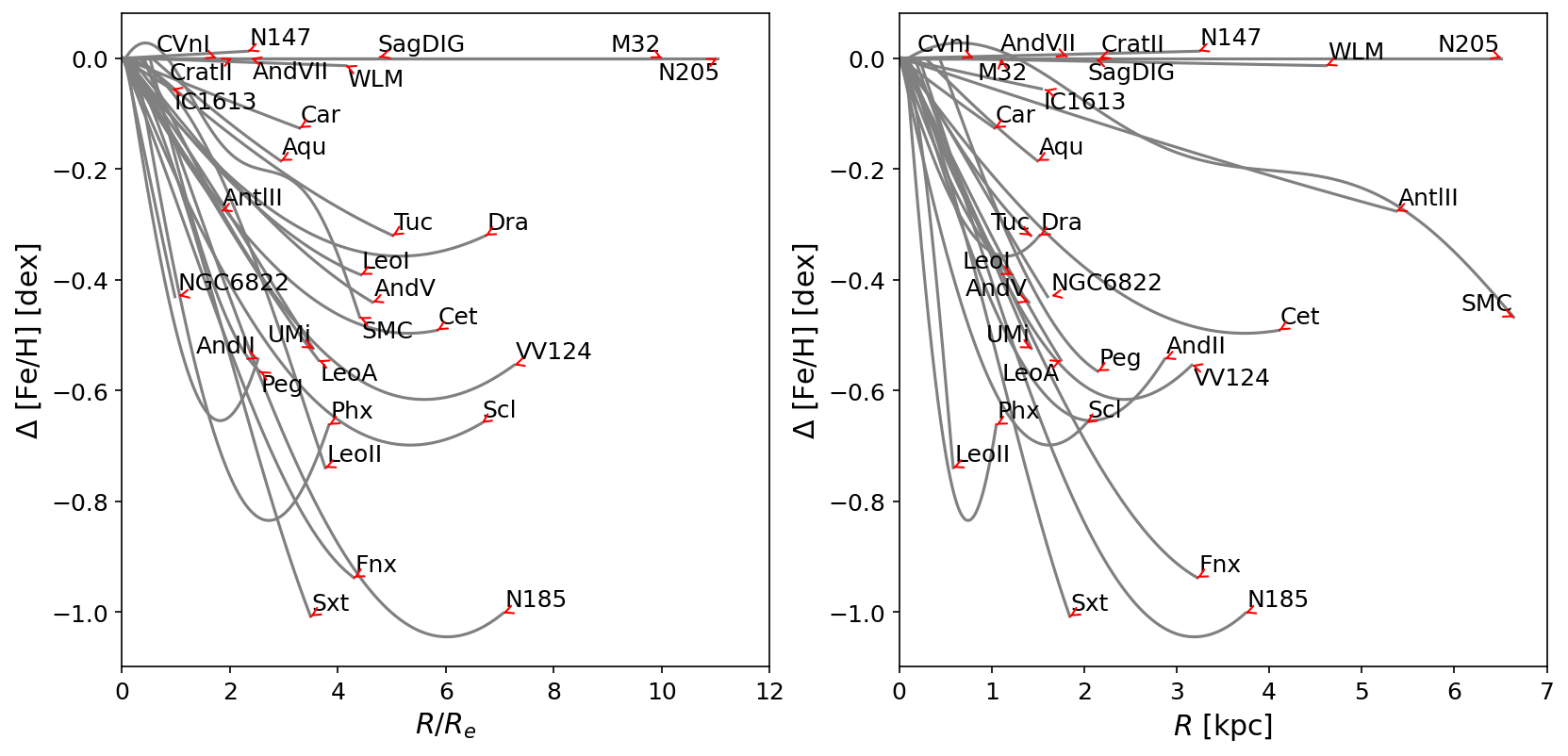}\\
    \includegraphics[width=\textwidth]{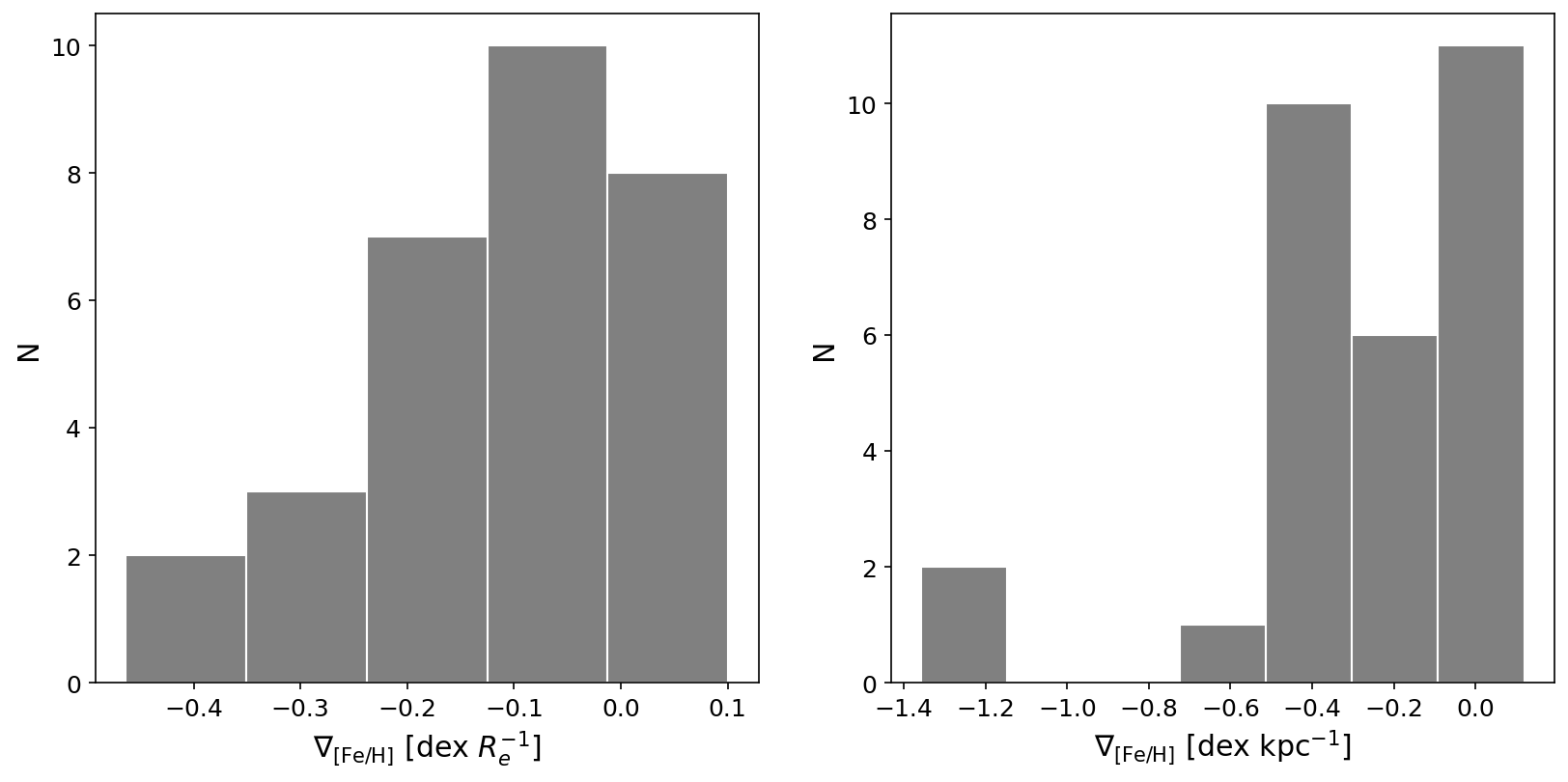}
    \caption{\textit{Top:} The smoothed [Fe/H] profiles as a function of SMA radius as obtained from the GPR fitting method, in units of the 2D half-light radius (left) and as a function of physical radius (right). The radial metallicity profiles have been normalised to their value at $R=0$. Dwarf galaxy names as in Table~\ref{tab:sample}. 
    \textit{Bottom:} distributions of the calculated metallicity gradients in the same units as the respective above panels; bin widths obtained following the Freedman-Diaconis normal reference rule.}
    \label{fig:mgrad_all}
\end{figure*}

Figure~\ref{fig:mgrad_all} shows the radial variation in the metallicity properties as a function of semi-major axis distance from the galaxy's centre (top right) and normalised by the galaxy's 2D half-light radius $R_{\rm e}$ (top left; $R_{\rm e}$ values from the compilation of \citealp{Battaglia2022}, and the references listed in Table~\ref{tab:data_ref}), given that the galaxies in our sample span three orders of magnitude in stellar luminosity and therefore have different sizes (from hundreds pc to few kpc). This is just one possible normalisation, for example other works in the literature have reported values normalised to the King's core radius $R_{\rm core}$ \citep[e.g.,][]{Leaman2013}, while for larger systems as spiral galaxies and ellipticals $R_{25}$ is sometimes used \citep[e.g.,][]{Ho-I-Ting2015}.

As can be appreciated from Fig.~\ref{fig:mgrad_all}, the profiles are either flat or declining with radius; a slight increase is often seen in the outskirts, but comparing the individual [Fe/H] as a function of radius with the GPR profile, this increase appears to be due to over-fitting of these sparse outer bins and can therefore be neglected. Therefore, the galaxies in our sample show either no metallicity gradients or negative metallicity gradients, out to the coverage of the spectroscopic data-sets. We refer to Appendix~\ref{sec:apx1} for a detailed view of the individual [Fe/H] data-sets and the corresponding GPR fits.

We proceeded to quantify the strength of the [Fe/H] gradient. Since we are interested in the average slope of the radial metallicity profile, we performed a linear fit on the smooth function returned by the GPR between $R=0$ and the projected SMA radius where the GPR curve presents a minimum (or at the last measured point, in absence of a minimum). This was motivated by the fact that the GPR curves are mostly linear or parabolic in the considered range. The gradient's value is therefore the slope of the linear fit. Its associated error is the error obtained on the slope parameter by performing a linear fit to the data instead\footnote{Our method for deriving metallicity gradients is mostly equivalent to calculating the derivative of a GPR curve, which is differentiable, over the same interval and obtaining the average of this derivative. This gives us the average slope of the radial metallicity profile, similar to that obtained with the linear fit. The associated error is therefore the error obtained from a linear least-squares fit, assuming the error of the data is that of the GPR curve, which takes into account the uncertainties on the individual metallicity measurements together with the inherent dispersion of the data. This is again equivalent to obtaining the slope error of a least-squares linear fit to the raw data.}.

Given that we are comparing galaxies of different size, we determine the gradient both in units of the 2D SMA half-light radius, $\nabla_{\rm [Fe/H]} (R/R_e)$, and in units of physical radius, $\nabla_{\rm [Fe/H]} (R)$. Results are given in Table~\ref{tab:met_grad}, with their distributions shown in the bottom panels of Fig.~\ref{fig:mgrad_all}, where the number of bins (and therefore their widths) was set following the Freedman-Diaconis normal reference rule\footnote{\url{https://docs.astropy.org/en/stable/visualization/histogram.html}.}. 

The distribution of gradient values cover the interval from $\sim$\,$-0.4$~dex\,$R_e^{-1}$ to $\sim$\,0~dex\,$R_e^{-1}$, or $\sim$\,$-1.3$~dex\,kpc$^{-1}$ to $\sim$\,0~dex\,kpc$^{-1}$, with the majority of the systems found in the range $\sim$\,$-0.2$~dex\,$R_e^{-1}$ to $\sim$\,0~dex\,$R_e^{-1}$, and $\sim$\,$-0.5$~dex\,kpc$^{-1}$ to $\sim$\,0~dex\,kpc$^{-1}$. Median values are $\nabla_{\rm [Fe/H]} (R/R_e) \sim -0.1$~dex\,$R_e^{-1}$ and $\nabla_{\rm [Fe/H]} (R) \sim -0.25$~dex\,kpc$^{-1}$, with median absolute deviation scatter of 0.14~dex\,$R_e^{-1}$ and 0.24~dex\,$R_e^{-1}$, respectively. 

The different sample sizes, spatial coverages and radial trends turn into rather different uncertainties in the recovered gradients, with $\nabla_{\rm [Fe/H]} (R/R_e)$ measures having associated errors ranging from $0.006$~dex\,$R_e^{-1}$ to $0.14$~dex\,$R_e^{-1}$, while for $\nabla_{\rm [Fe/H]} (R)$ errors go from $0.004$~dex\,kpc$^{-1}$ to $0.43$~dex\,kpc$^{-1}$. In particular, the large $\nabla_{\rm [Fe/H]} (R/R_e)$ uncertainty for NGC~6822, despite its high number of individual metallicity measurements, is due to the spectroscopic sample covering an area just inside its half-light radius, even though this spans more than $10\arcmin$ across the sky. 

Finally, to verify that our method of deriving metallicity gradients does not rely too much on the position of the minimum on the GPR curves, we compared our results with those obtained by calculating the gradient as the difference in metallicity between the centre and $2\times R_e$ of the GPR curves. We found a good general agreement (see Fig.~\ref{fig:mgrad-vs} in Appendix~\ref{sec:apx0}), with the only outlier (i.e., at $>3$-$\sigma$) being Sculptor, which was found to have a steeper gradient. Even if we were to assume this value, our conclusions would still remain unchanged.

\section{Comparison of the metallicity gradients}
\label{sec:analysis}

We show how the metallicity gradient strengths compare to different galaxy's properties (host distance, stellar mass, star formation timescales) for the considered sample of LG dwarf galaxies.

\subsection{Trends with environment and morphological types}
\label{subsec:met-grad-dist}

\begin{figure*}
    \centering
    \includegraphics[width=\textwidth]{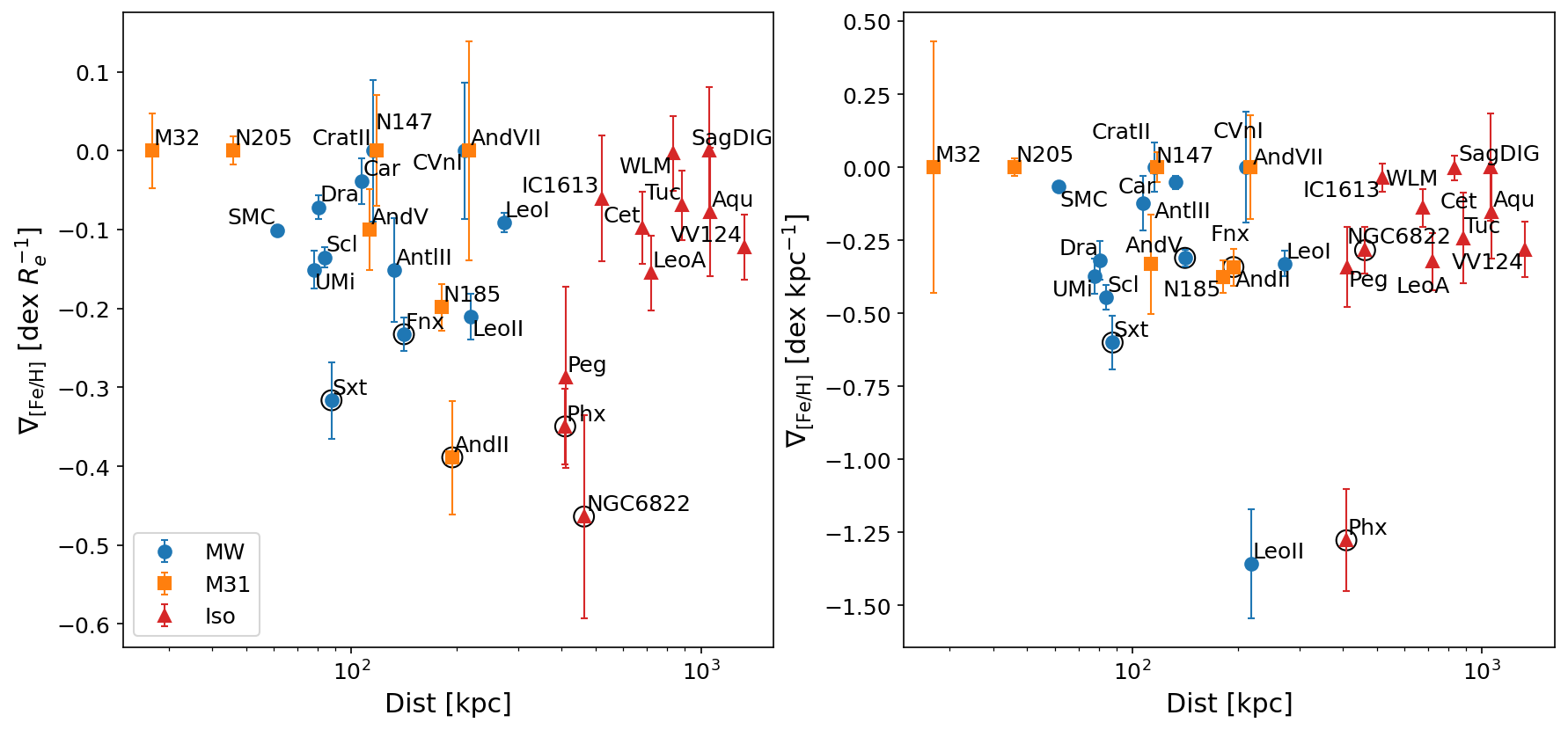}\\
    \includegraphics[width=\textwidth]{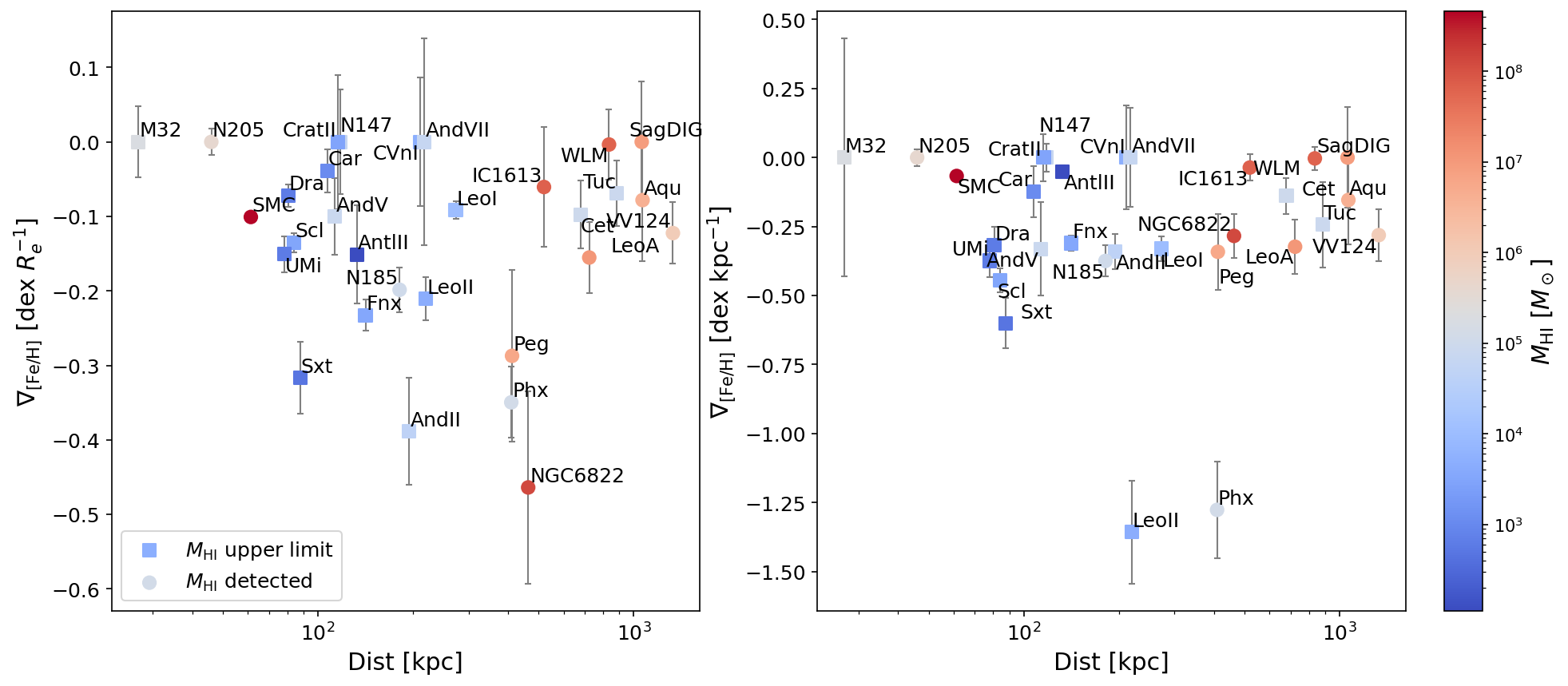}
    \caption{Distribution of metallicity gradients as a function of distance from the nearest large LG spiral. \textit{Top:} satellites of the Milky Way and of M31, and the isolated systems, are marked as blue circles, orange squares, and red triangles, respectively. Dwarf galaxies that have possibly experienced a past merger event are marked with an open circle. \textit{Bottom:} same as above, but with data colour-coded according to their HI content; circles mark a full detection, while squares mark the 5-$\sigma$ upper limit \citep[data from][and references therein]{Putman2021}.
    On the y-axes, the left panels show the metallicity gradients in units of the 2D SMA half-light radius, while on the right panels they are in units of the physical radius.}
    \label{fig:mgrad_Dist}
\end{figure*}

In order to examine the role of the environment, we looked for possible trends with the distance to the nearest large spiral. As shown in Fig.~\ref{fig:mgrad_Dist}, satellite dwarfs as well as isolated systems do not show on visual inspection any particular differences depending on their distance from the MW or M31 centres (see values in Table~\ref{tab:met_grad})\footnote{Distances obtained using distance modules from \citet{Battaglia2022} and the references in Table~\ref{tab:data_ref}. We assume for M31 coordinates and heliocentric distance from \citet{Conn2012}; for the MW, a heliocentric distance of 8.122~kpc \citep{GravityCo2018}, while the Galactic centre coordinates are from \citet{Reid+Brunthaler2004}.}. 
In general, it would appear that the distribution of metallicity gradients (either as $\nabla_{\rm [Fe/H]} (R/R_e)$ or $\nabla_{\rm [Fe/H]} (R)$) of satellite systems has a slightly larger scatter than that of isolated dwarfs. However, by comparing them using the Kolmogorov-Smirnov and Anderson-Darling statistical tests\footnote{Using \texttt{scipy.stats} functions \texttt{ks\_2samp} and \texttt{anderson\_ksamp}.}, the distributions do not show a statistically significant difference. 

We note that the systems with the strongest $\nabla_{\rm [Fe/H]} (R/R_e)$ (i.e., $\lesssim$\,$-0.25$~dex\,$R_e^{-1}$), are present both among the isolated dwarf galaxies and among the satellites; removing them from our analysis does not change our conclusions.
We also point out that the majority of these systems may have experienced a merger event in their past (\citealp[see e.g.,][for the cases of And~II, Phoenix, and Sextans, respectively]{Amorisco2014,Kacharov2017,Cicuendez+Battaglia2018}; \citealp[but also][for the case of Fornax]{Battaglia2006,Amorisco+Evans2012,deBoer2012,delPino2015,Leung2020}; \citealp[and][]{deBlok2000,deBlok2006,Demers2006,Battinelli2006,Hwang2014}, \citealp[but also][]{Zhang2021}, for the case of NGC~6822), and potentially mergers may play a role in driving the formation of strong gradients \citep{Benitez-Llambay2016}. We refer to Sect.~\ref{subsec:met-grad-mergers} for more details on this aspect.

We further looked for differences between the metallicity gradients by grouping our sample according to the galaxies's HI-gas content (see Fig.~\ref{fig:mgrad_Dist}, bottom panels; $M_{\rm HI}$ data from \citealp{Putman2021}). In particular, we looked for statistical differences between the gas-poor systems (i.e., dSph, cE/dE; see Table~\ref{tab:sample}) and the gas-rich ones (the dIrr and dTr), finding none. This is in contrast to the results reported by \citet{Leaman2013}, where they found a dichotomy between the dSphs and dIrrs, with the former showing a radially decreasing metallicity profile, while the latter had a flat profile. We attribute this discrepancy mainly to the limited sample of dwarf galaxies they analysed, which we tripled here.

The above results indicate a limited role of the prolonged environmental interaction between a satellite and its host galaxy as the main actor in the formation of strong radial metallicity gradients in LG dwarf galaxies. 
Whether tidal and ram-pressure stripping eventually have a role in shaping metallicity gradients could depend on several factors, such as the satellite's internal properties and orbital history, the build-up stage of the host, and whether the satellite fell into the host potential before or after its star formation had shut down \citep[based on][]{Mayer2007,Sales2010,Hausammann2019,DiCintio2021}. Although any process in general related to the mixing of stellar orbits could potentially weaken an existing gradient \citep[e.g.,][]{Sales2010}, depending on the orbit, an encounter with an host could trigger central star formation in a gas-containing dwarf galaxy, leading instead to a strengthening of its metallicity gradient (\citealp[e.g.,][]{Hausammann2019,DiCintio2021}; \citealp[see also discussion in][]{Koleva2011}). Overall, it seems that the environment has the potential to increase the scatter of the distribution of metallicity gradients in satellites,  a hint of which is seen in Fig.~\ref{fig:mgrad_Dist}, but so far the statistics are such that we cannot firmly confirm this conclusion.

We further looked for possible correlations between the metallicity gradients of the MW satellites and their quenching timescale (i.e., the time spent by a dwarf galaxy to form 90\% of its stars after its first infall into the MW potential, as calculated by \citealp[]{Fillingham2019arXiv} and \citealp[]{Miyoshi+Chiba2020} using \textit{Gaia}-DR2 proper motions), finding none. We also examined the orbital parameters (i.e. eccentricity, pericentric distance, period, and time since the last pericentric passage) provided by \citet{Battaglia2022} considering different MW-potentials, but our sample is too small (12 dwarf galaxies) to reach definitive conclusions. For reference, we display our results in Appendix~\ref{sec:apx_orb}.

\subsection{Trends with stellar mass/luminosity}
\label{subsec:met-grad-lum}

\begin{figure*}
    \centering
    \includegraphics[width=\textwidth]{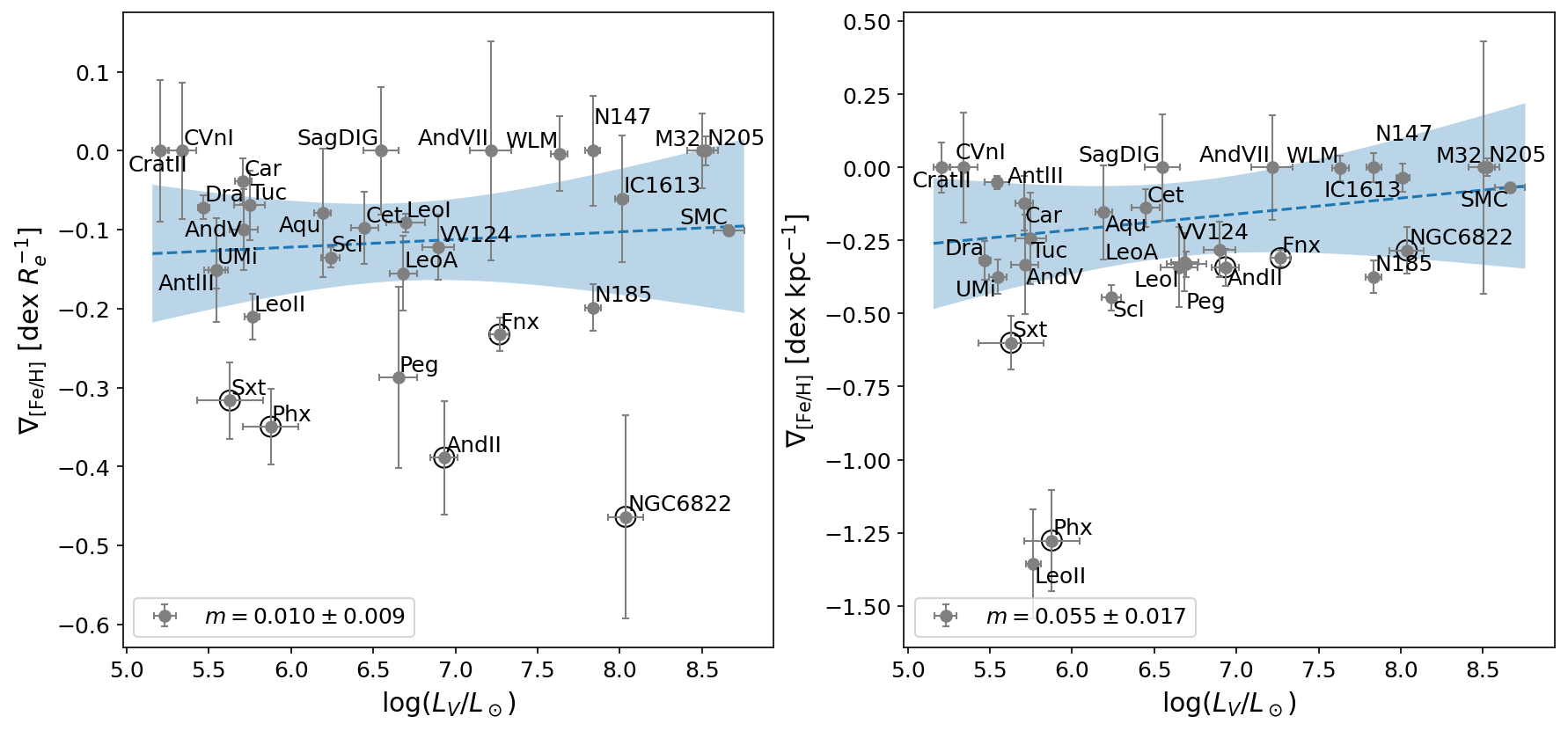}
    \caption{Distribution of metallicity gradients as a function of the galaxy's luminosity in V-band, with the respective uncertainties. The blue dashed lines represent the linear LSQ-fit to the data and the blue shaded areas the 95\% confidence interval. 
    In the legend boxes are reported the slope's value from the LSQ-fits. On the y-axes, the left panel shows the metallicity gradients in units of the 2D SMA half-light radius, while on the right panel they are in units of the physical radius. Dwarf galaxies that have possibly experienced a past merger event are marked with an open circle.}
    \label{fig:mgrad_Lstar}
\end{figure*}

Figure~\ref{fig:mgrad_Lstar} shows an overview of the strength of the metallicity gradient as a function of the galaxy's luminosity in the V-band (see values in Table~\ref{tab:met_grad}). The distribution of $\nabla_{\rm [Fe/H]} (R/R_e)$ displays a large scatter (left panel); on the other hand, $\nabla_{\rm [Fe/H]} (R)$ shows a hint of a relation with $L_V$ (right panel) as shown by the linear fit (with a slope $m=0.05\pm0.02$) and the 95\% confidence interval. This seems to be a manifestation of the fact that more luminous galaxies are in general larger than fainter ones \citep[e.g.,][]{Tolstoy2009}, as it is reflected by the fact that the relation weakens significantly for $\nabla_{\rm [Fe/H]} (R/R_e)$, that is when normalising for the half-light radius, with the confidence interval being consistent with no linear relation ($m=0.01\pm0.01$). 

We reached the same conclusion by replacing $L_V$ with the average metallicity $\left \langle {\rm [Fe/H]} \right \rangle$, recovering a correlation with $\nabla_{\rm [Fe/H]} (R)$ due to the luminosity-metallicity relation \citep[e.g.,][]{Kirby2013}, and with the velocity dispersion $\sigma_v$ -- which is a proxy of the dynamical mass inside the half-light radius $M_{1/2}$ \citep{Walker2009c,Wolf2010}. We are aware that $\sigma_v$ could be a biased mass-indicator in those cases where significant internal rotation is present (i.e., $V_{\rm rot}/\sigma_v \gtrsim0.5$). Accounting for this effect \citep[see e.g.,][]{Kirby2014}, however, would have a negligible effect on the relation between $\nabla_{\rm [Fe/H]} (R)$ and $M_{1/2}$.

Making a qualitative comparison, stellar metallicity gradients have also been detected in early- and late-type field galaxies (having $8.0<{\rm log}(M_*/M_\odot)<11.5$ and distances $z\lesssim0.03$), with only the former showing no clear correlation between $\nabla_{\rm [Fe/H]} (R/R_e)$ and their stellar masses \citep[see e.g.,][]{Koleva2011,Goddard2017}.

On the other hand, zoom-in hydrodynamic simulations of isolated dwarf galaxies are also inconclusive about the presence of a correlation between $\nabla_{\rm [Fe/H]} (R/R_e)$ and stellar mass. \citet{Schroyen2013} and \citet[][hereafter RJ18]{Revaz+Jablonka2018} found that additional factors could have an impact on the formation of metallicity gradients, for example their SFH and the presence of internal rotation.
The absence of a correlation with stellar mass, average metallicity, but also with stellar rotation (although they had few systems with $V_{\rm rot}/\sigma_v>0.5$) was instead reported by \citet[][see their Fig.~A2]{Mercado2021}, who conversely found a tight correlation with the median stellar age of their simulated systems (see Sects.~\ref{subsec:met-grad-sfh}, \ref{subsec:met-grad-rotation} and \ref{sec:met-grad-simul}).

\subsection{Correlation with star formation timescales?}
\label{subsec:met-grad-sfh}

\begin{figure*}
    \centering
    \includegraphics[width=\textwidth]{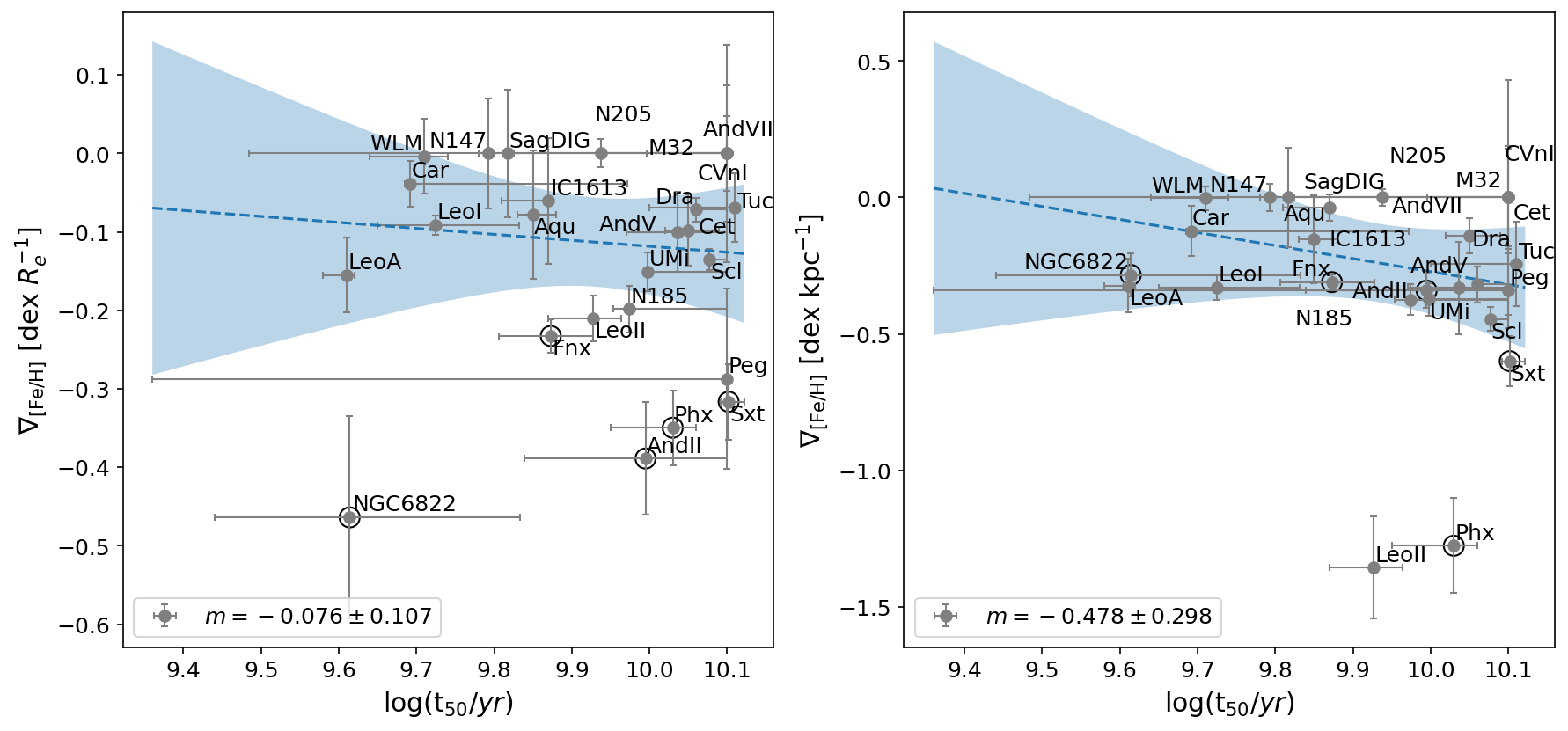}
    \includegraphics[width=\textwidth]{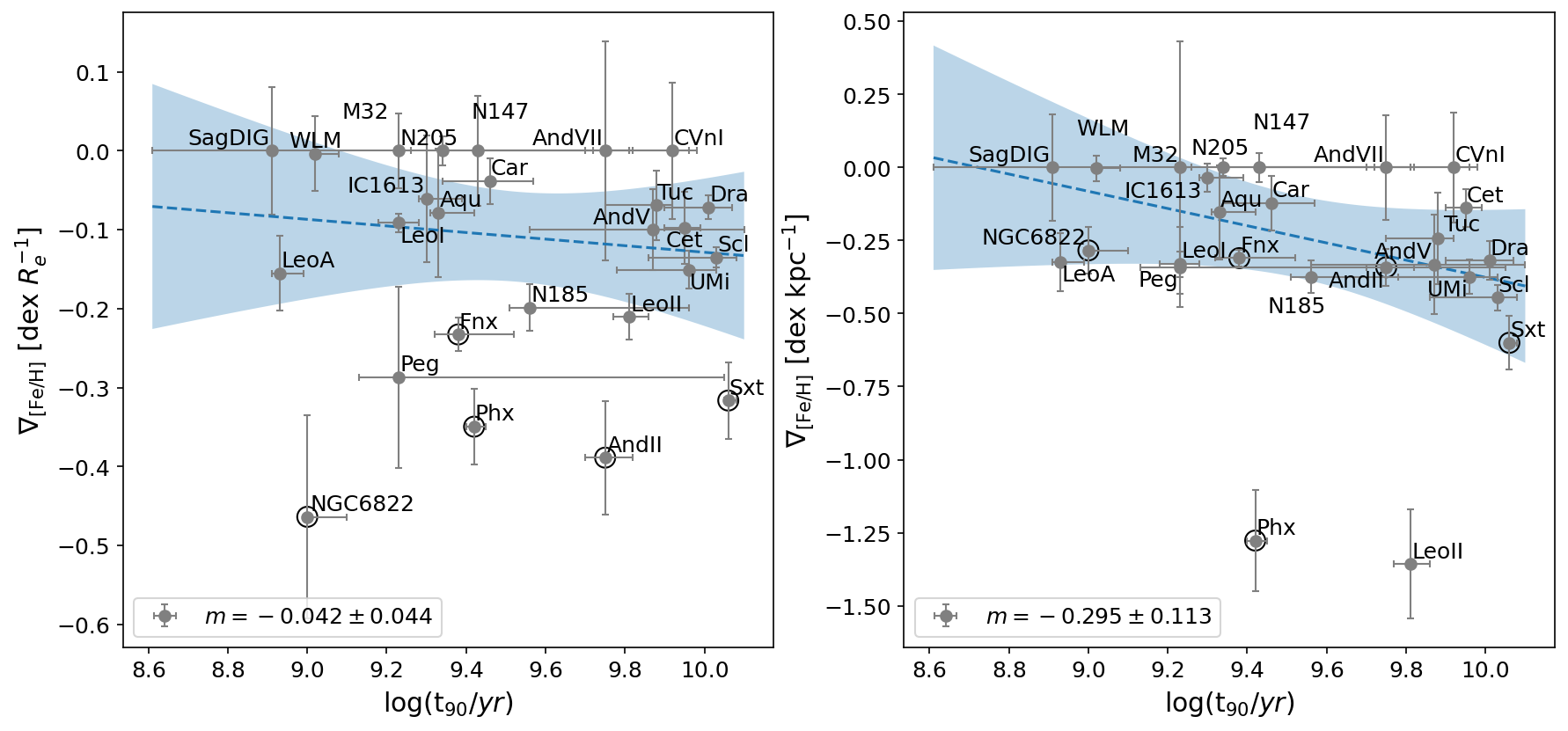}
    \caption{Distribution of metallicity gradients as a function of the $t_{50}$ (\textit{top} panels), and $t_{90}$ (\textit{bottom} panels), for our sample of dwarf galaxies. The blue dashed lines represent the LSQ-fit to the data and the blue shaded areas the associated 95\% confidence interval. On the y-axes, the metallicity gradients in units of the 2D SMA half-light radius (left panels), and in units of the physical radius (right panels). In the legend boxes are reported the slope’s value from the LSQ-fits. Dwarf galaxies that have possibly experienced a past merger event are marked with an open circle.}
    \label{fig:mgrad_t90}
\end{figure*}

To explore possible relations with star formation timescales, we turn to consider the $t_{50}$ and $t_{90}$ quantities that indicate the look-back time at which, respectively, the cumulative stellar mass inferred by the measured SFH reaches 50\% and 90\% of its current total value (see values in Table~\ref{tab:met_grad}).
We used the compilations of \citet{Weisz2014,Weisz2015}, except for those dwarfs analysed by \citet{Bettinelli2018} and \citet{Albers2019}.

Looking at Fig.~\ref{fig:mgrad_t90}, no clear correlations emerge when metallicity gradients are plotted per unit of the 2D half-light radius, either as a function of ${\rm log}(t_{50})$ (top left) or ${\rm log}(t_{90})$ (bottom left). When we consider instead metallicity gradients per units of physical radius (right panels), a stronger relation appears with both $t_{50}$ and $t_{90}$. In particular, for the latter quantity we obtain a significant slope value of $m=-0.29\pm0.11$ from the linear LSQ-fit.
The correlations observed for $\nabla_{\rm [Fe/H]} (R)$ are again a manifestation of the scaling-relations among dwarf galaxies, as seen in the previous section. In this case they are in large part a consequence of the fact that less luminous/massive galaxies (which have smaller $R_e$) show on average shorter SFHs than more luminous/massive galaxies (which have larger $R_e$; \citealp[see e.g.,][]{Weisz2015}). This is reflected by the weak trends observed for $\nabla_{\rm [Fe/H]} (R/R_e)$ (i.e., normalising for $R_e$) that are consistent with no linear relation within the confidence intervals. The same conclusions are reached by expressing $t_{50}$ and $t_{90}$ values in a linear scale instead of a logarithmic one. It is worth noting that some of these measures have large associated errors that could smear any possible correlation, so it would be particularly useful in the future to have access to SFHs that are both homogeneous and more accurate and precise than currently available compilations.

An important result of this analysis is the lack of a clear and strong correlation between $\nabla_{\rm [Fe/H]} (R/R_e)$ and $t_{50}$, in contrast with results obtained by \citet{Mercado2021}. Analysing a set of simulated dwarf galaxies with stellar masses comparable to those of our data-set, \citet{Mercado2021} found a linear relation between the median age and the metallicity gradient of their simulated systems. This result was supported by a restricted sample of 10 LG dwarf galaxies following a similar relationship. As shown in Fig.~\ref{fig:mgrad_t90} (upper left panel) and Fig.~\ref{fig:mgrad_t50_mercado} (right panel), by increasing the observed sample size the clear correlation found by \citet{Mercado2021} practically disappears. 

The link between the length of the star formation period and the emergence of metallicity gradients in dwarf galaxies, on the other hand, has been explored by RJ18, analysing a different set of simulated dwarf galaxies. 

We refer to Sects.~\ref{subsec:met-grad-simul-M21} and \ref{subsec:met-grad-simul-RJ18} for a detailed comparison between these sets of simulations and our analysis.

\subsection{On the role of angular momentum}
\label{subsec:met-grad-rotation}

\begin{figure}
    \centering
    \includegraphics[width=.49\textwidth]{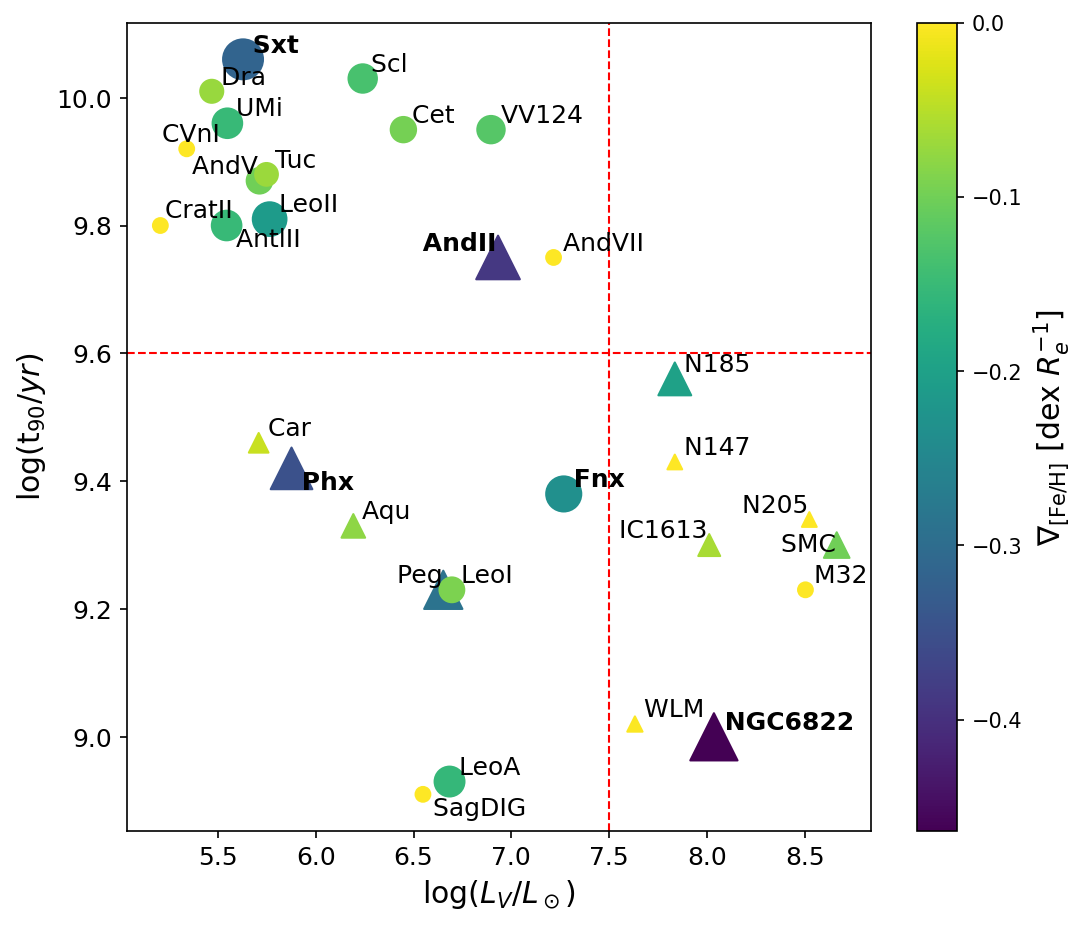}
    \caption{Luminosity versus $t_{90}$ for our sample of dwarf galaxies. Systems that shows significant stellar rotation are marked with triangles, all the others with circles; merger candidates are highlighted with bold text; the colour-coding and marker sizes are according to the value of their metallicity gradients $\nabla_{\rm [Fe/H]} (R/R_e)$. Dashed red lines help the reader to identify the sub-regions considered in Sect.~\ref{subsec:met-grad-rotation}.}
    \label{fig:Lv_T90_mgrad}
\end{figure}

Cosmological zoom-in simulations have highlighted the role played by angular momentum in preventing a radial metallicity profile in dwarf galaxies. \citet{Schroyen2013} and RJ18 have shown how the presence of a significant angular momentum may produce a centrifugal barrier that prevents the gaseous component from flowing towards the centre of the galaxy, thus preventing star formation from becoming more spatially concentrated and a metallicity gradient from forming. 
In particular, RJ18 found that internal rotation has a significant impact in preventing the formation of a radial metallicity profile in the highest-mass systems of their sample (i.e., with $M_* \gtrsim 10^8 M_\odot$), which are also those with the longest SFHs.

In our sample, significant stellar rotation (i.e., with $0.5 \lesssim V_{\rm rot}/\sigma_v \lesssim 2.0$) has been detected in the SMC \citep{DeLeo2020}, the dEs \citep{Geha2006,Geha2010}, WLM \citep{Leaman2012}, IC~1613 (\citealp{Wheeler2017}; Taibi et al.~in prep.), NGC~6822 \citep{Demers2006,Wheeler2017}, and at lower masses in Aquarius \citep{Hermosa2020}, Pegasus \citep{Kirby2014}, and probably Carina \citep{Martinez-Garcia2021}, while as prolate rotation in And~II \citep{Ho2012} and Phoenix \citep{Kacharov2017}\footnote{We do not include in this list M32 and Antlia~II for which clear signs of internal stellar rotation have been observed, but that may have been caused by tidal perturbations from their hosts \citep{Howley2013,Ji2021}. The Carina $V_{\rm rot}/\sigma_v$ value reported by \citet{Martinez-Garcia2021} is significant at the 2-$\sigma$ level.}. 

To better compare the properties of the rotating systems with the rest of the sample, we show in Fig.~\ref{fig:Lv_T90_mgrad} the observed stellar luminosity of our systems compared to their SFH's duration (i.e., their log($t_{90}/yr)$), with marker sizes and colours proportional to their $\nabla_{\rm [Fe/H]} (R/R_e)$; rotating dwarfs are highlighted with a different symbol. Since the SMC, VV~124, Antlia~II and Crater~II have no calculated $t_{90}$ from the literature, we gave them fiducial values based on the length of their SFH \citep{Jacobs2011,Rubele2018,Walker2019,Ji2021}.

We observe that 7/12 of the rotating systems have high stellar masses and extended SFHs (i.e., log$(L_V/L_\odot) > 7.5$ and  log$(t_{90}/yr) < 9.6$), with four of them showing $\nabla_{\rm [Fe/H]} (R/R_e)$ consistent with 0, within the 1-$\sigma$ error. 
On the other hand, among the rest of the rotating systems with lower stellar masses, just one shows a flat [Fe/H] profile within the errors (i.e., Carina, which however has an uncertain $V_{\rm rot}/\sigma_v$ detection). Thus, although milder gradients along with significant stellar rotation are likely to be observed among massive dwarf galaxies with an extended SFH, the same does not necessarily hold at low stellar masses.

It is interesting to further explore the properties of the low-mass systems. Those that have extended SFHs (i.e., log$(t_{90}/yr) < 9.6$), are characterised by a strong presence of rotating systems and have $\nabla_{\rm [Fe/H]} (R/R_e)$ values with median and scatter of $-0.12$~dex\,$R_e^{-1}$ and 0.14~dex\,$R_e^{-1}$, respectively.

Fainter systems that have formed stars for a shorter period (i.e., having log$(L_V/L_\odot) < 7.5$ and log$(t_{90}/yr) > 9.6$), and that in general are not rotating \citep[e.g.,][]{Wheeler2017}, show negative gradients with a similar median of $-0.11$~dex\,$R_e^{-1}$, but a smaller scatter of 0.06~dex\,$R_e^{-1}$).
It is difficult, however, to ascertain if SFH and angular momentum play a role in driving the scatter of $\nabla_{\rm [Fe/H]} (R/R_e)$ values between the two sub-sample of low-mass systems (apart from the presence of peculiar systems, such as prolate rotators), mainly due to their limited sizes.

\subsection{On the role of mergers}
\label{subsec:met-grad-mergers}
In our sample, several systems may have experienced a dwarf-dwarf merger in their past, namely: Andromeda~II \citep{Amorisco2014,Lokas2014,delPino2015}, Phoenix \citep{Kacharov2017}, Sextans \citep{Cicuendez+Battaglia2018}, Fornax \citep{Battaglia2006,Amorisco+Evans2012,deBoer2012,delPino2015,Leung2020,Pace2021}, and NGC~6822 \citep{deBlok2000,deBlok2006,Demers2006,Battinelli2006,Hwang2014,Zhang2021}.
Evidence supporting a past ($z\lesssim1$) major merger in these dwarf galaxies includes the existence of stellar substructures (including the presence of shell-/stream-like features), and peculiar structural or internal kinematic properties (e.g., prolate rotation).

We find in our analysis that the merger candidates are all characterised by a strong metallicity gradient, with $\nabla_{\rm [Fe/H]} (R/R_e)$ values narrowly distributing around $-0.35$~dex\,$R_e^{-1}$ (Figs.~\ref{fig:mgrad_Dist}-\ref{fig:Lv_T90_mgrad}).

Simulations have shown that such strong gradients can be the result of a past merger event. \citet{Benitez-Llambay2016}, analysing a set of dwarf galaxies from the CLUES simulation, report two main conditions for steep gradients to occur: a dwarf-dwarf merger must take place to disperse the old, metal-poor stellar components formed before the event; in-situ and centrally concentrated star formation must follow the subsequent accretion of residual gas. 
\citet{Cardona-Barrero2021}, analysing the set of simulations of RJ18, have also shown how a major merger can be responsible for producing a prolate stellar rotation together with a strong metallicity gradient, both features observed in And~II and Phoenix \citep{Ho2012,Kacharov2017}.

According to \citet{Deason2014}, the majority of dwarf-dwarf mergers have occurred at high-$z$, with an expected 10\% ($15-20$\%) of dwarf satellite (isolated) systems with $M_*\gtrsim10^6M_\odot$ having experienced a major merger since $z=1$. Specifically for the LG, they expect this to be the case for $\sim$\,$1-3$ ($\sim$\,$2-5$) MW or M31 satellite dwarfs (isolated). 
These numbers are in good agreement with those of our merger candidates, considering that they cover the same stellar mass range (where our full sample is almost complete, especially considering MW-satellites and isolated systems). 

If we were to exclude the merger candidates from the full distribution of metallicity gradients, the median and scatter of the sample would reduce to $-0.08$~dex\,$R_e^{-1}$ and 0.11~dex\,$R_e^{-1}$, respectively. The intrinsic scatter (i.e., taking into account measurement errors on the gradient values) about the median of such distribution would be only 0.05~dex\,$R_e^{-1}$. This result indicates that the values of $\nabla_{\rm [Fe/H]} (R/R_e)$, once merger candidates are excluded, are remarkably similar to each other despite the large differences in the properties of the analysed systems, which we recall span three orders of magnitude in luminosity, have formed stars on various time scales, and live in different environments. This may also indicate that the formation of mild metallicity gradients is intrinsic to the evolution of dwarf galaxies.

\section{Comparison with simulations}
\label{sec:met-grad-simul}

Several works in the literature have followed the chemical evolution of dwarf galaxies in cosmological zoom-in simulations, exploring the interplay between gravity, angular momentum and stellar feedback, the result of which may eventually lead to the formation or not of metallicity gradients. 

In this section, we examine in more detail whether the predictions of such simulations are in agreement with our observations. In particular, we compare our results with those of \citet{Mercado2021}, while we analyse, applying the same method used for the observed sample, the systems simulated by \citet{Revaz+Jablonka2018}, and those presented in \citet{Bellovary2019,Akins2021,Munshi2021}. 
The considered simulations cover a stellar mass range similar to that of the observations ($10^{5}\lesssim M_*/M_\odot\lesssim10^{9}$), but were generated using distinct hydrodynamic codes and different stellar feedback recipes.

\subsection{Comparison with \citet{Mercado2021}}
\label{subsec:met-grad-simul-M21}

\begin{figure*}
    \centering
    \includegraphics[width=0.45\textwidth]{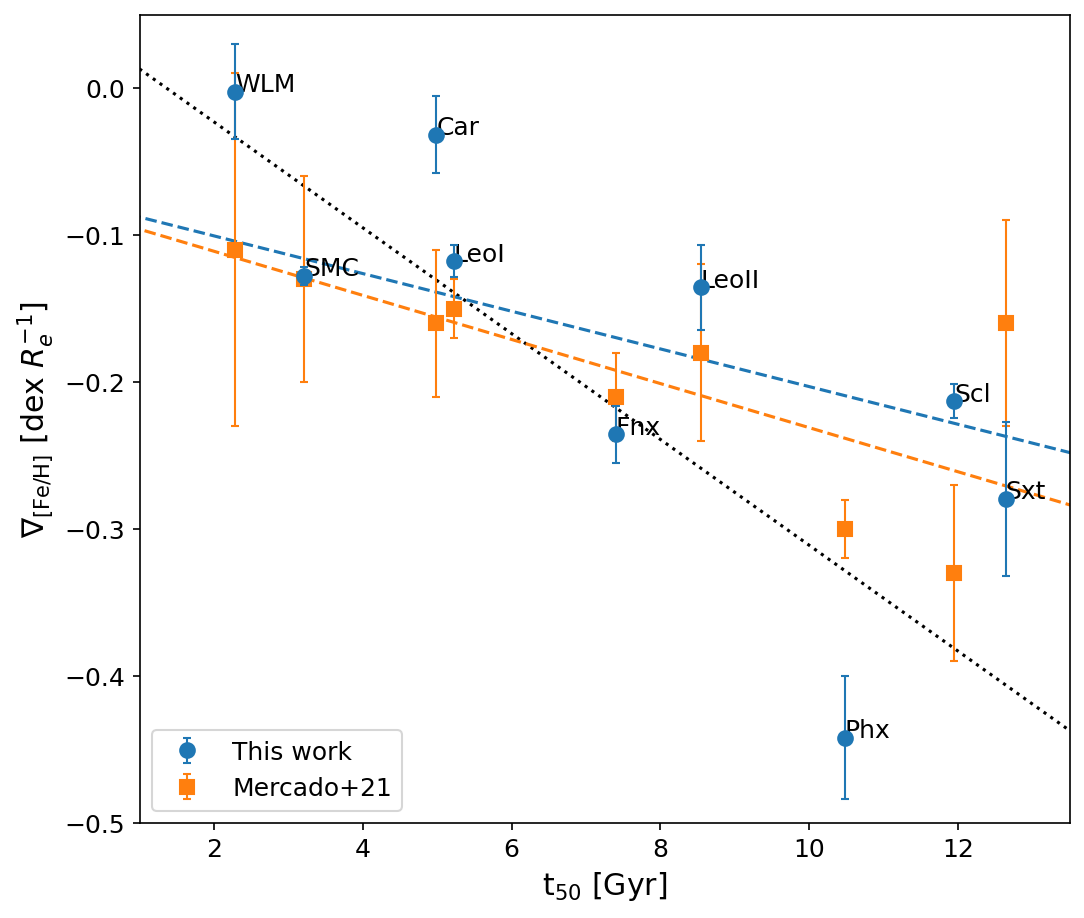}
    \includegraphics[width=0.45\textwidth]{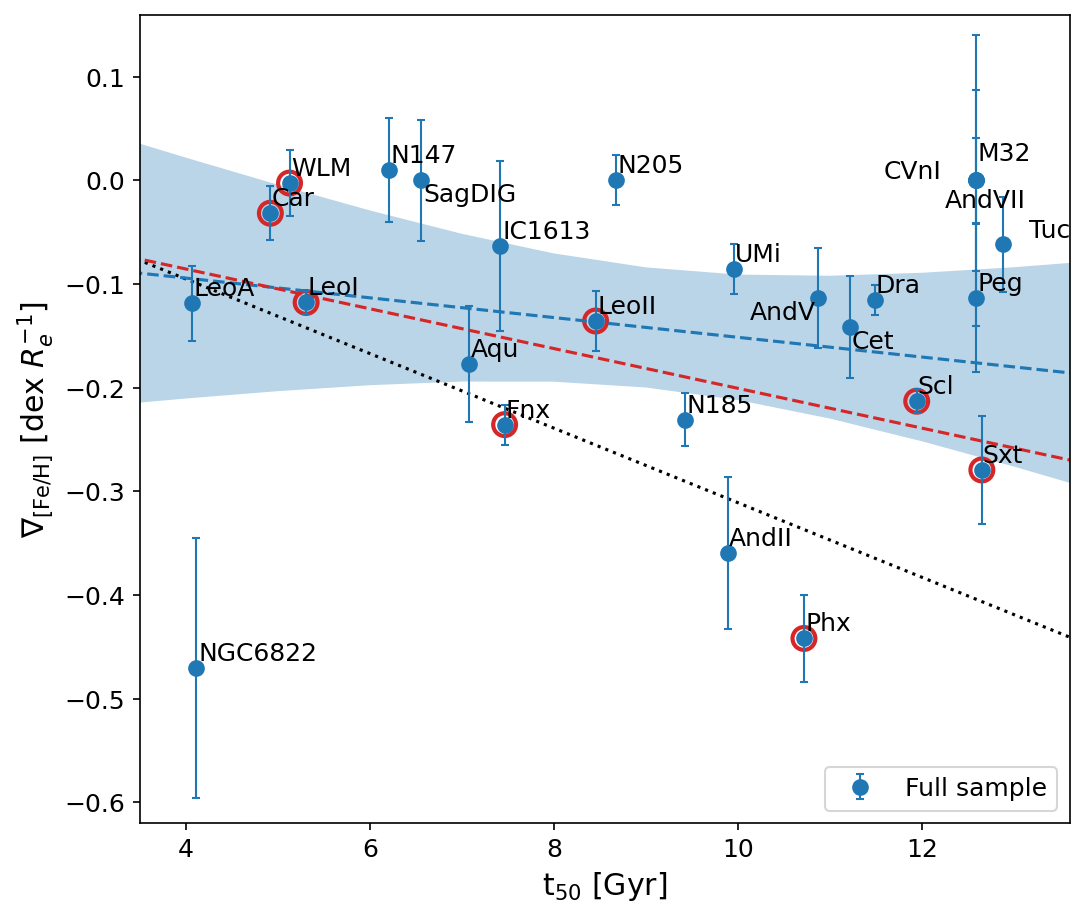}
    \caption{\textit{Left}: metallicity gradients as a function of $t_{50}$ obtained by \citet{Mercado2021} for a sample of LG dwarf galaxies marked by orange squares with error-bars, compared with our determinations for the same sample obtained following the Mercado et al.\, criteria (i.e., assuming circular radii and with gradients calculated within $2\times R_{\rm e, circ}$) marked by blue circles with error-bars. The orange and blue dashed lines are the linear least-square fit to the described samples, respectively, while the black dotted line (in both panels) is the linear relation found by Mercado et al.\, for their set of simulated systems. 
    \textit{Right}: metallicity gradients obtained again following the Mercado et al.\, criteria but for the full sample of galaxies analysed in this work, as a function of their $t_{50}$ adopted from \citet{Weisz2014,Bettinelli2018,Albers2019}; the blue dashed line and shaded area are respectively the linear least-square fit and the associated 95\% confidence interval of the sample. As reference, red circles mark the sub-sample considered by \citet{Mercado2021}, with the red dashed line being the linear fit to it.
    }
    \label{fig:mgrad_t50_mercado}
\end{figure*}

\citet[][hereafter M21]{Mercado2021}, analysing a set of 26 isolated dwarf galaxies from the FIRE-2 simulation with stellar masses in the range $10^{5.5}<M_*/M_\odot<10^{8.3}$, found a significant linear relation between $\nabla_{\rm [Fe/H]} (R/R_e)$ and $t_{50}$. 

The authors explain this correlation as due to the combined effect of two different mechanisms. Repeated episodes of stellar feedback ``puff-up'' the spatial distribution of stars: since older, more metal-poor populations experience more of these cycles than younger and metal-richer populations, the puffing-up will be stronger on the former than on the latter, creating a metallicity gradient. Then, whether this gradient remains strong or weakens depends on the extent of the late star-forming spatial region: in those galaxies that experience late gas accretion, this (recycled/metal-enriched) gas tends to be deposited at large radii, feeding the formation of metal-richer stars in the outskirts, thus weakening the metallicity gradient. 

Although with a shallower slope, the authors obtain a similar dependence for a sample of 10 LG dwarf galaxies, both isolated and satellites, based on $t_{50}$ and $\nabla_{\rm [Fe/H]} (R/R_e)$ derived, respectively, from SFHs and radial metallicity profiles from spectroscopic observations available in the literature.

For sake of comparison, Fig.~\ref{fig:mgrad_t50_mercado} (left) shows the relationship between the strength of metallicity gradients with respect to $t_{50}$ obtained by M21 for simulated galaxies (black dotted line), compared to the observed relation they obtained for their selected sub-sample of LG dwarf galaxies (orange squares and dashed line). We overplot our determination of the [Fe/H] gradient for the same sub-sample (blue circles and dashed line). In order to be consistent, we calculated the metallicity gradients as in M21, that is using the 2D ``circular'' radii\footnote{We use the terminology circular radius to indicate simply $R= \sqrt{x^2 + y^2}$, where $x$ and $y$ are the coordinates projected on the tangent plane to the dwarf galaxy centre.} and limiting the analysis to twice the 2D projected geometric half-light radius. Values for our sample of galaxies are reported in Table~\ref{tab:met_grad_circ}. We  also report in the figure the same $t_{50}$ values tabulated by M21.

From the figure, it can be appreciated that the different spectroscopic samples and/or methodology we use cause differences in the values of the gradient strength of the individual galaxies, but overall the trend with $t_{50}$ remains, even though our relation is slightly shallower that theirs ($m=-0.013\pm0.005$). Our calculated values further highlight the discrepancy with the theoretical trend (black dashed line) obtained by M21 from their simulated sample ($m_{\rm M21}=-0.036\pm0.005$). 

In Fig.~\ref{fig:mgrad_t50_mercado} (right) we instead show the metallicity gradients again assuming 2D circular radii, but this time for our full sample of galaxies as a function of their $t_{50}$ values listed in Table~\ref{tab:met_grad} (see Sect.~\ref{subsec:met-grad-sfh} for further details). Increasing the sample size further flattens the relation with $t_{50}$ ($m=-0.009\pm0.006$), with the 95\% confidence interval fully consistent with there being no relation between the two quantities. The sub-sample of LG dwarf galaxies selected by M21 are marked in red; they show a linear decreasing trend comparable to the one obtained in the left panel, despite the slight difference in the $t_{50}$ values considered. 

The observations, therefore, do not seem to back up the theoretical prediction of a clear relation between strength of the metallicity gradient and $t_{50}$. This might be a consequence of the observed uncertainties related to the determination of $t_{50}$ smudging the relation, or simply point to a lower impact of stellar feedback in the evolution of LG dwarf galaxies with respect to that involved in the FIRE-2 simulations.

\subsection{Comparison with \citet{Revaz+Jablonka2018}}
\label{subsec:met-grad-simul-RJ18}

\begin{figure*}
    \centering
    \includegraphics[width=0.49\textwidth]{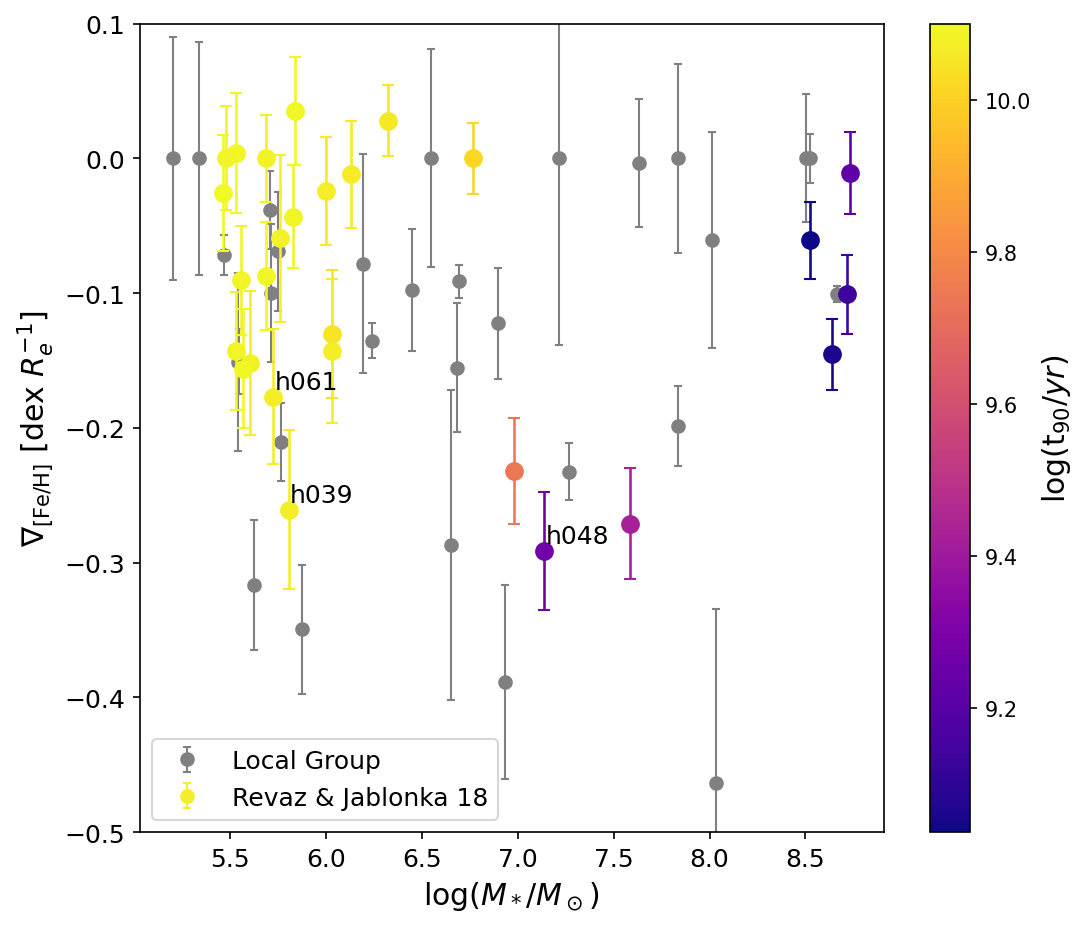}
    \includegraphics[width=0.49\textwidth]{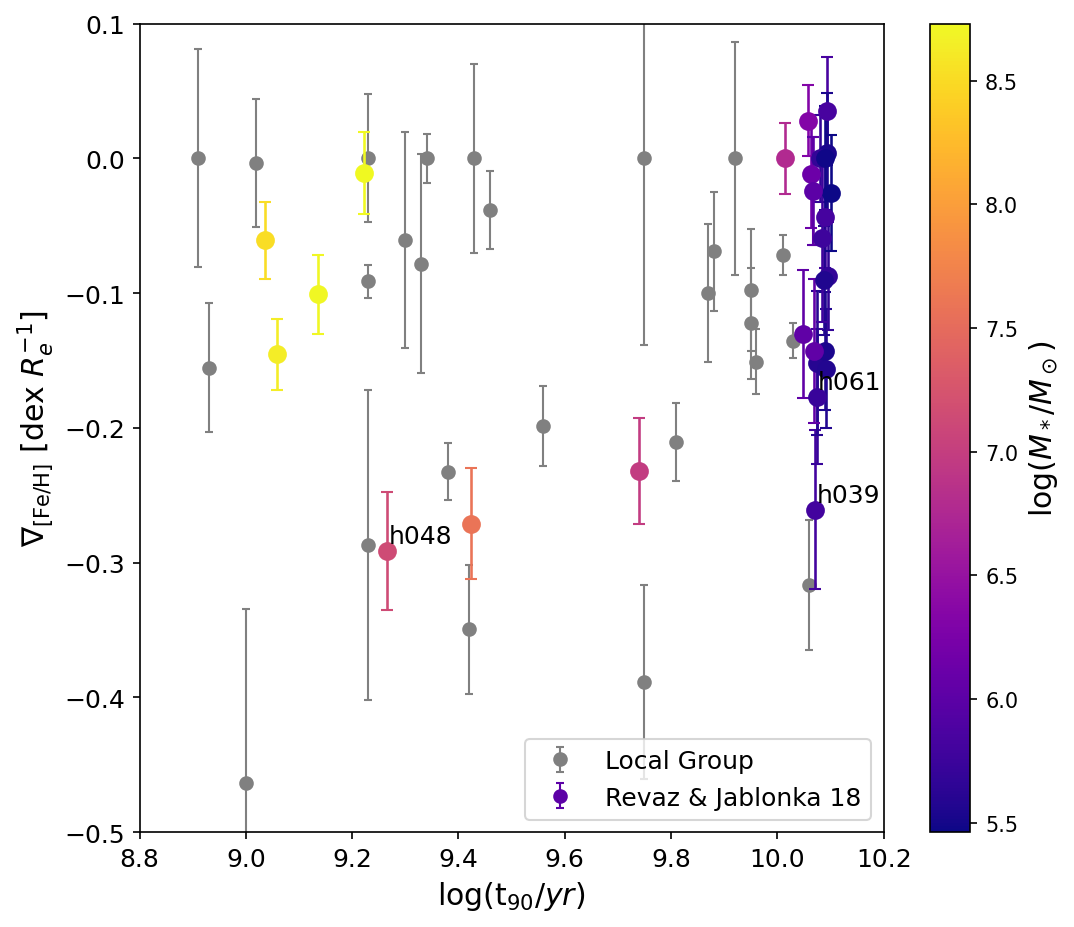}
    \caption{Distribution of the metallicity gradients calculated for the simulated systems presented in \citet{Revaz+Jablonka2018}, on the left panel as a function of their stellar mass and color-coded according to their $t_{90}$, while inverting these two variables in the right panel.
    Numbers identify the individual halos mentioned in the main text, while the gradient values for the Local Group dwarf galaxies presented in this work are shown as grey circles with their associated error bars. We have assumed in this case a stellar mass-to-light ratio of one. 
    }
    \label{fig:sim_mgrad_RJ18}
\end{figure*}

The link between the full length of the star formation period and metallicity gradients in dwarf galaxies has been explored by RJ18, who analysed a set of cosmological zoom-in simulations of 27 isolated dwarf galaxies with stellar masses between $10^{5.5}-10^{8.75} M_\odot$. They showed that as long as star formation is active, metallicity gradients build-up gradually, since the gas reservoir shrinks towards the galaxy's centre. In fainter systems (i.e., with $M_*\leq10^6 M_\odot$), star formation is soon quenched by the UV background heating and thus gradients tend to be marginal. Similarly, at higher masses the presence of a significant angular momentum counteract their sustained SFH producing flatter metallicity profiles. A major merger, however, could lead to the formation of a relatively strong and long-lasting gradient at any mass \citep[see also][]{Benitez-Llambay2016,Cardona-Barrero2021}. We note that the RJ18 simulations use a different star formation recipe that leads to a lower impact of stellar feedback than the FIRE-2 simulations analysed by M21.

We have analysed here the simulated systems presented in RJ18 applying the same method used for our observed data-set in order to calculate their metallicity gradients and perform a direct comparison with the LG systems. We adopted a fiducial [Fe/H] error of 0.2~dex for the stellar particles in the simulations, which is a typical observational uncertainty. 

As shown in Fig.~\ref{fig:sim_mgrad_RJ18}, the scenario described by RJ18 produces something like a U-shape in the diagrams where the $\nabla_{\rm [Fe/H]} (R/R_e)$ is seen against the stellar mass or the $t_{90}$. 
Indeed, the strongest gradients are observed for systems at intermediate masses (i.e., log$(M_*/M_\odot) \sim 7.0$) with extended SFHs; systems at the high-mass end, characterised by a continuous and sustained SFH, have marginal gradients due to their high-angular momentum; at the low-mass end instead, where the SFH is soon quenched, gradients show a large scatter.

\begin{figure}
    \centering
    \includegraphics[width=0.49\textwidth]{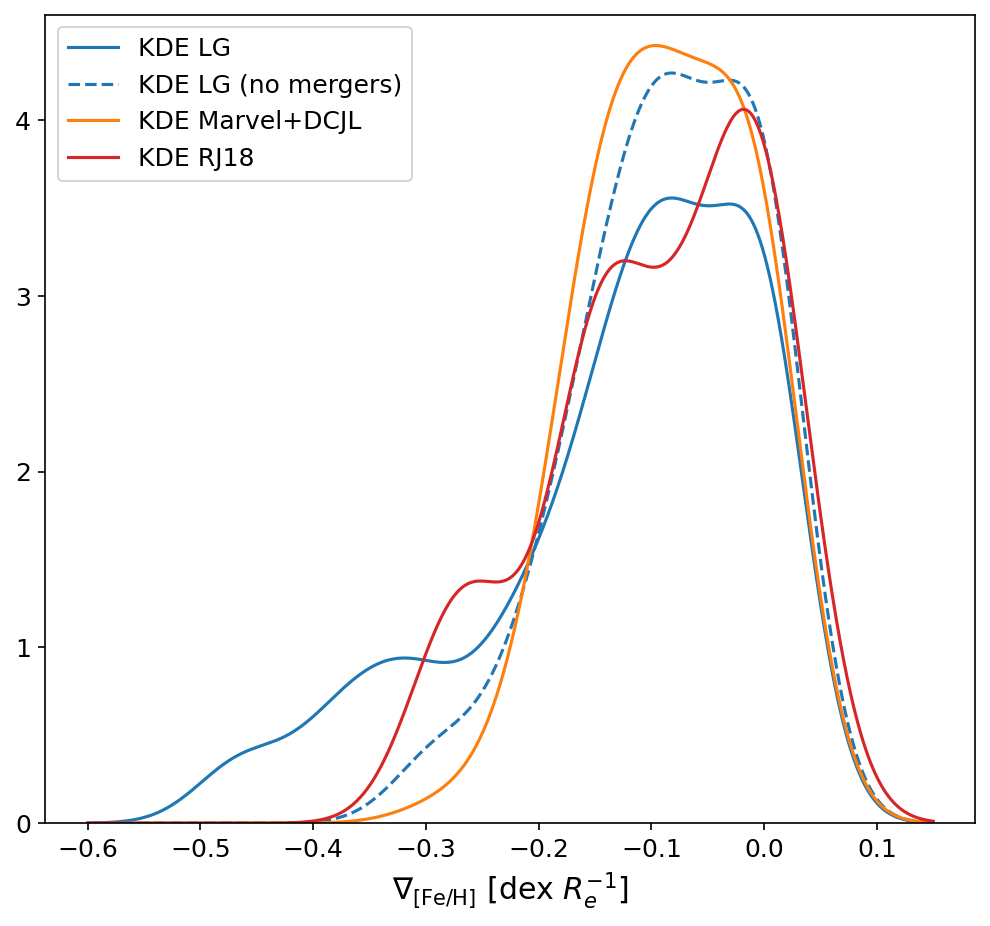}
    \caption{Kernel density estimation of the metallicity gradient distribution for the Local Group sample (with and without merger candidates in blue solid and dashed lines, respectively), compared with the simulated sets of \citet{Revaz+Jablonka2018} (red line) and the MARVEL/DCJL (orange line) ones analysed in Sects.~\ref{subsec:met-grad-simul-RJ18} and \ref{subsec:met-grad-simul-MARVEL}, respectively. We assumed in all cases a Gaussian kernel of bandwidth 0.04~dex\,$R_e^{-1}$, which is comparable to the average error of the metallicity gradients.}
    \label{fig:sim_mgrad_KDE_RJ18_DCJL+MARVEL}
\end{figure}

The simulations cover a similar range of $\nabla_{\rm [Fe/H]} (R/R_e)$ values as the observed galaxies, with a median and scatter of $-0.09$~dex\,$R_e^{-1}$ and 0.10~dex\,$R_e^{-1}$, respectively. These values are comparable to the observed ones, excluding the merger candidates, as shown in Fig.~\ref{fig:sim_mgrad_KDE_RJ18_DCJL+MARVEL} where we performed a kernel density estimation (assuming in all cases a Gaussian kernel of bandwidth 0.04~dex\,$R_e^{-1}$, i.e., comparable to the average error of the metallicity gradients). To compare the different distributions we used the \textsc{python} package \texttt{KernelDensity} in \texttt{scikit-learn} \citep{scikit-learn}. 

The simulated systems, indeed, do not show $\nabla_{\rm [Fe/H]} (R/R_e)$ values which are in absolute terms as extreme as the observational ones, although taking error bars into account these may be plausible. On the other hand, two of the strongest gradients in the simulations (i.e., for \texttt{h061} and \texttt{h048}) are for systems which experienced a major merger at $\sim$\,6~Gyr \citep[see also][]{Cardona-Barrero2021}. 
We note that the strong gradient observed for \texttt{h039} is instead caused by its large half-light radius, artificially enhanced due to the presence of small ultra-faint satellites orbiting around it, with some of them in a process of being merged/destroyed at $z=0$.

Despite the general agreement, we do not recover a strong dependence between the values of $\nabla_{\rm [Fe/H]} (R/R_e)$ and the stellar mass or SFH of the observed systems as for the simulated ones. In particular, we observe a larger scatter of observed systems at log$(M_*/M_\odot) \sim 7.0$ and log$(t_{90}/yr) \sim 9.5$, where the formation of strong gradients should be favoured by their extended SFH. This may be due to the limited sample of simulated galaxies analysed by RJ18, which does not include analogues of observed dwarf galaxies such as Carina, Leo~I and Fornax, but also Aquarius and Leo~A, characterised by  a significant fraction of intermediate age stars \citep[e.g.,][]{Gallart1999,deBoer2012,deBoer2014,Cole2014,Ruiz-Lara2021}.
Such limitations could also be due to the specific galaxy formation model implemented in RJ18, in particular regarding the impact of the UV-background heating and/or the efficiency of hydrogen self-shielding. However, as shown in RJ18, their simulated galaxies are able to reproduce over several orders of magnitudes the observed properties of LG systems in terms of $L_V$, $\sigma_v$, $R_e$, and $M_{\rm HI}$. Therefore, it is difficult to attribute the lack of intermediate-mass dwarf galaxies with an extended SFH in the RJ18 sample to a limitation of their model or to a simple lack of sufficient statistics.

We recall that RJ18 simulations follow the evolution of dwarf galaxies in isolation. The influence of a MW-host on the same set was explored by \citet{Hausammann2019}, who found that indeed ram-pressure and tidal interactions could help star formation last longer, with a possible impact on the formation of metallicity gradients. However, this would require very specific conditions (a very late entry with a small pericenter in the MW-potential, or a very early accretion) which seems to limit the actual role of a MW-like host. We have shown in Sect.~\ref{subsec:met-grad-dist} that we find no significant differences between the metallicity gradient properties of satellite and isolated systems.

\subsection{Comparison with DCJL/MARVEL sets}
\label{subsec:met-grad-simul-MARVEL}

\begin{figure*}
    \centering
    \includegraphics[width=0.49\textwidth]{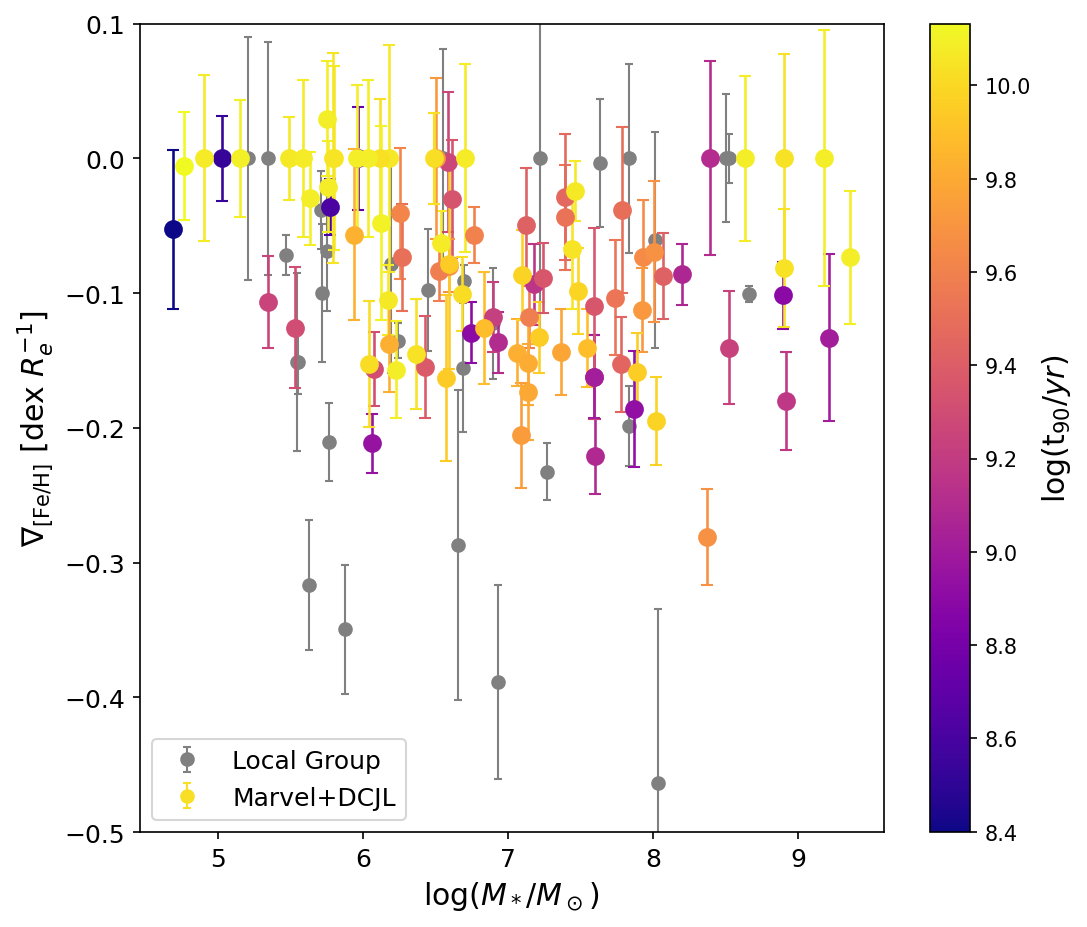}
    \includegraphics[width=0.49\textwidth]{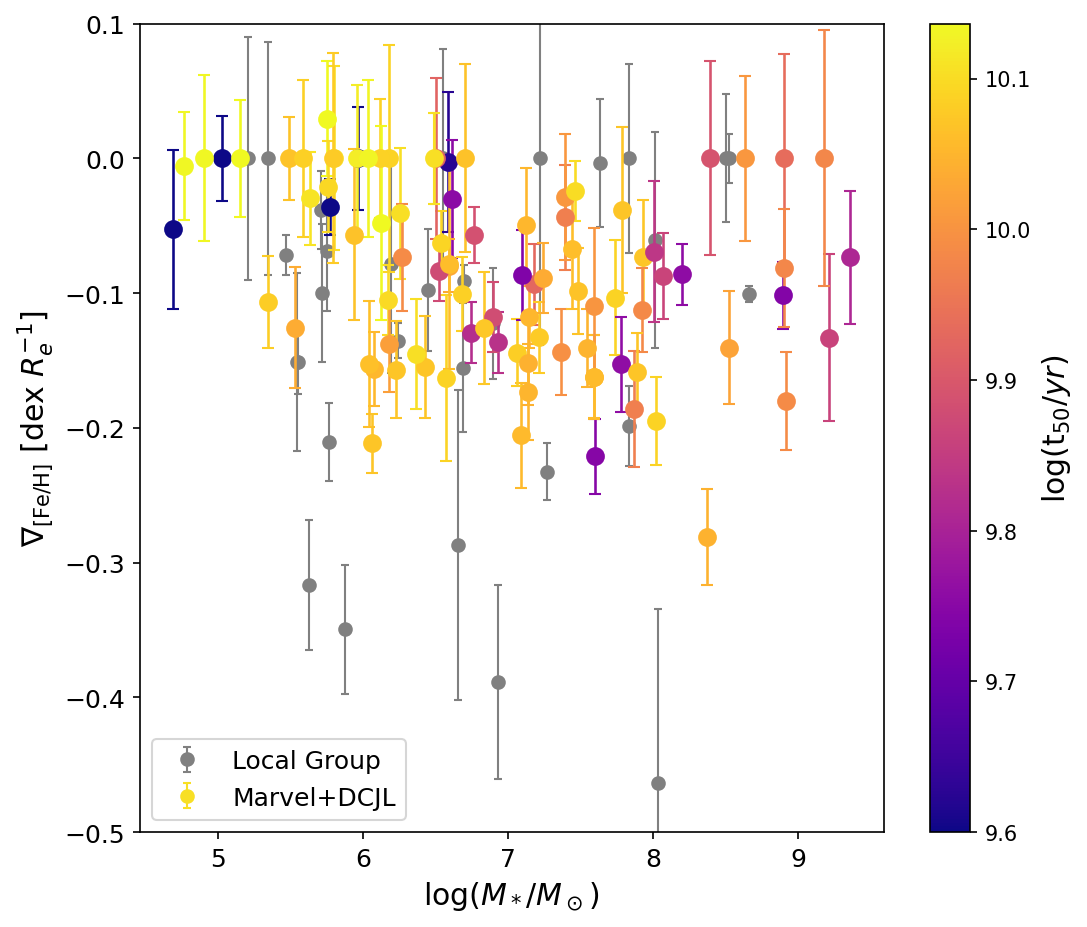}\\
    \includegraphics[width=0.49\textwidth]{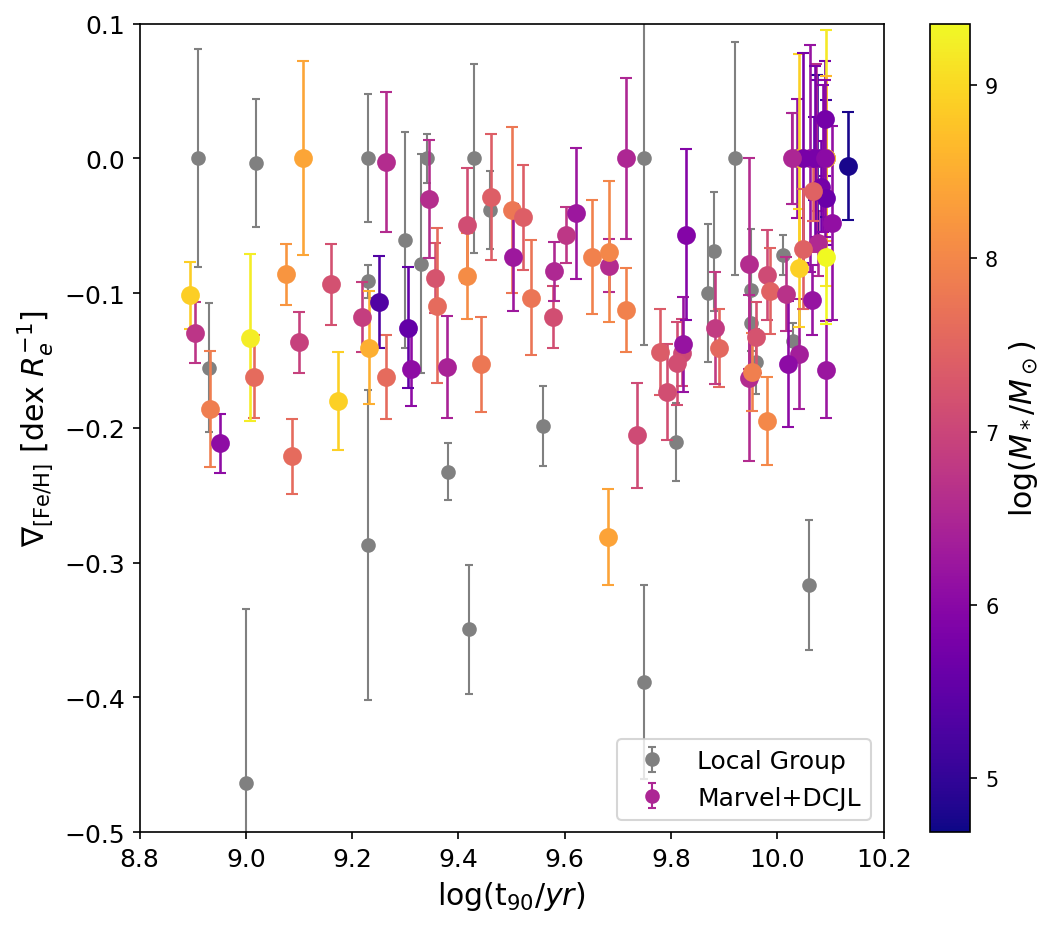}
    \includegraphics[width=0.49\textwidth]{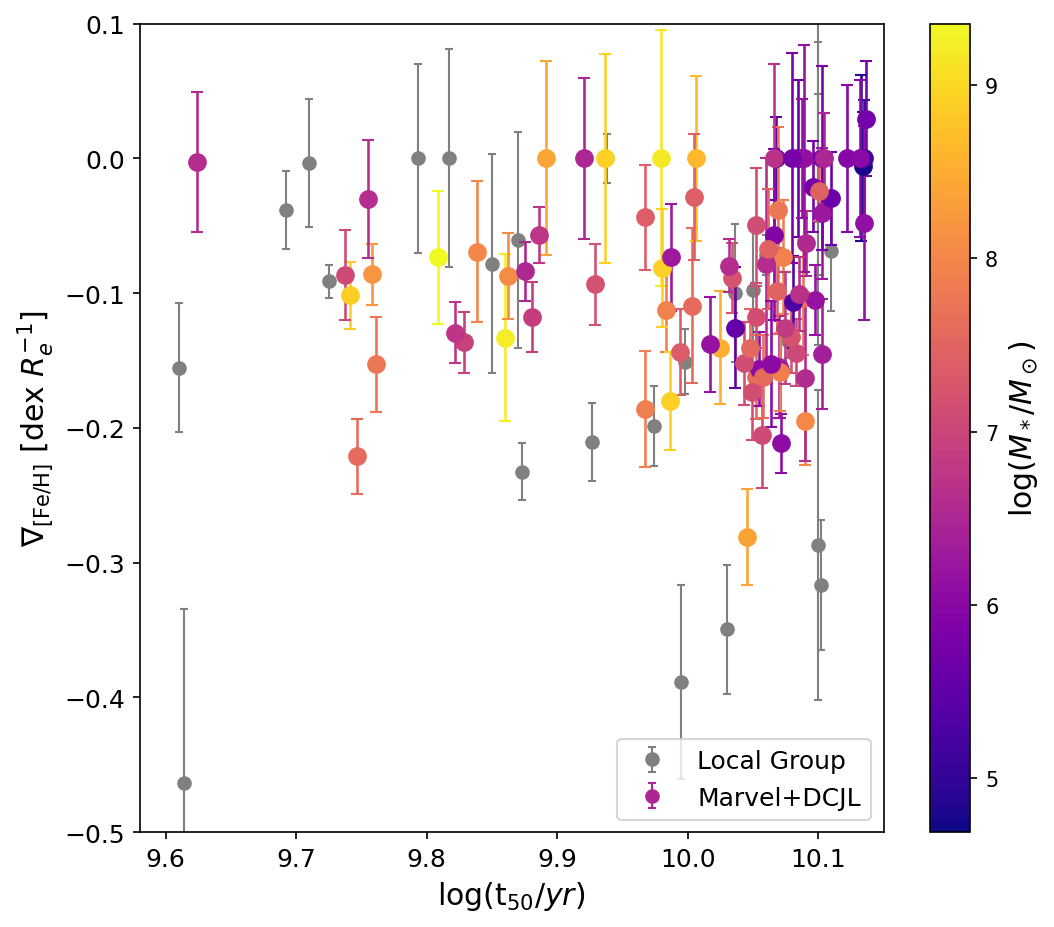}
    \caption{Distribution of the metallicity gradients calculated for the simulated systems of both the DCJL and MARVEL sets, as a function of their stellar mass color-coded according to their $t_{90}$ (top left) and $t_{50}$ (top right), and as a function of $t_{90}$ (bottom left) and $t_{50}$ (bottom right) in both cases color-coded according to their stellar mass.
    Also shown are the gradient values for the Local Group dwarf galaxies presented in this work, shown as grey circles with their associated error bars. We have assumed in this case a stellar mass-to-light ratio of one. 
    }
    \label{fig:sim_mgrad_Mstar_DCJL+MARVEL}
\end{figure*}

We further analysed a sample of dwarf galaxies selected from two sets of zoom-in cosmological simulations: the ``near-mint DC Justice League'' (DCJL) presented in \citet{Bellovary2019} and \citet{Akins2021}, and the ``MARVEL-ous Dwarfs'' (MARVEL) from \citet{Munshi2021}. 
The DCJL set consists of $\sim$\,100 dwarf galaxies from four MW-like environments (namely Sandra, Ruth, Sonia, and Elena) that mimic the MW in terms of host halo-mass but with different formation histories. The MARVEL set instead is formed by $\sim$\,50 field dwarfs (with some of them being satellites of dwarf galaxies) from four low-density environments (named CptMarvel, Elektra, Rogue, and Storm).
Both sets of simulations were run using the same code and underlying physics.

The analysed systems cover a similar range of stellar masses as the RJ18 and M21 simulations, which, however, implemented different stellar feedback recipes with, simplifying, the DCJL/MARVEL sets halfway between the RJ18 set, characterised by low feedback, and M21 \citep[see][for a comparison between the `burstiness' of the DCJL/MARVEL and FIRE-2 sets]{Iyer2020}. We refer to \citet{Bellovary2019}, \citet{Akins2021}, and \citet{Munshi2021} for further details on the DCJL/MARVEL simulations.  

The analysis of these simulated systems was conducted applying the same method developed for our observed data-set to calculate their metallicity gradients and also used for the RJ18 set. The individual metallicity values initially had no associated uncertainties, hence we adopted a fiducial [Fe/H] error of 0.2~dex for each stellar particle.
Our results show that the obtained $\nabla_{\rm [Fe/H]} (R/R_e)$ range in all cases between $-0.25$~dex\,$R_e^{-1}$ and $0$~dex\,$R_e^{-1}$, with medians and scatter values around $-0.10$~dex\,$R_e^{-1}$ and $0.09$~dex\,$R_e^{-1}$ (see Figs.~\ref{fig:sim_mgrad_KDE_RJ18_DCJL+MARVEL} and \ref{fig:sim_mgrad_Mstar_DCJL+MARVEL}). 
We also find that both DCJL and MARVEL systems show similar $\nabla_{\rm [Fe/H]} (R/R_e)$ distributions which suggests that in these simulations the environment plays a limited role in driving the scatter of metallicity gradients.
The obtained results are again in good agreement with our observations, excluding for the steepest gradients we observed, which have no match in these simulations. In addition, similar to our observations, the simulated systems do not show clear correlations between their $\nabla_{\rm [Fe/H]} (R/R_e)$ and the stellar mass, nor with their $t_{50}$ or $t_{90}$ (see again Fig.~\ref{fig:sim_mgrad_Mstar_DCJL+MARVEL}).

In a forthcoming paper, we will provide details on the chemical and star formation history of these simulated systems, together with an analysis of their past merger history and a thorough comparison with other sets of simulations carried out with different initial conditions and stellar feedback recipes (e.g., the aforementioned RJ18 simulations).

\section{Summary and conclusions}
\label{sec:end}

In this work we conduct a homogeneous search and measurement of stellar metallicity gradients in Local Group dwarf galaxies. We use publicly available spectroscopic catalogues of red giant stars in 30 systems, either isolated or satellites of the Milky Way and M31. To date, this is the largest compilation of its kind.

We analyse the radial variation of the metallicity [Fe/H] values using a Gaussian Process Regression analysis. From the resulting profiles, we determine a value for the metallicity gradient, both in units of the physical radius in kpc, $\nabla_{\rm [Fe/H]} (R)$, and in units of the 2D SMA half-light radius, $\nabla_{\rm [Fe/H]} (R/R_e)$.
We find that the majority of systems analysed show negative gradients, with median values $\nabla_{\rm [Fe/H]} (R/R_e) \sim -0.1$~dex\,$R_e^{-1}$, and $\nabla_{\rm [Fe/H]} (R) \sim -0.25$~dex\,kpc$^{-1}$.

Our systems do not show any correlation between their gradients and their morphological types (or internal gas content), nor with the distance from the nearest host galaxy (i.e., the MW or M31). These results point to a limited role of the environment as the main actor in the formation of strong radial metallicity gradients in LG dwarf galaxies. An important fact here is the presence of negative gradients in the isolated systems.

We find mild correlations between $\nabla_{\rm [Fe/H]} (R)$ and the stellar luminosity of our systems, as well as with their $t_{50}$ and $t_{90}$ values (i.e., the look-back times at which, respectively, the cumulative stellar mass inferred by the SFH reaches 50\% and 90\% of its total value), as a result of the scaling relations among LG systems (i.e., stellar mass-size, stellar mass-metallicity, and stellar mass-SFH relations). Regarding $\nabla_{\rm [Fe/H]} (R/R_e)$, no clear correlations are found with the same parameters. 

In particular, we inspect the dependency of $\nabla_{\rm [Fe/H]} (R/R_e)$ with the $t_{50}$, which was found by \citet{Mercado2021} to exhibit a strong anti-correlation with the gradient strength in their simulated dwarfs. While we recover the shallow dependency that they found for a sub-sample of 10 LG dwarf galaxies, when considering the full sample we analysed the trend between the strength of $\nabla_{\rm [Fe/H]} (R/R_e)$ and $t_{50}$ is consistent with no correlation between the two quantities. 

We also looked for differences in the values of $\nabla_{\rm [Fe/H]} (R/R_e)$ of those systems with significant stellar rotation. Simulations have shown that a high angular momentum could prevent the formation of strong gradients, particularly at higher masses \citep{Schroyen2013, Revaz+Jablonka2018}. 
Among dwarf galaxies with $L_V>10^{7.5} L_\odot$, characterised by an extended SFH and significant stellar rotation, we observe some systems with flat metallicity gradients. However, taking the rest of the rotating systems into account, a high angular momentum does not necessarily imply a flat gradient.

The strongest gradients in our sample, $\nabla_{\rm [Fe/H]} (R/R_e)\lesssim-0.25$~dex\,$R_e^{-1}$, are observed in And~II, Phoenix, Sextans, Fornax, and NGC~6822, all systems that in the literature have been proposed as likely to have experienced a past merger event, given the chemo-kinematic properties of their stellar component. Simulations have shown that a dwarf-dwarf merger event followed by a re-ignition of star formation activity in the galaxy's centre would indeed produce strong metallicity gradients \citep{Benitez-Llambay2016,Cardona-Barrero2021}. 

Excluding the merger candidates from the full distribution of metallicity gradients, the median $\nabla_{\rm [Fe/H]} (R/R_e)$ reduce to $-0.08$~dex\,$R_e^{-1}$ with an associated intrinsic scatter of only 0.05~dex\,$R_e^{-1}$. This result indicates that metallicity gradients of LG dwarf galaxies are remarkably self-similar despite the large differences in their observed properties (i.e., in terms of stellar mass, star formation history, environment).

We complement our study with the analysis of simulated dwarf galaxies from the literature by applying the same method used for the observed sample to calculate their metallicity gradients. In particular, we examine the sample of \citet{Revaz+Jablonka2018} ($\sim$\,30, isolated dwarf galaxies) recovering their previous results and finding that the average and scatter of their $\nabla_{\rm [Fe/H]} (R/R_e)$ values are in good agreement with those of the observations. However, differently from their work, we do not recover any correlation between $\nabla_{\rm [Fe/H]} (R/R_e)$ and the luminosity and SFH of the observed systems. We note that our sample contains low-mass systems with an extended SFH, which have no counterparts in their simulations.

Similarly, we analyse a sample of simulated dwarf galaxies ($\sim$\,150, either central or satellites of MW-like hosts) selected from two sets of cosmological simulations: DCJL \citep{Bellovary2019,Akins2021}, and MARVEL \citep{Munshi2021}. We obtain comparable $\nabla_{\rm [Fe/H]} (R/R_e)$ distributions between the different simulated samples, whose values are again in good agreement with those of the observed systems, once we exclude the steepest gradients that have no match in the simulations. Similarly to the observations, we further find that the simulated systems do not show clear correlation between their gradients and the stellar mass, nor with the $t_{50}$ or $t_{90}$.

In general, it seems that the interplay between the several internal factors that could led to the formation of metallicity gradients is more complex than what is shown by the simulations with which we have compared our results. The role of the environment, on the other hand, seems limited, although we cannot exclude that effects such as tidal and ram-pressure stripping may have an impact in increasing the scatter of metallicity gradient values among the satellite systems.

The study of radial metallicity gradients in LG dwarf galaxies will benefit in the future of further investigations. From an observational point of view, it would be important to extend the surveyed area of some systems, in particular NGC~6822. To confirm that its radial metallicity profile continues to be so steep even at larger radii would be of particular relevance in the context of the connection between strong gradients and major merger events. 

On the other hand, in order to better understand the impact that star formation time scales have on the formation of metallicity gradients in dwarf galaxies, it would be useful to have measurements of their SFHs that are both homogeneous and more accurate and precise than those currently available. In this context, it would also be interesting to explore how the metallicity gradients compare to the radial age gradients that could be obtained from the analysis of individual stars.

Finally, we foresee in the future to perform a thorough comparison between the sets of simulations considered in this work, which differ in their initial conditions, environment, and implemented stellar feedback recipes.

With our results we wish to provide the astronomical community with constraints on the processes that shape the formation and evolution of dwarf galaxies in the Local Group.

\begin{acknowledgements}
The authors would like to thank M.~Pawlowski, S.~Cardona-Barrero, and A.~Di Cintio for useful discussions and comments at different stages of this work.

ST acknowledges funding of a Leibniz-Junior Research Group (PI: M.~Pawlowski; project number J94/2020) via the Leibniz Competition.
ST and GB acknowledge support from: the Agencia Estatal de Investigaci{\'o}n del Ministerio de Ciencia e Innovaci{\'o}n (AEI-MICIN) and the European Regional Development Fund (ERDF) under grant with reference AYA2017-89076-P, while the AEI-MICIN under grant with reference PID2020-118778GB-I00/10.13039/501100011033; GB acknowledge the AEI-MICIN under grant with reference CEX2019-000920-S.
AMB and FDM acknowledge support from HST AR-13925 provided by NASA through a grant from the Space Telescope Science Institute, which is operated by the Association of Universities for Research in Astronomy, Incorporated, under NASA contract NAS5-26555.  AMB acknowledges support from HST AR-14281. AMB and CLR acknowledge support from NSF grant AST-1813871.  Resources supporting this work were provided by the NASA High-End Computing (HEC) Program through the NASA Advanced Supercomputing (NAS) Division at Ames Research Center.

This research has made use of NASA’s Astrophysics Data System, VizieR catalogue access tool, and extensive use of Python, including Numpy, Scipy, Astropy and Scikit-learn packages.
\end{acknowledgements}

%
%

\bibliographystyle{bibtex/aa} 
\bibliography{bibtex/LG_grad.bib} 

\begin{appendix}

\section{Sanity checks}
\label{sec:apx0}

\subsection{Using a different [Fe/H]-CaT equivalent width calibration for Carina and the SMC}

Here we recalculate the individual [Fe/H] measurements for the Carina and SMC data-sets, by applying respectively the \citet{Starkenburg2010} and \citet{Carrera2013} calibrations, and we compare them to the estimates given in literature studies from which the spectroscopic samples were extracted (see Table~\ref{tab:sample}), which used the \citet{Carretta+Gratton1997} scale, not optimal at low metallicities. 

\begin{figure*}
    \centering
    \includegraphics[width=0.49\textwidth]{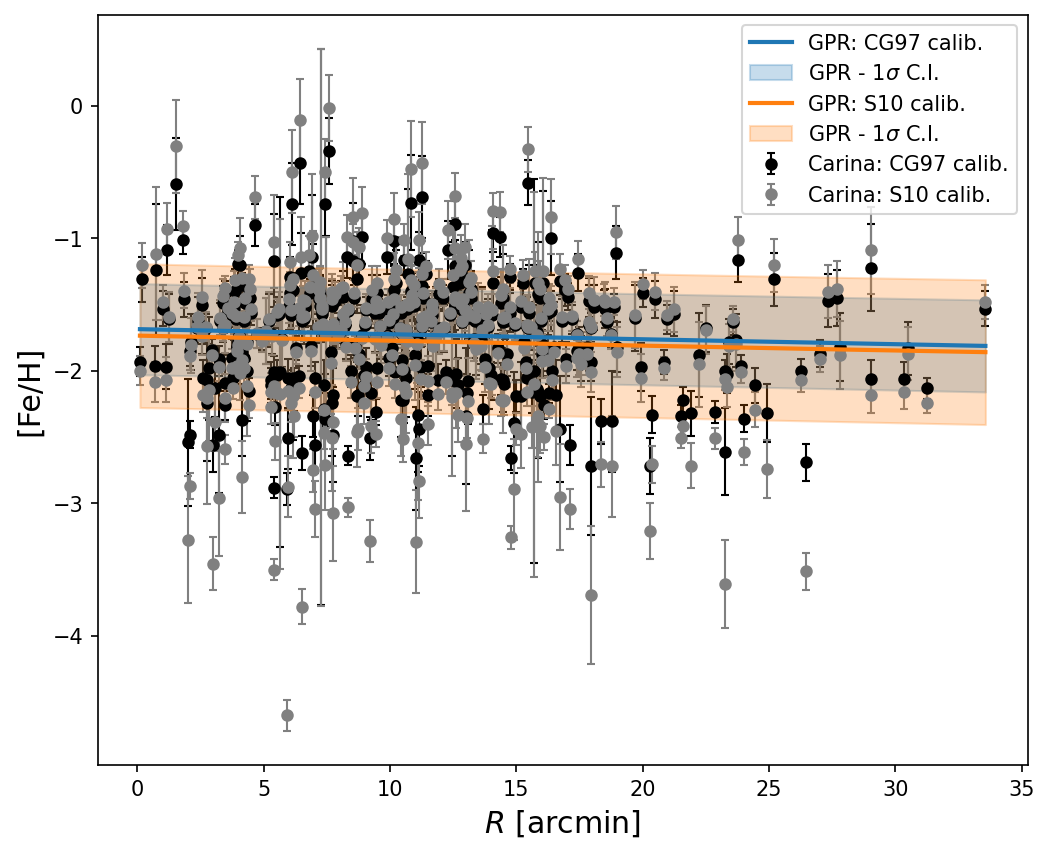}
    \includegraphics[width=0.49\textwidth]{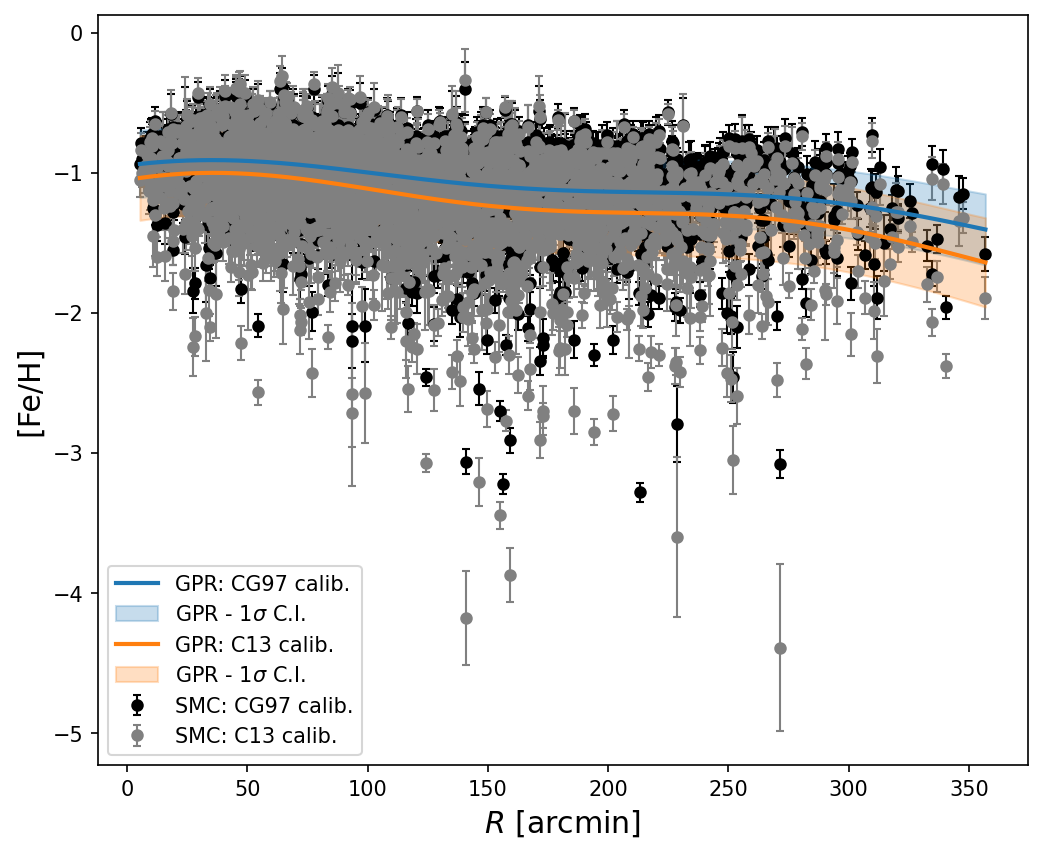}
    \caption{
    Comparison of the distribution of individual [Fe/H] measurements and their errors, as a function of SMA radius, for Carina (left) and the SMC (right), between the original values (i.e., respectively from \citealp{Koch2006} and \citealp{Dobbie2014}, in black) calculated using the \citet[][CG97]{Carretta+Gratton1997} scale, and those applying respectively the \citet[][S10]{Starkenburg2010} and \citet[][C13]{Carrera2013} calibrations (in gray). The blue (orange) solid line and shaded area represent the result of the Gaussian process regression analysis, and the associated 1-$\sigma$ confidence interval, to the original (re-calibrated) samples.
    }
    \label{fig:mgrad_GPR_test}
\end{figure*}

For Carina (Fig.~\ref{fig:mgrad_GPR_test}, left panel), the [Fe/H] distribution resulted slightly metal-poorer than the original by $\sim$~0.05~dex, but the calculated gradient remained basically unchanged within the errors. On the other hand, for the SMC (Fig.~\ref{fig:mgrad_GPR_test}, right panel), the overall [Fe/H] distribution resulted more metal-poor by $\sim$~0.1~dex and the gradient resulted slightly steeper ($-0.130\pm0.007$~dex\,$R_e^{-1}$, vs $-0.101\pm0.006$~dex\,$R_e^{-1}$, i.e., different at a 3-$\sigma$ level). However, we warn that in this case we needed to use absolute $K_s$-band magnitudes for the calibration, while apparent magnitudes were provided in the original source \citep[i.e.,][]{Dobbie2014}, so differential reddening may be an additional issue here considering its significance in the SMC. Nevertheless, the change in the gradient value for the SMC is small (i.e., of the order of 0.05~dex\,$R_e^{-1}$), and does not change the conclusions we have reached in the main text.

\subsection{Average metallicity properties}

Here we discuss the results of the sanity checks performed on the average metallicity properties of our galaxy sample.

In particular, we verified that the global values of the galaxy metallicities we calculate (i.e., their weighted average [Fe/H]) were in good agreement with those expected according to the \citet[][]{Kirby2013} stellar luminosity-metallicity relation.

\begin{figure*}
    \centering
    \includegraphics[width=0.49\textwidth]{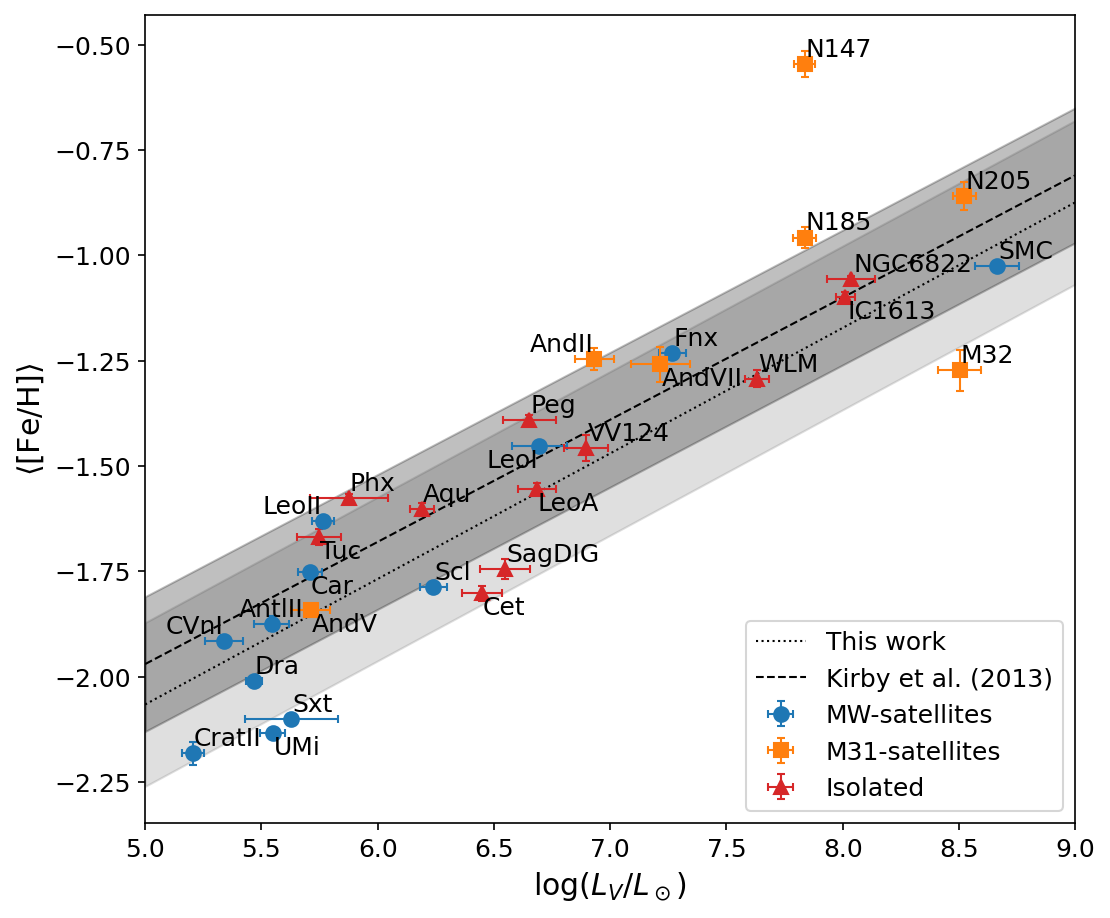}
    \includegraphics[width=0.49\textwidth]{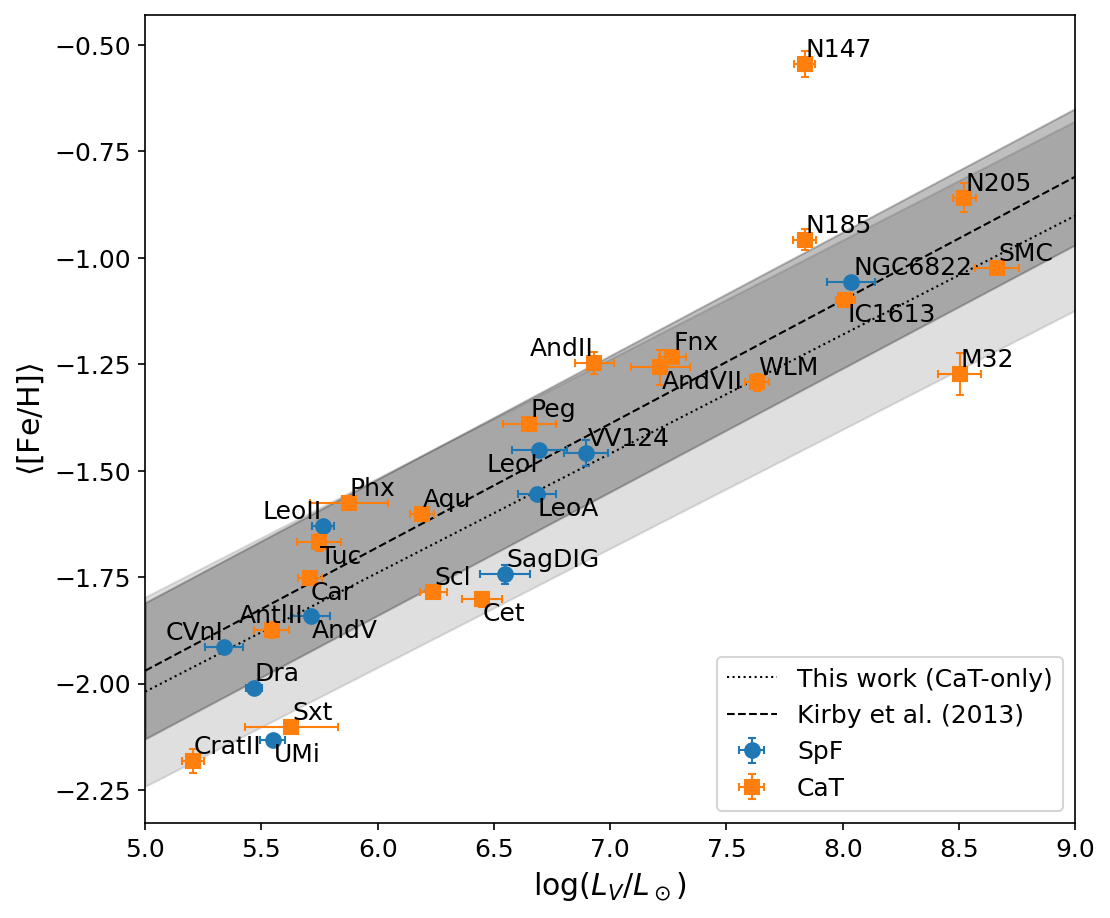}\\
    \includegraphics[width=0.49\textwidth]{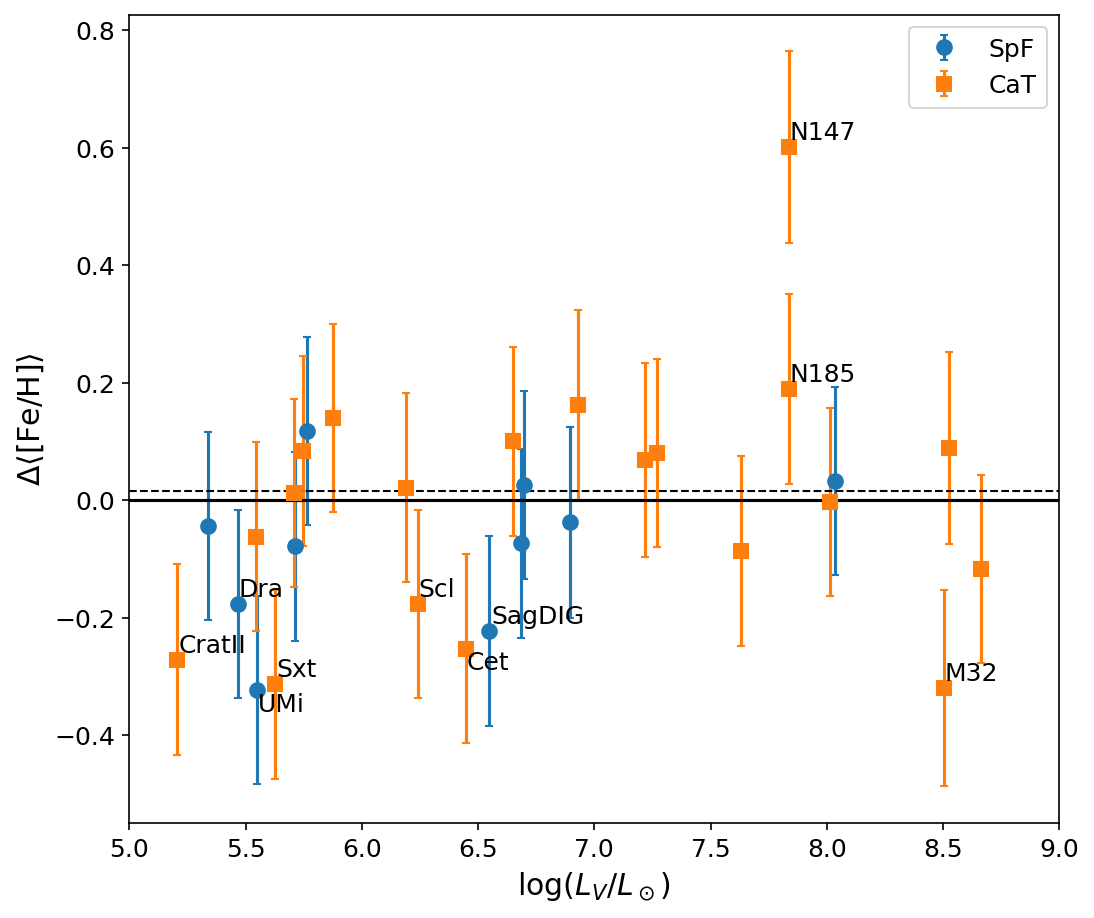}
    \includegraphics[width=0.49\textwidth]{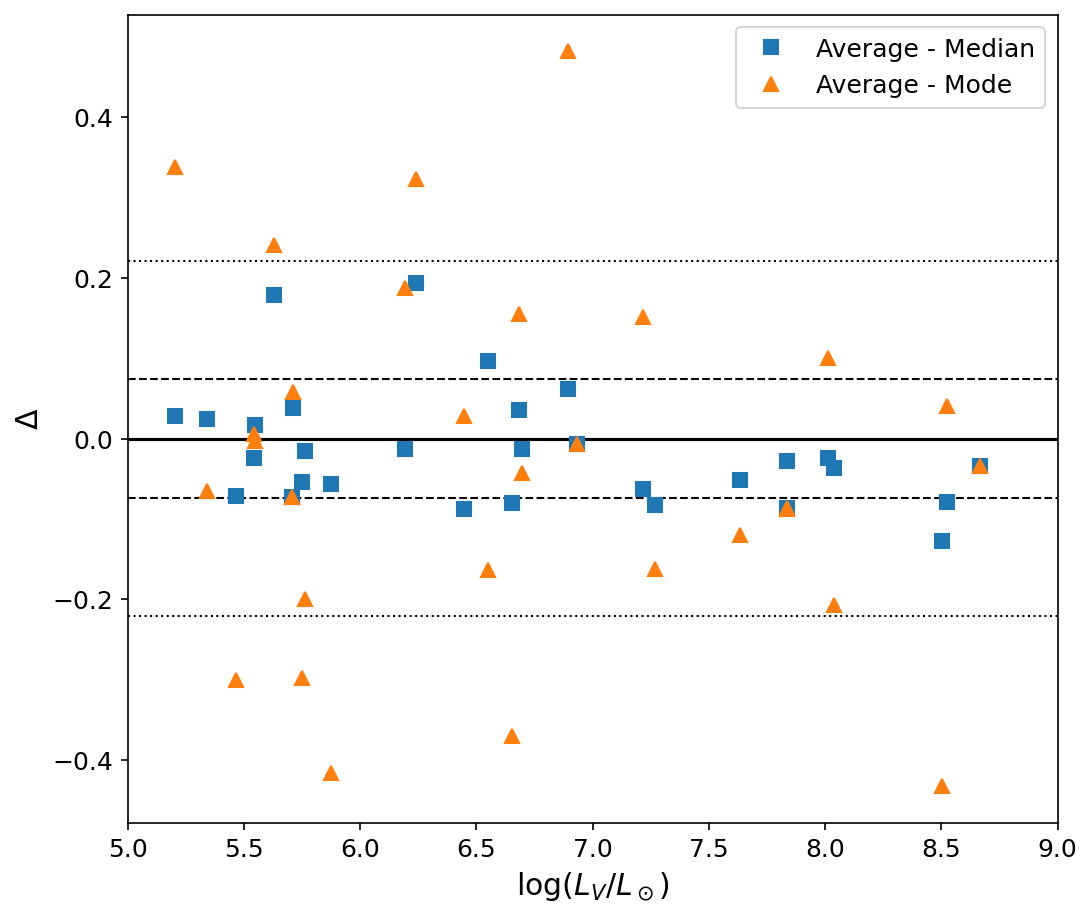}
    \caption{\textit{Top left:} stellar luminosity-metallicity diagram for our sample of dwarf galaxies; MW and M31 satellites are indicated with blue circles and orange squares, while the isolated systems as red triangles; overplotted the linear relationship recovered for our sample, indicated by a dotted line and a \textit{rms}-band in light-grey, while a dashed line and \textit{rms}-band in dark-grey indicate the \cite{Kirby2013} relation. \textit{Top right:} same as the left panel, but with symbols indicating those systems whose metallicities have been obtained with the CaT calibration method (orange squares) or using a spectral fitting technique (blue circles); overplotted the linear relationship recovered for the sub-sample using the CaT-method, indicated by a dotted line and a \textit{rms}-band in light-grey.
    \textit{Bottom left:} difference between the weighted-average of the metallicity values of our galaxies and those obtained from the \citet{Kirby2013} relation at their luminosity; symbols as in the top right panel; error bars take into account the individual errors on the weighted-average metallicity values and the \textit{rms} value of 0.16~dex of the Kirby et al.\, relation; the dashed line is the median offset of $0.02$~dex for the sub-sample using the CaT-method.
    \textit{Bottom right:} difference between the weighted-average and the median (mode) of the metallicity values indicated with blue squares (orange triangles) as a function of stellar luminosity; dashed (dotted) lines indicate the \textit{rms} scatter.
    }
    \label{fig:lum_met}
\end{figure*}

Figure~\ref{fig:lum_met} (top left) shows that the stellar luminosity-metallicity relation obtained by performing a linear fit to our determinations of the global [Fe/H] is in good agreement with that by \citet[][their Eq.~3]{Kirby2013}, with the slope and intercept being within $\sim$\,2-$\sigma$ from each other determinations. Considering the root-mean-square about the best-fitting line, we found a higher value than that reported by Kirby et al.~(i.e., 0.19~dex vs 0.16~dex), probably due to the heterogeneity of methods to derive metallicities in our sample. Nevertheless, the two relationships show considerable overlap between them, demonstrating the absence of large systematic offset in the data obtained by different methods.
Performing the linear fit only to those galaxies with [Fe/H] measurements derived with the CaT calibration method, the derived parameters show even better agreement with the Kirby et al.~relation (i.e., at $\sim$\,1-$\sigma$ level), despite a larger scatter ($rms=0.22$~dex) compared to the full sample (see the top-right panel). The absence of significant deviations between the global [Fe/H] of our galaxies and those obtained from Kirby et al.~relation at their luminosity is further evidenced in the bottom left panel, where we recover a median difference value of only $-0.02$~dex for the full sample. Considering the CaT-based measurements, we also recover a small value of $0.02$~dex. Such result shows how eventual variations in the [Ca/Fe] ratio, inherent to those data-set relying on the CaT-method, do not have a significant impact in the recovery of the average metallicity properties.

Finally, in the bottom-right panel, we investigate whether using different ways of quantifying the galaxies’ global metallicity may cause significant changes on the stellar luminosity-metallicity relation: we find that using the median [Fe/H], the weighted average on [Fe/H] or the mode of the metallicity distribution function results in very similar relations, at least for the sample of galaxies here considered, but with the mode introducing a larger scatter, as clearly shown in the figure. We provide the values per galaxy in Table~\ref{tab:met_grad_circ}.

\subsection{Alternative method to calculate the gradient}

\begin{figure*}
    \centering
    \includegraphics[width=0.49\textwidth]{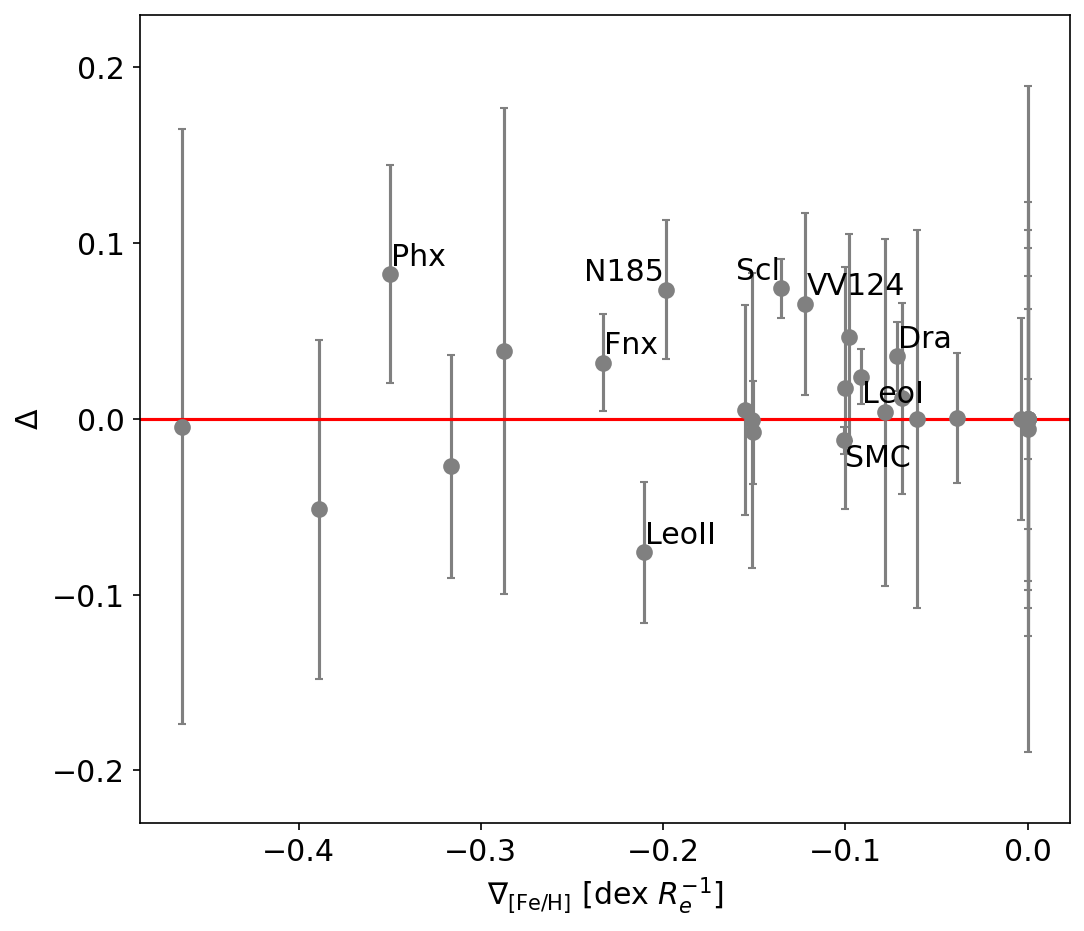}
    \includegraphics[width=0.49\textwidth]{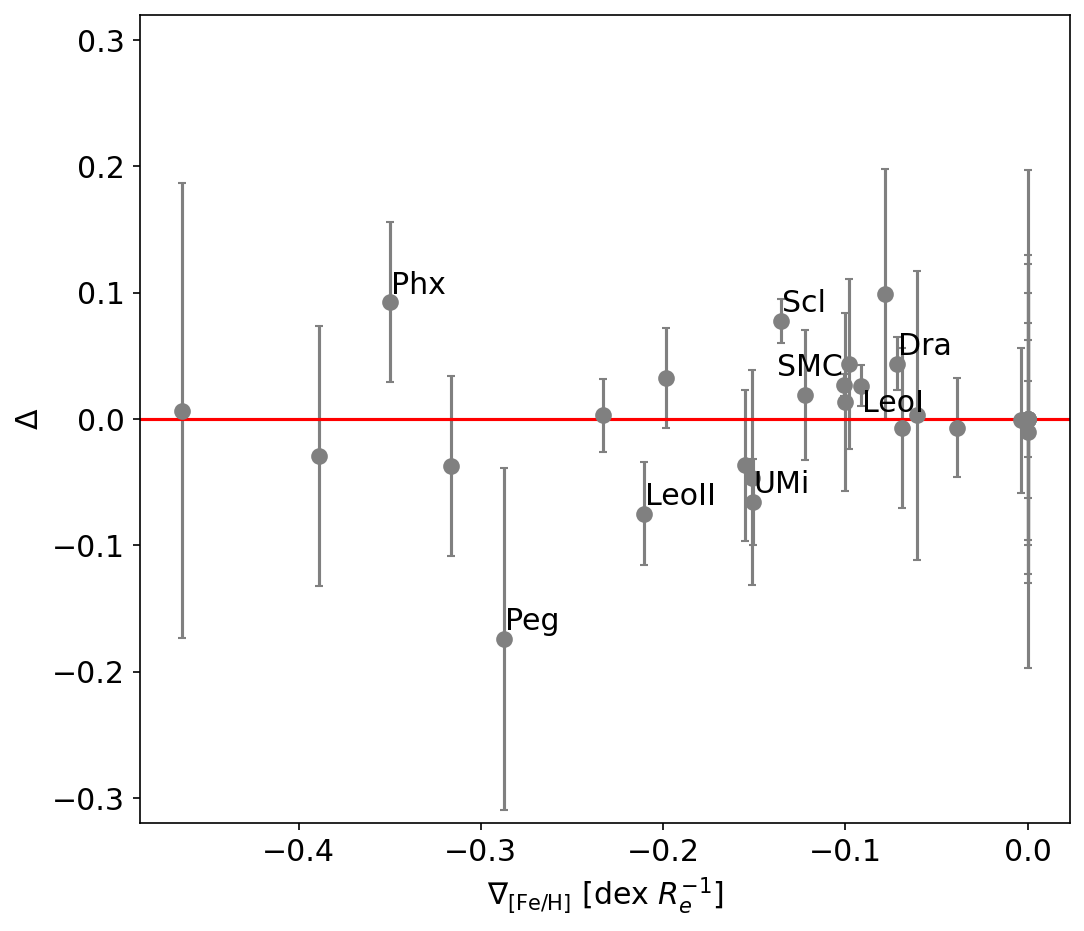}
    \caption{\textit{Left panel:} difference between the assumed metallicity gradients (on the x-axis) and those calculated between the centre and $2\times R_e$ of the GPR curves. \textit{Right panel:} as the left one, but showing the difference with values calculated assuming circular radii (see Sect.~\ref{subsec:met-grad-simul-M21}). Galaxy names in both panels indicate those systems for which a difference $>1$-$\sigma$ is recovered.
    }
    \label{fig:mgrad-vs}
\end{figure*}

We finally report in Fig.~\ref{fig:mgrad-vs} results on the comparison between the calculated metallicity gradients $\nabla_{\rm [Fe/H]} (R/R_e)$ used in our analysis and those obtained with an alternative method, namely by calculating the gradient as the difference between the initial value of the GPR curves and that at $2\times R_e$ (on the left panel, while on the right one assuming circular radii, as done in Sect.~\ref{subsec:met-grad-simul-M21}). We observe a general good agreement, with 21 (22, for the circular radii case) out of 30 measurements in agreement within 1-$\sigma$, while 29 out of 30 agreeing within 2-$\sigma$.

\section{Individual GPR profiles}
\label{sec:apx1}

We report here, as supplementary material, the plots of the individual [Fe/H] measurements (which sources are listed in Table~\ref{tab:sample}) as a function of the projected SMA radius in units of the 2D half-light radius, and their associated Gaussian process regression fits. Each panel report in a box the name of the corresponding dwarf galaxy.
We further provide Table~\ref{tab:data_ref} with the references to the parameters adopted for those systems not included in the \citet{Battaglia2022} compilation.

\begin{table}
\caption{References to the parameters adopted for those systems not included in the \citet{Battaglia2022} compilation (i.e., SMC, the M31 satellites and the isolated systems Cetus, Tucana, and Aquarius). For each system, the numeric code indicates the reference for the central coordinates, distance modulus, 2D SMA half-light radius, position angle, ellipticity ($e=1-b/a$), and velocity dispersion, respectively.
\textbf{References:} (1) \citet{McConnachie2012}; (2) \citet{DeLeo2020}; (3) \citet{Subran+Subram2012}; (4) \citet{Conn2012}; (5) \citet{Crnojevic2014}; (6) \citet{Geha2010}; (7) \citet{Kirby2020}; (8) \citet{Taibi2018}; (9) \citet{Taibi2020}; (10) \citet{Higgs2021}; (11) \citet{Hermosa2020}}
\label{tab:data_ref}
\centering
\begin{tabular}{cccccc}
\hline
\hline
  \multicolumn{1}{c}{Galaxy} &
  \multicolumn{1}{c}{(RA, Dec)} &
  \multicolumn{1}{c}{D.~M.} &
  \multicolumn{1}{c}{$R_e$} &
  \multicolumn{1}{c}{(P.~A.; $e$)} &
  \multicolumn{1}{c}{$\sigma_v$} \\
\hline
  SMC        & 1 & 1 & 2  & 3  & 2  \\
  M32        & 1 & 1 & 1  & 1  & 1  \\
  NGC205     & 1 & 1 & 1  & 1  & 6  \\
  NGC185     & 1 & 4 & 5  & 5  & 6  \\
  NGC147     & 1 & 4 & 5  & 5  & 6  \\
  AndVII     & 1 & 1 & 1  & 1  & 7 \\
  AndV       & 1 & 4 & 1  & 1  & 7 \\
  AndII      & 1 & 4 & 1  & 1  & 1  \\
  Cetus      & 1 & 1 & 1  & 1  & 8 \\
  Tucana     & 1 & 1 & 1  & 1  & 9 \\ 
  Aquarius   & 1 & 1 & 10 & 10 & 11 \\
\hline
\hline
\end{tabular}
\end{table}

\begin{figure*}
    \centering
    \includegraphics[width=0.49\textwidth]{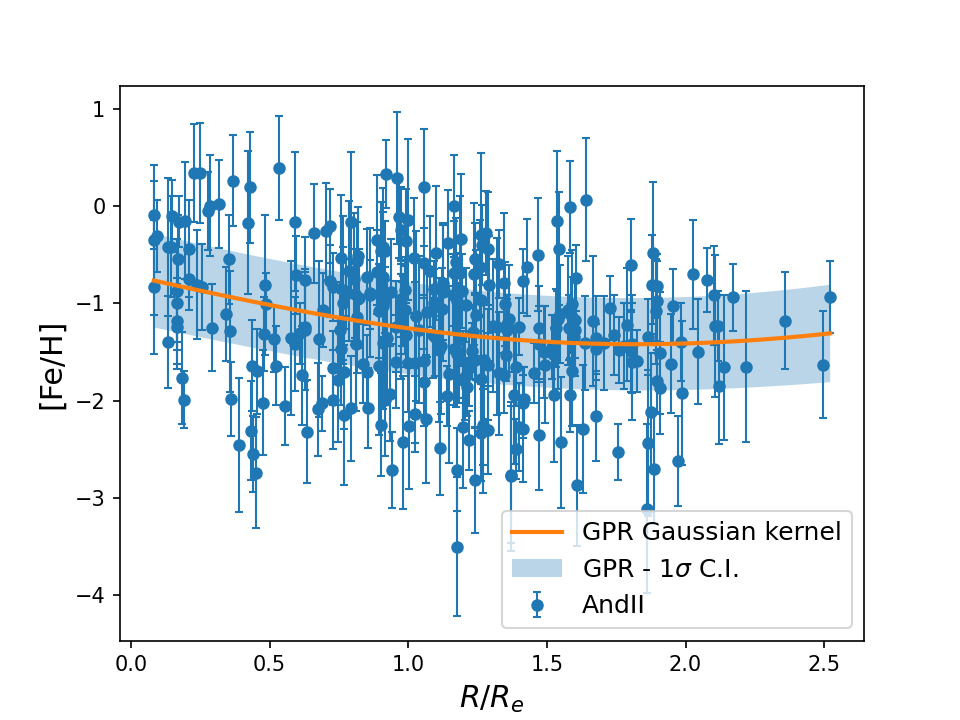}
    \includegraphics[width=0.49\textwidth]{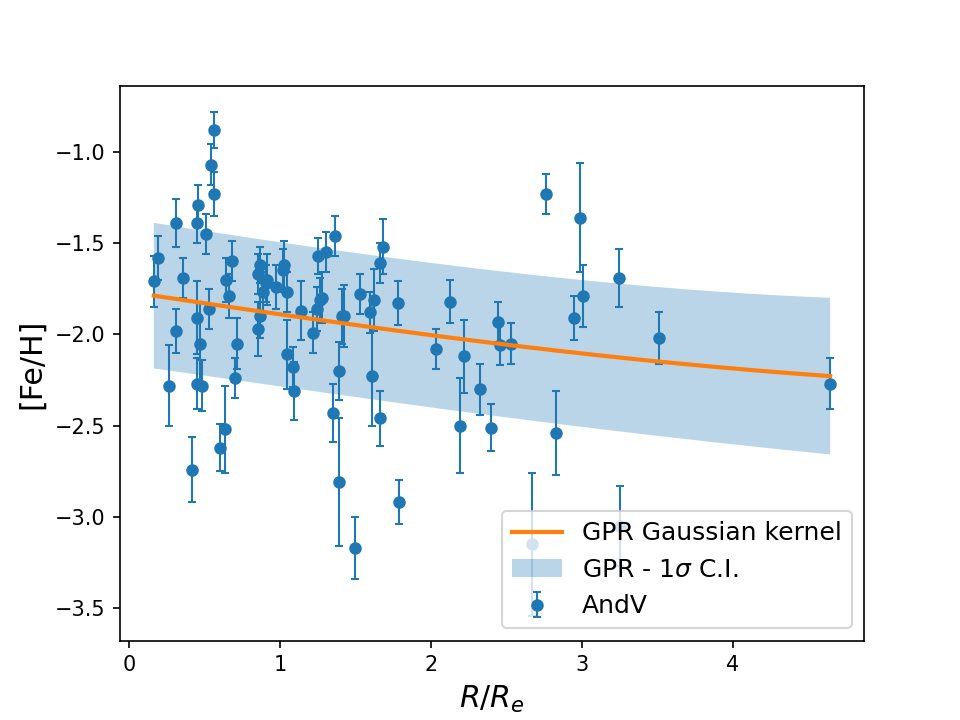}\\
    \includegraphics[width=0.49\textwidth]{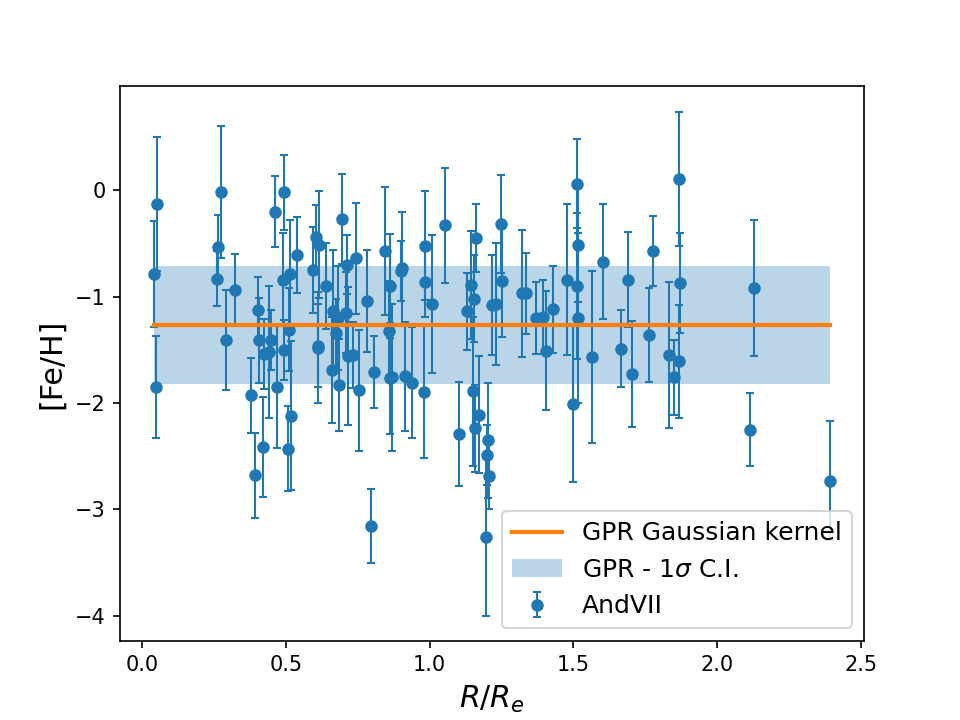}
    \includegraphics[width=0.49\textwidth]{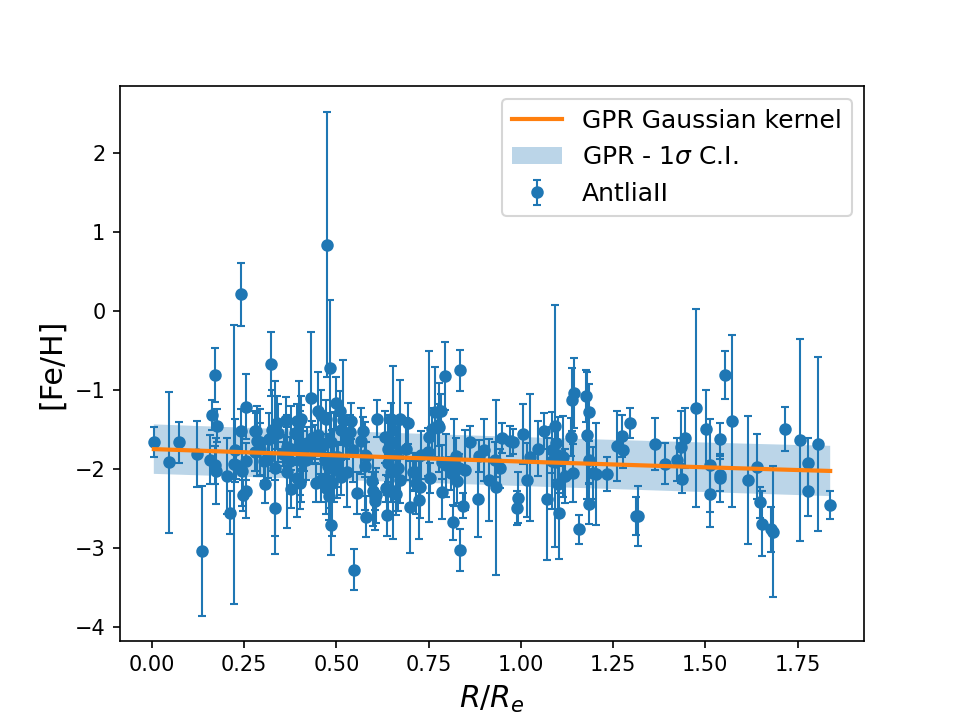}\\
    \includegraphics[width=0.49\textwidth]{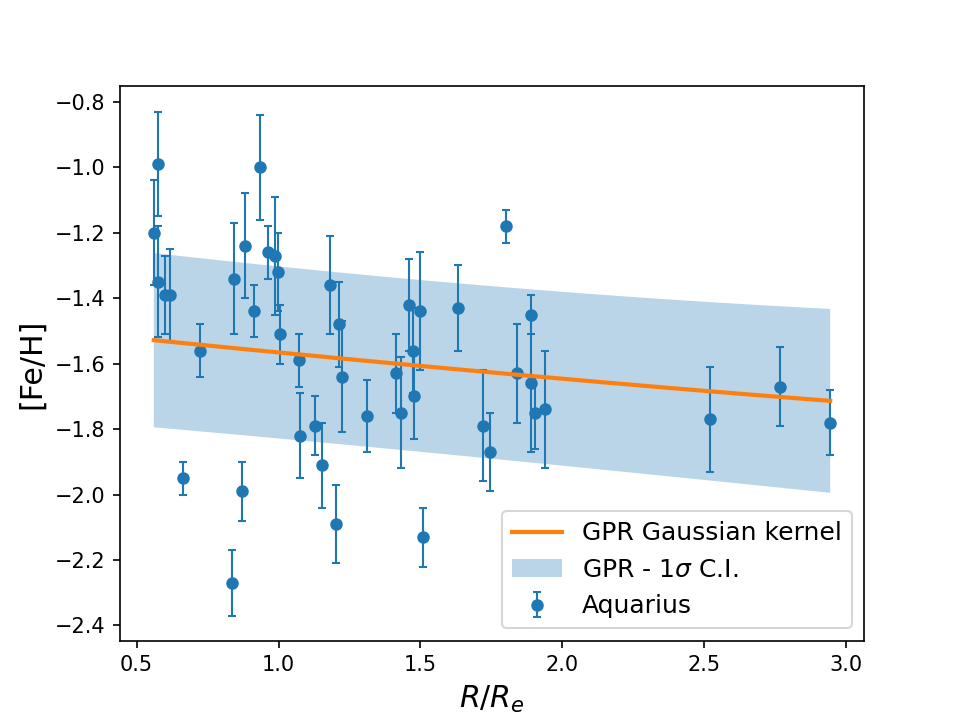}
    \includegraphics[width=0.49\textwidth]{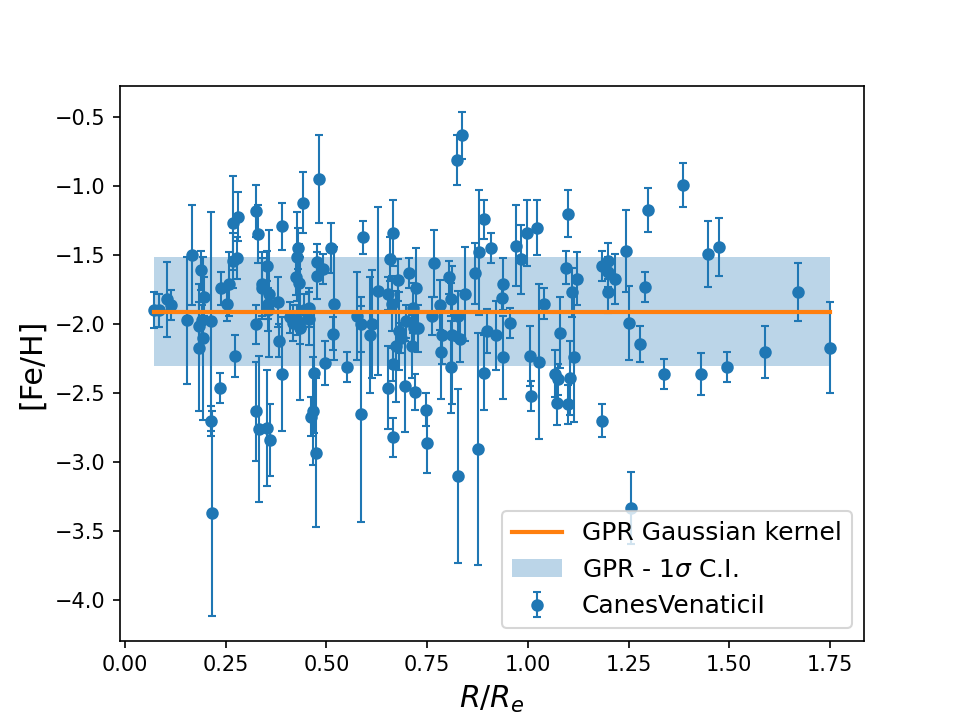}
    \caption{Distribution of individual [Fe/H] measurements and their errors, as a function of SMA radius in units of the 2D half-light radius, for the dwarf galaxies considered in this work. The orange solid line and blue shaded area represent the result of the Gaussian process regression analysis (using a Gaussian kernel) and the associated 1-$\sigma$ confidence interval. In the legend box of each panel is reported the name of the corresponding dwarf galaxy.
    }
    \label{fig:mgrad_GPR_1}
\end{figure*}

\begin{figure*}
    \centering
    \includegraphics[width=0.49\textwidth]{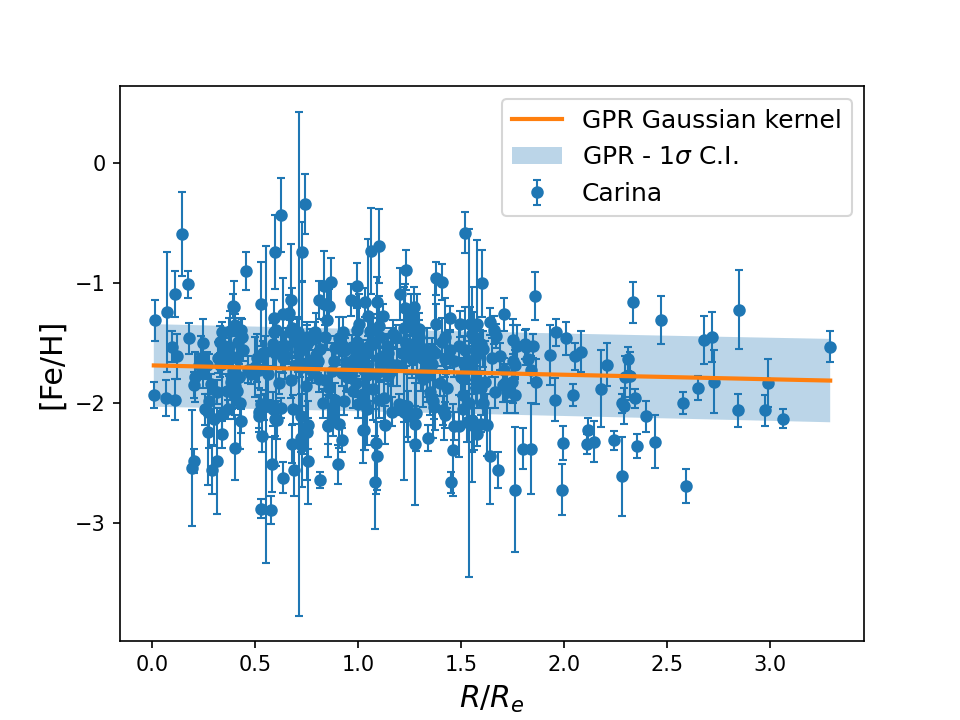}
    \includegraphics[width=0.49\textwidth]{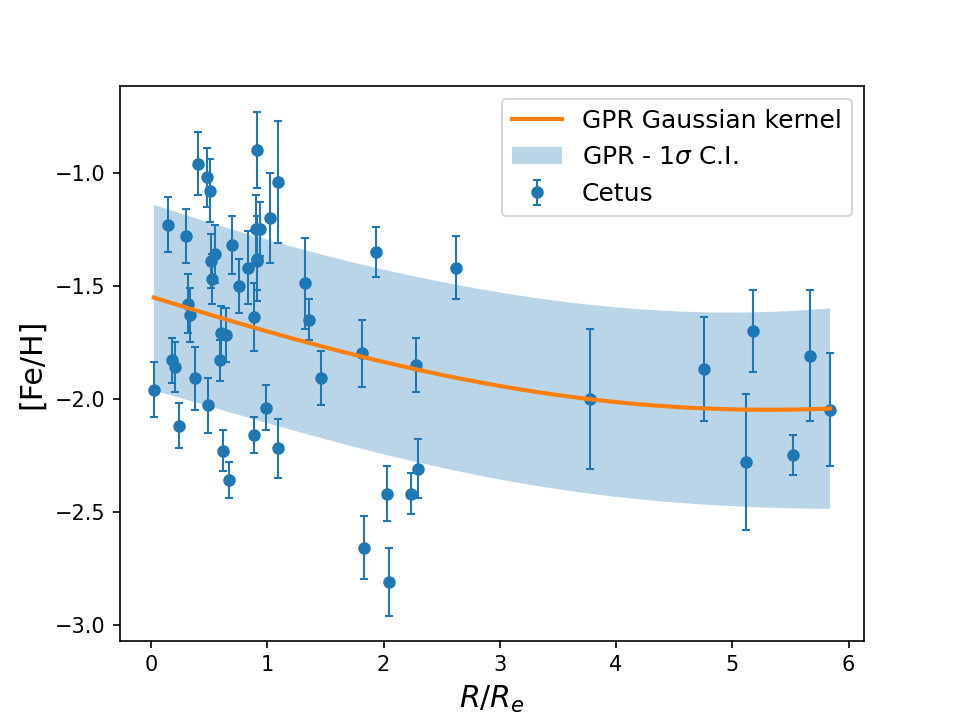}\\
    \includegraphics[width=0.49\textwidth]{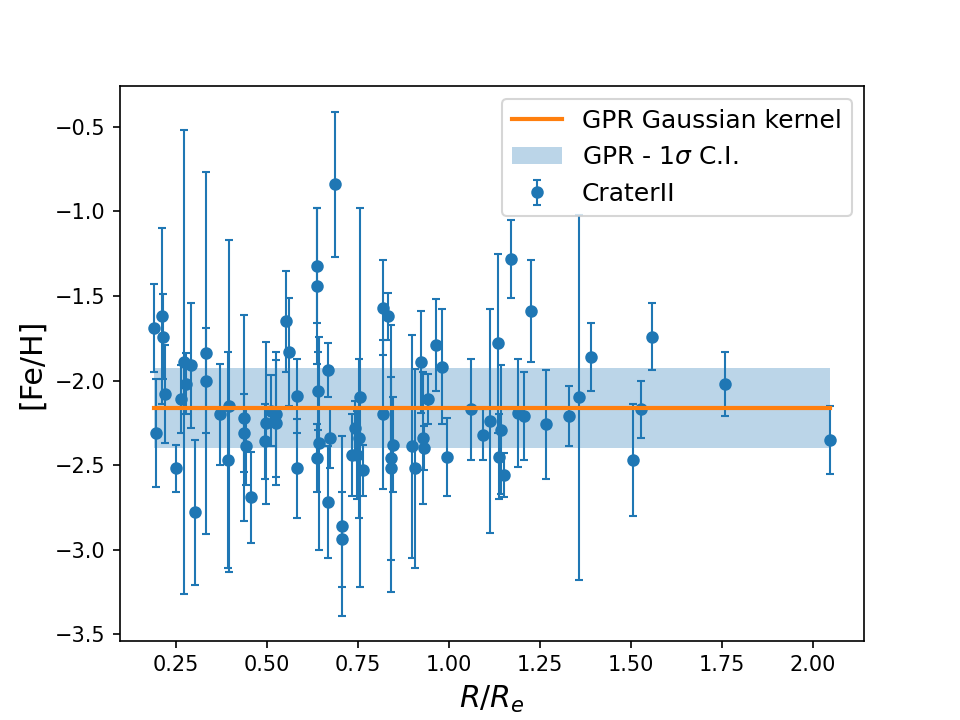}
    \includegraphics[width=0.49\textwidth]{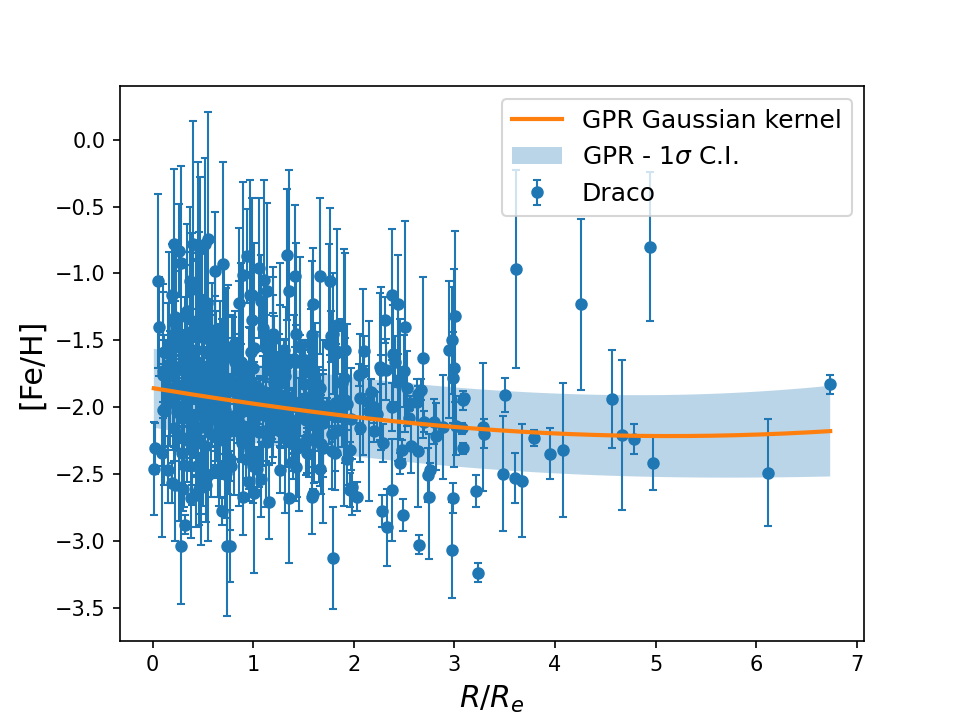}\\
    \includegraphics[width=0.49\textwidth]{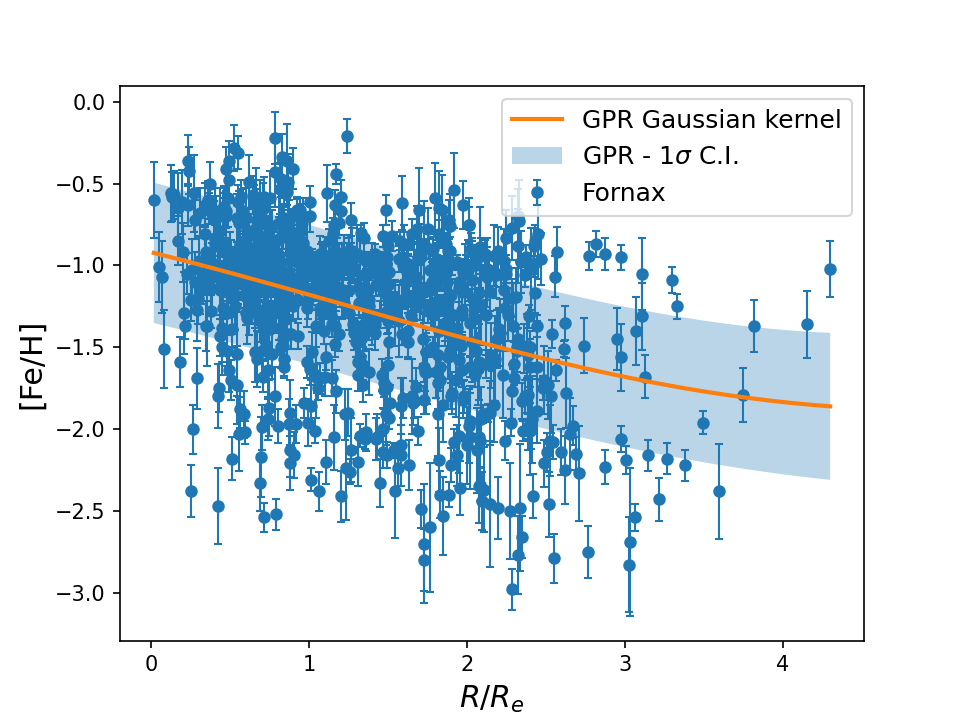}
    \includegraphics[width=0.49\textwidth]{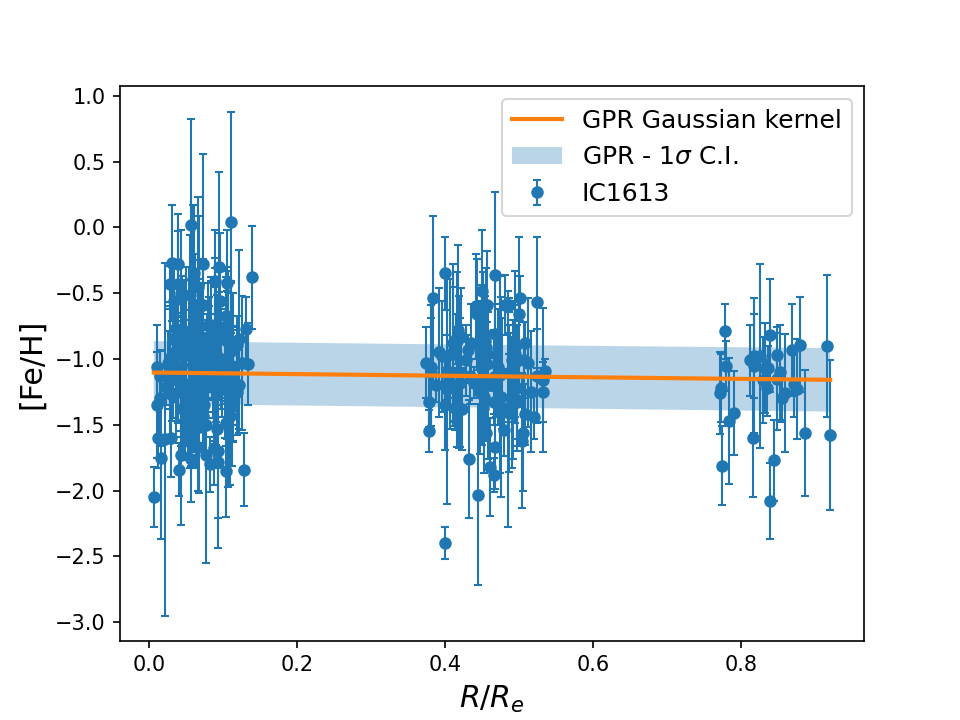}
    \caption{Follow as in Fig.~\ref{fig:mgrad_GPR_1}.
    }
    \label{fig:mgrad_GPR_2}
\end{figure*}

\begin{figure*}
    \centering
    \includegraphics[width=0.49\textwidth]{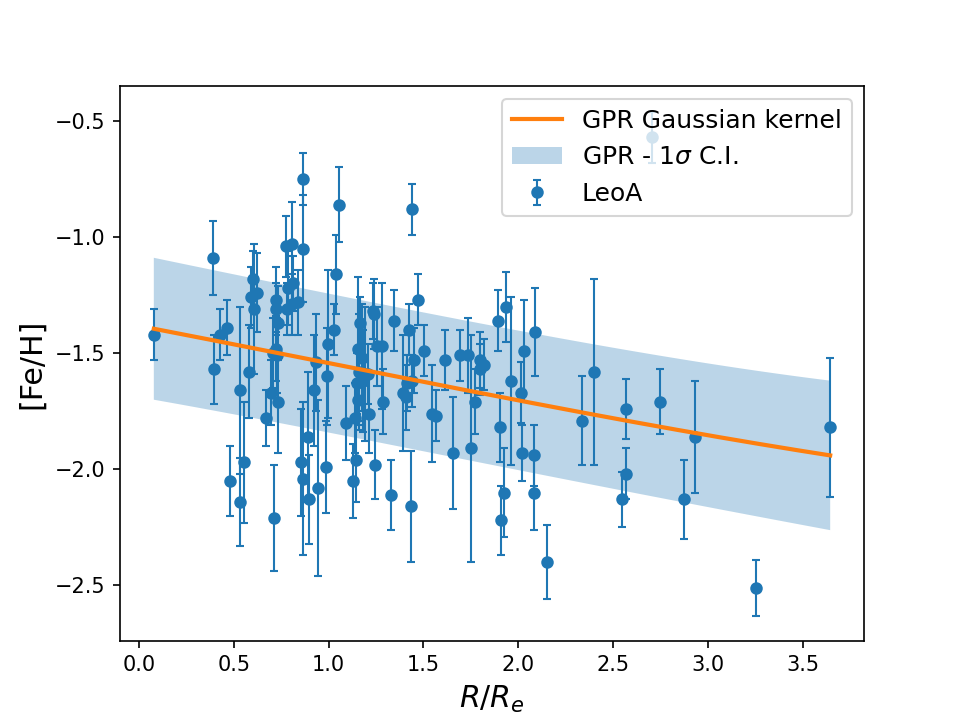}
    \includegraphics[width=0.49\textwidth]{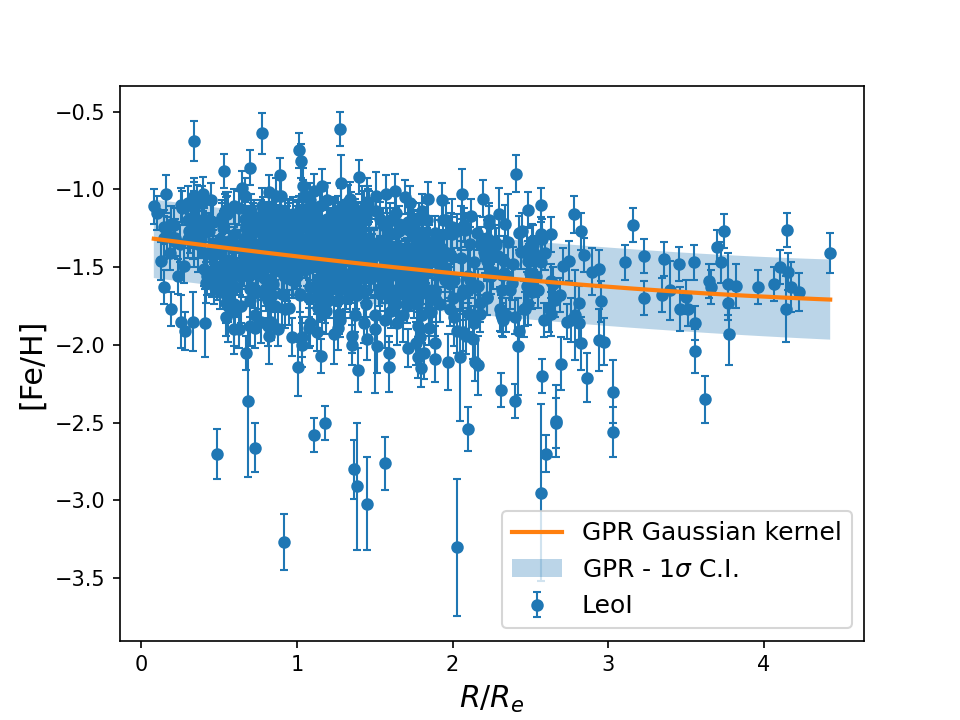}\\
    \includegraphics[width=0.49\textwidth]{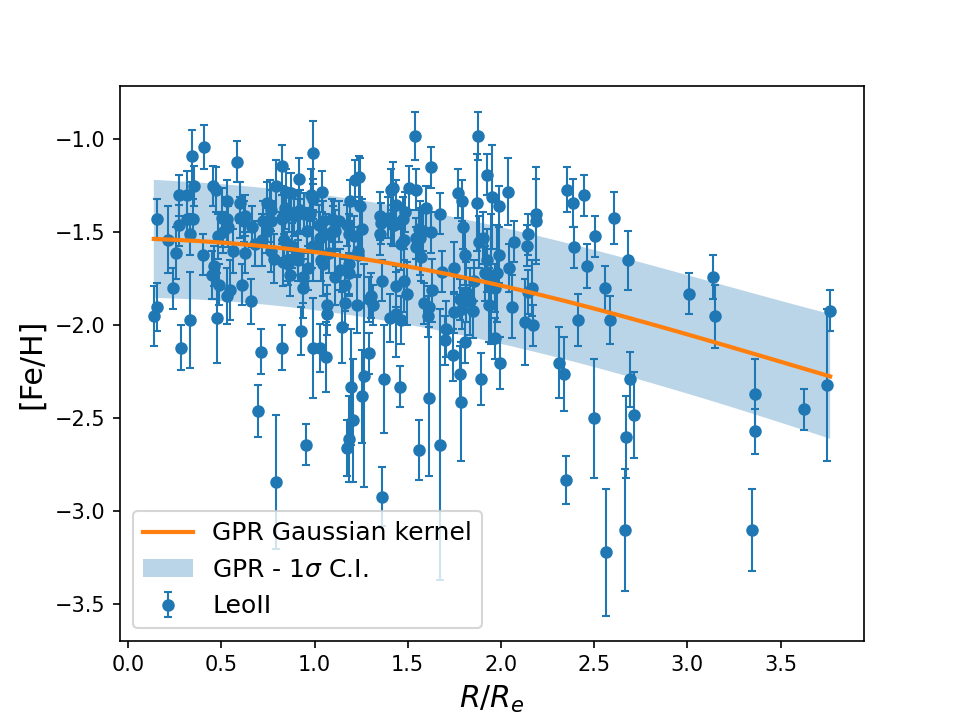}
    \includegraphics[width=0.49\textwidth]{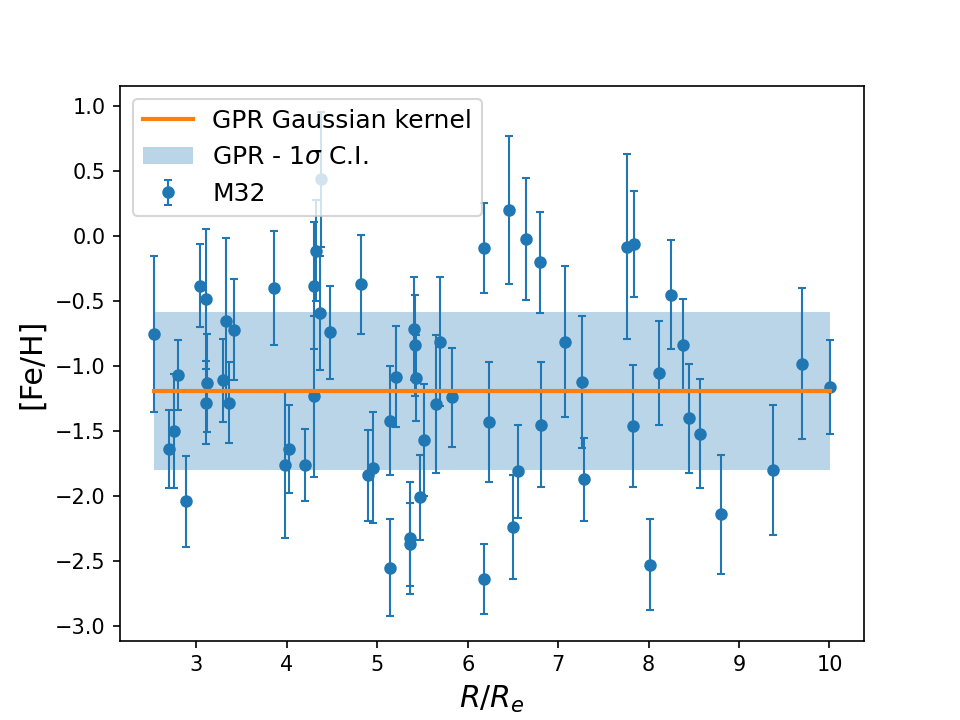}\\
    \includegraphics[width=0.49\textwidth]{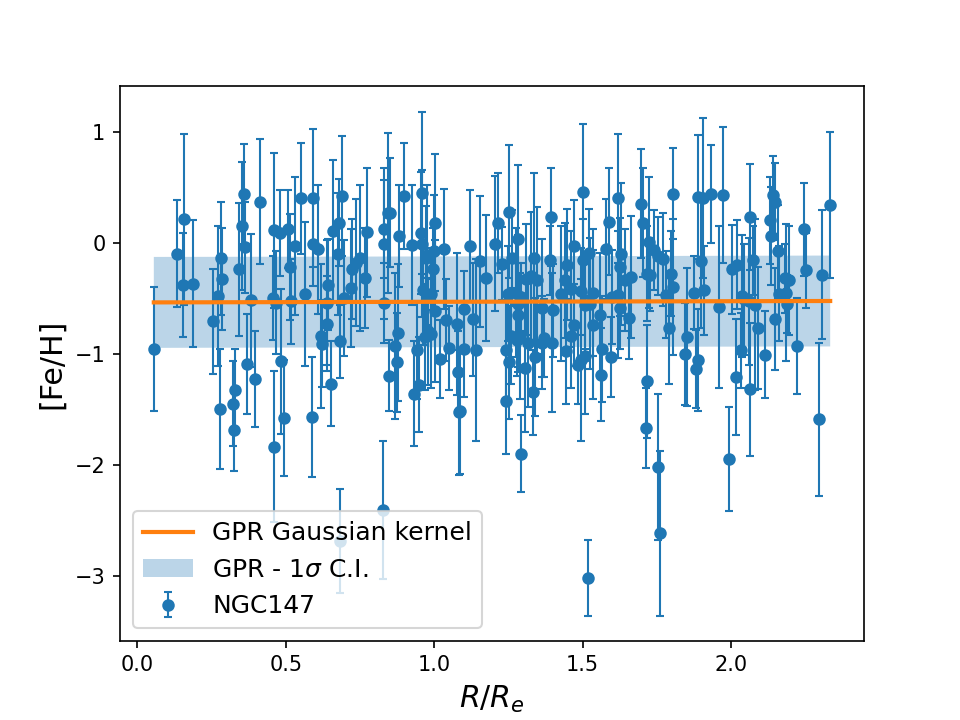}
    \includegraphics[width=0.49\textwidth]{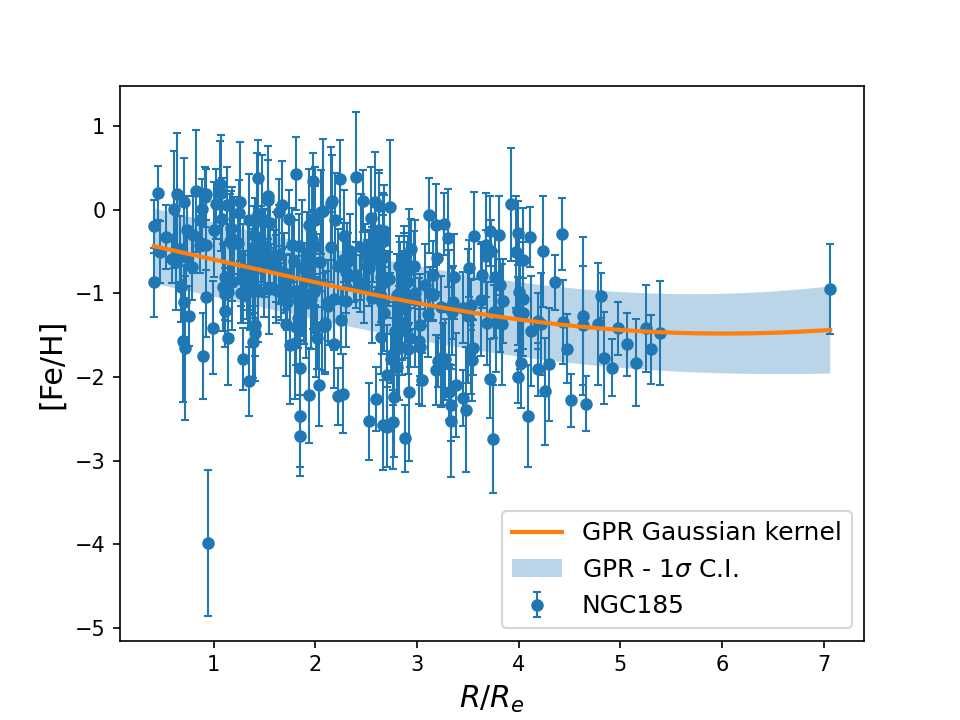}
    \caption{Follow as in Fig.~\ref{fig:mgrad_GPR_1}.
    }
    \label{fig:mgrad_GPR_3}
\end{figure*}

\begin{figure*}
    \centering
    \includegraphics[width=0.49\textwidth]{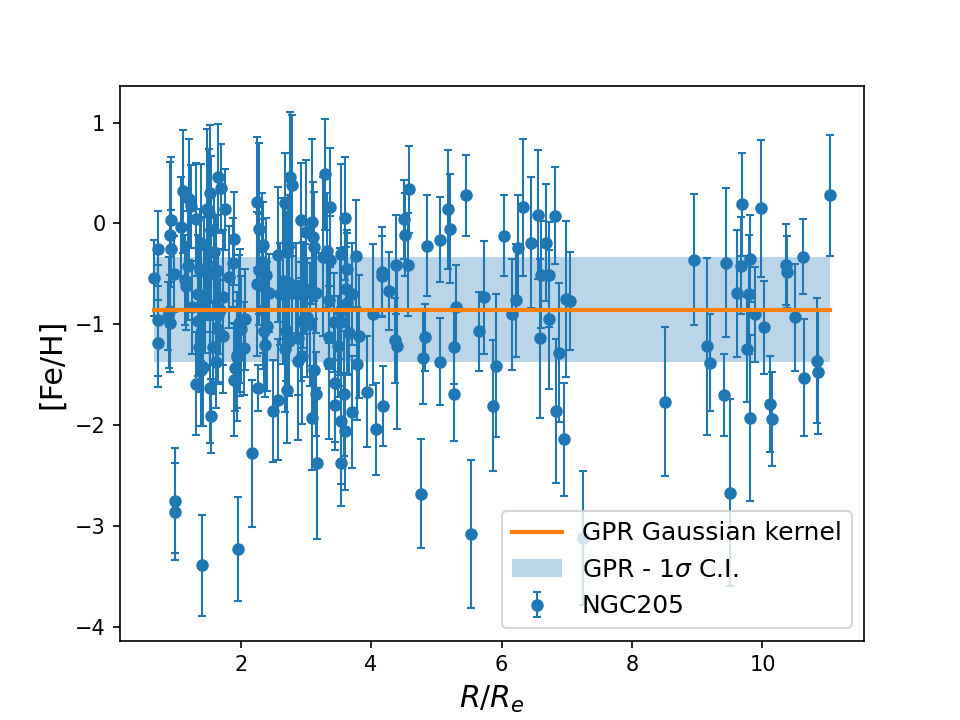}
    \includegraphics[width=0.49\textwidth]{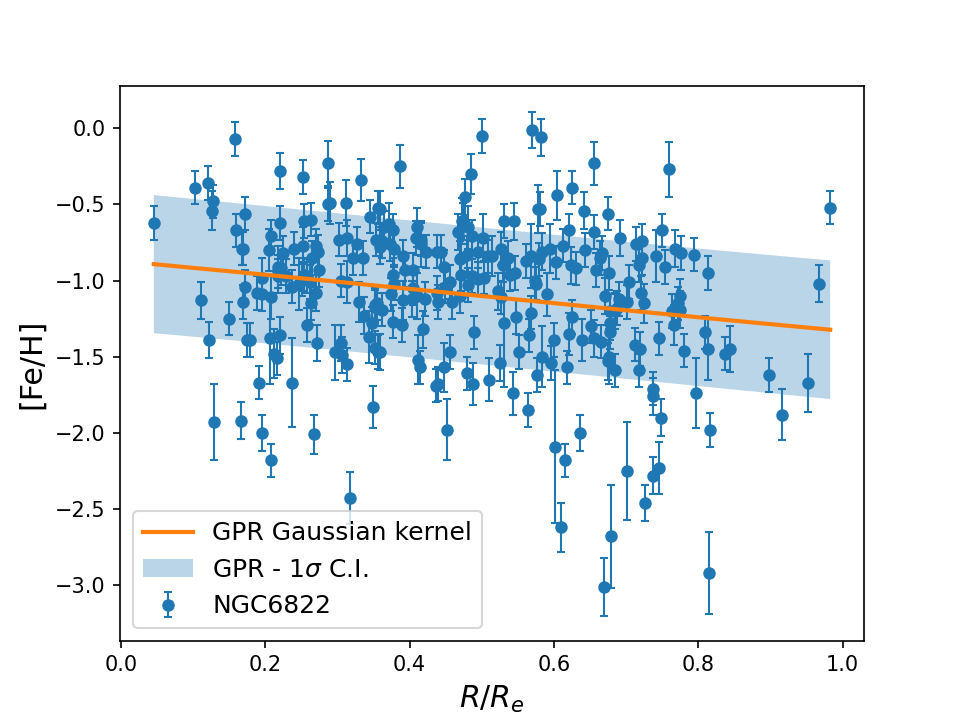}\\
    \includegraphics[width=0.49\textwidth]{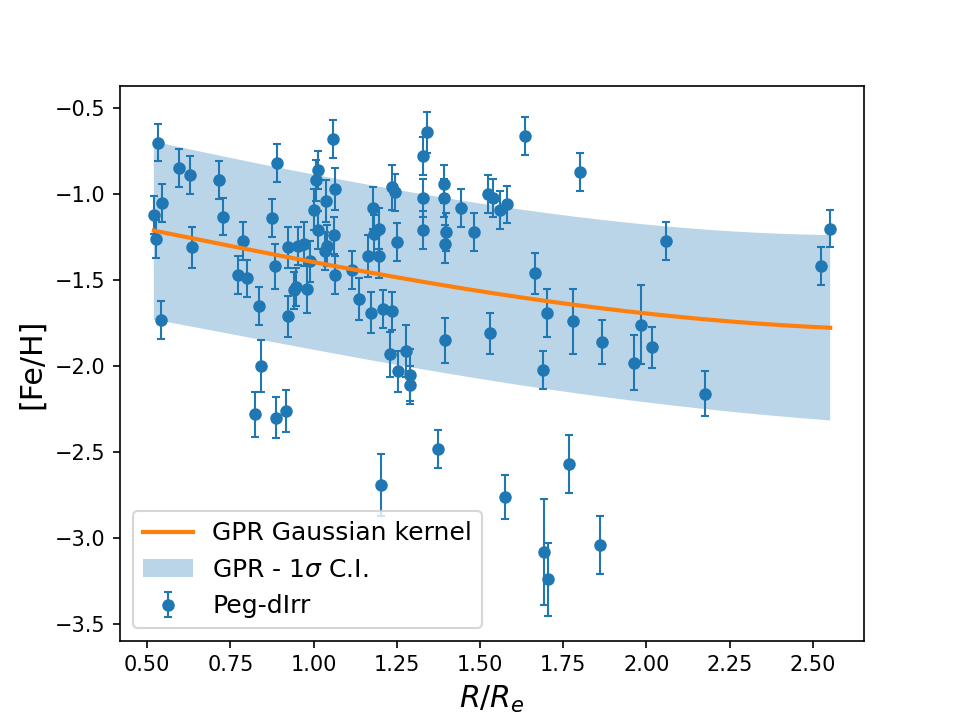}
    \includegraphics[width=0.49\textwidth]{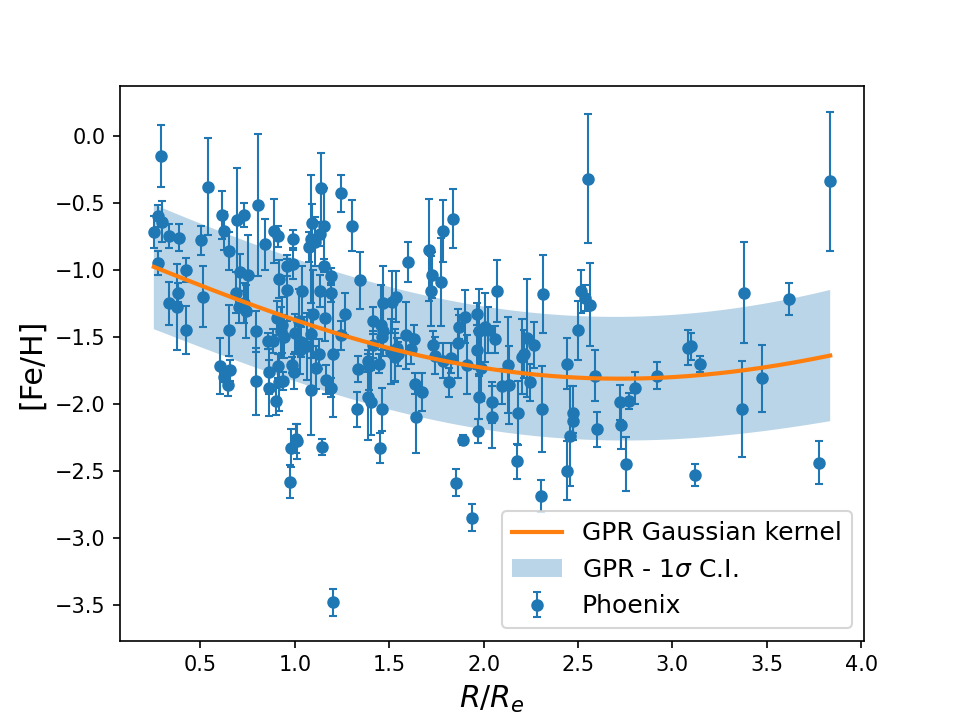}\\
    \includegraphics[width=0.49\textwidth]{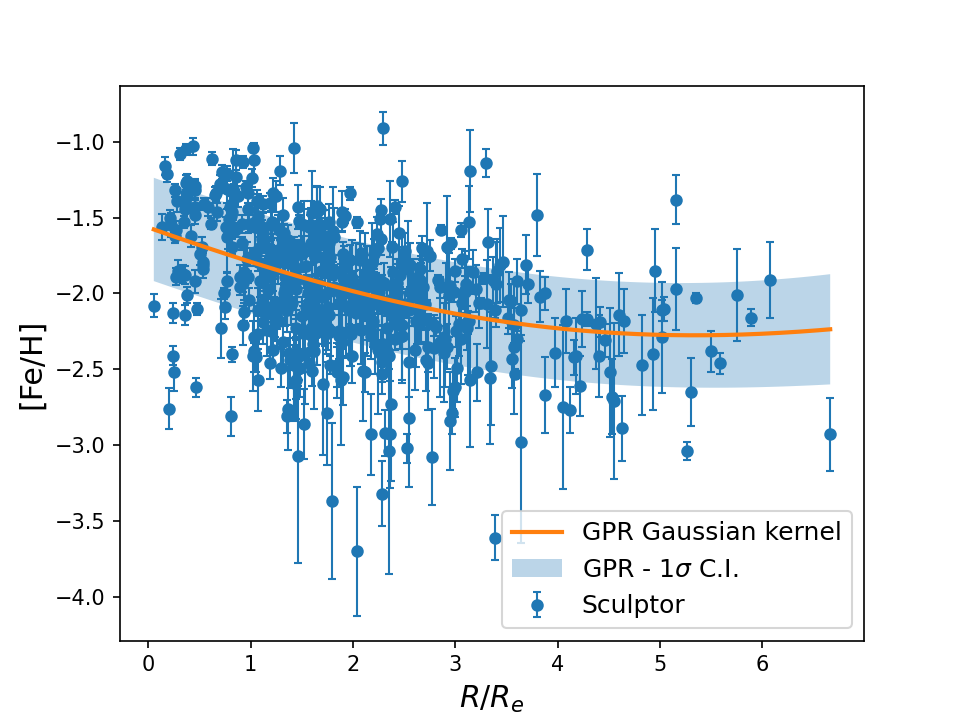}
    \includegraphics[width=0.49\textwidth]{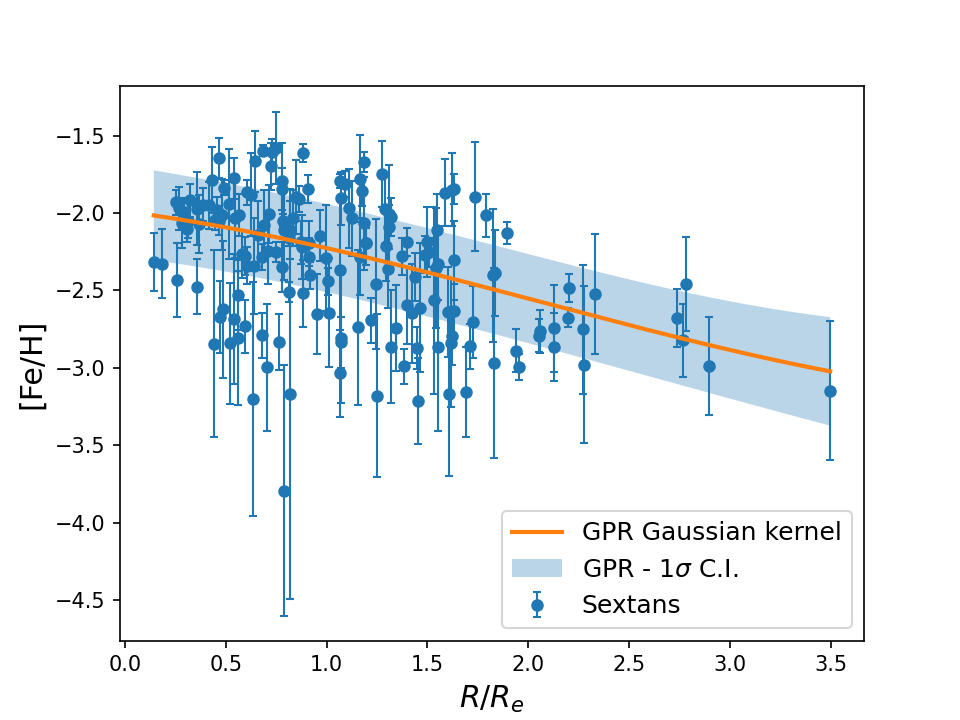}
    \caption{Follow as in Fig.~\ref{fig:mgrad_GPR_1}.
    }
    \label{fig:mgrad_GPR_4}
\end{figure*}

\begin{figure*}
    \centering
    \includegraphics[width=0.49\textwidth]{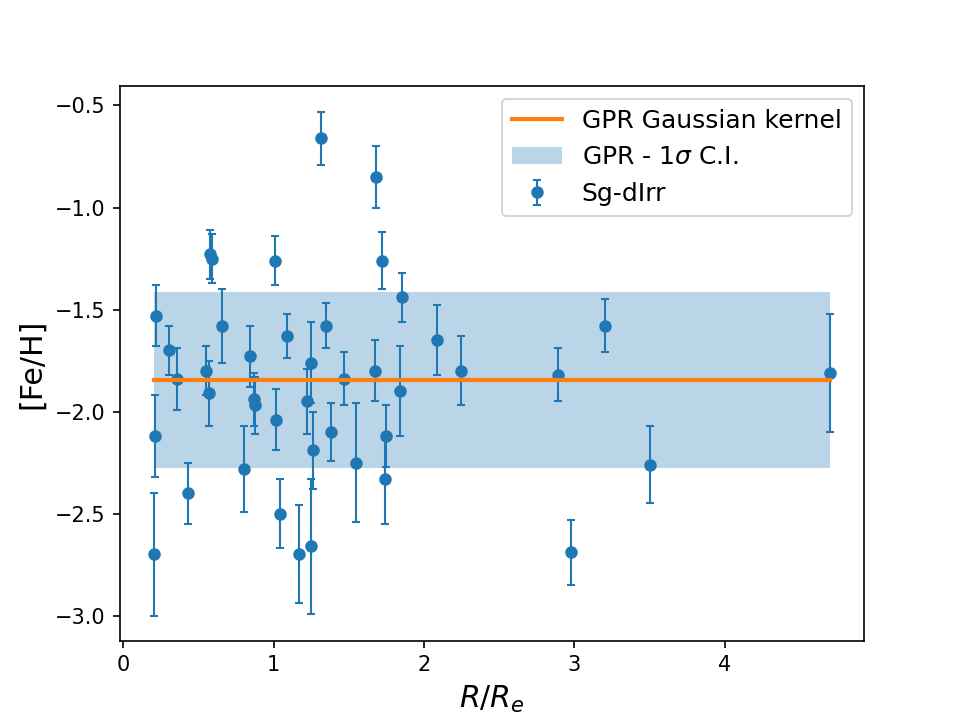}
    \includegraphics[width=0.49\textwidth]{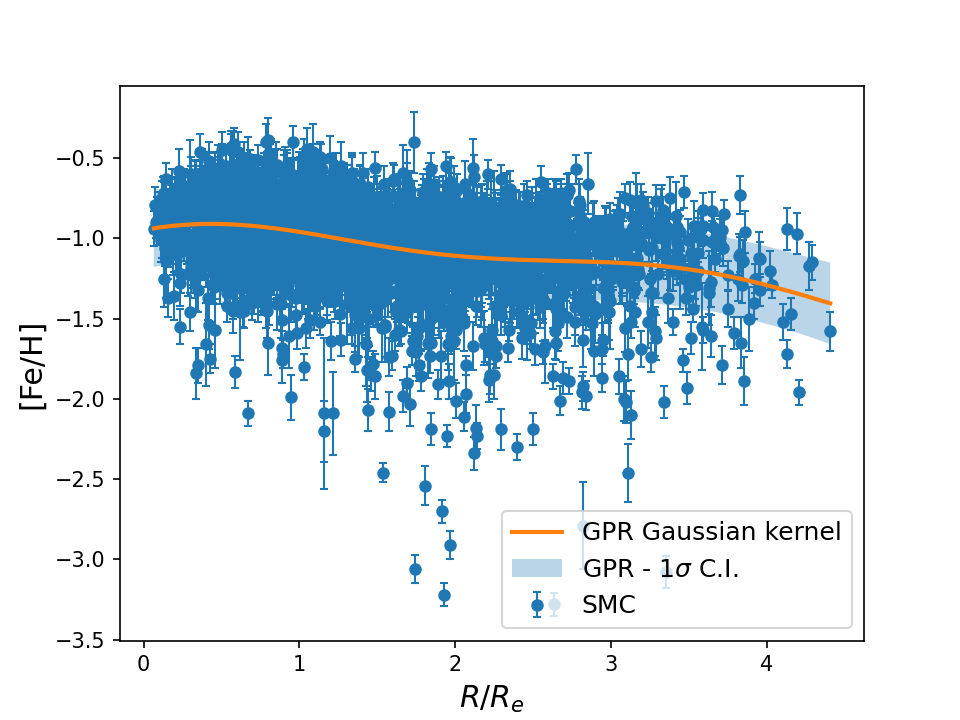}\\
    \includegraphics[width=0.49\textwidth]{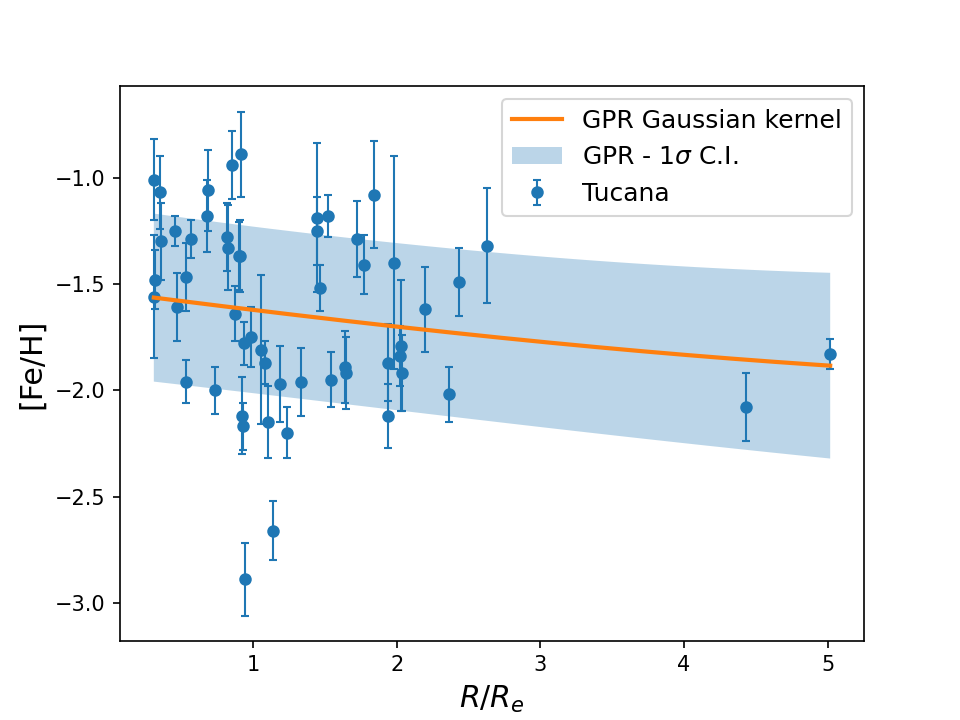}
    \includegraphics[width=0.49\textwidth]{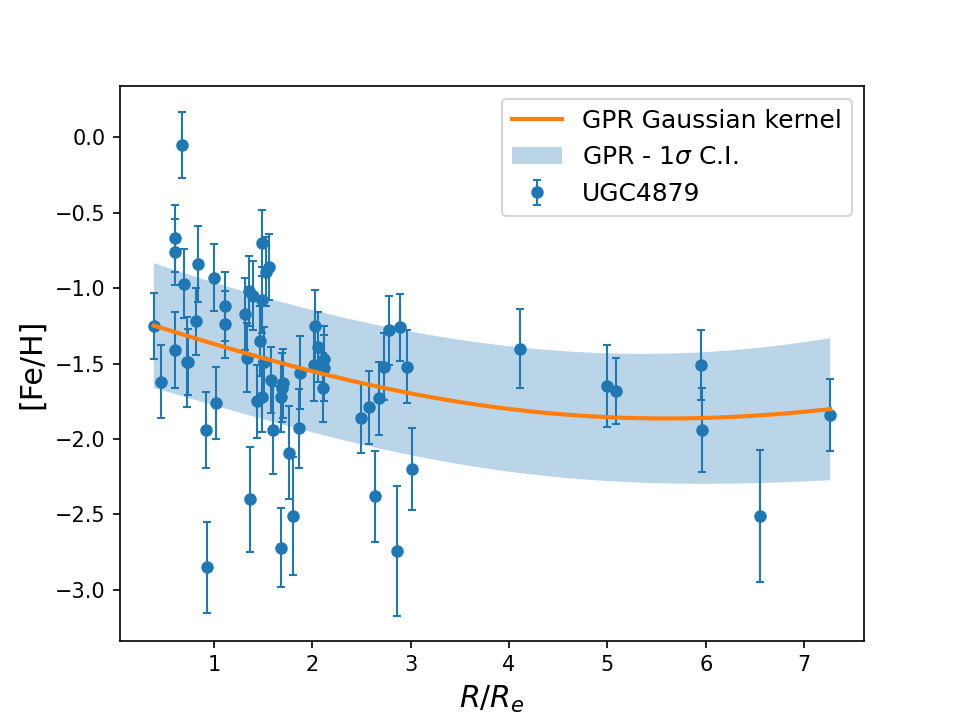}\\
    \includegraphics[width=0.49\textwidth]{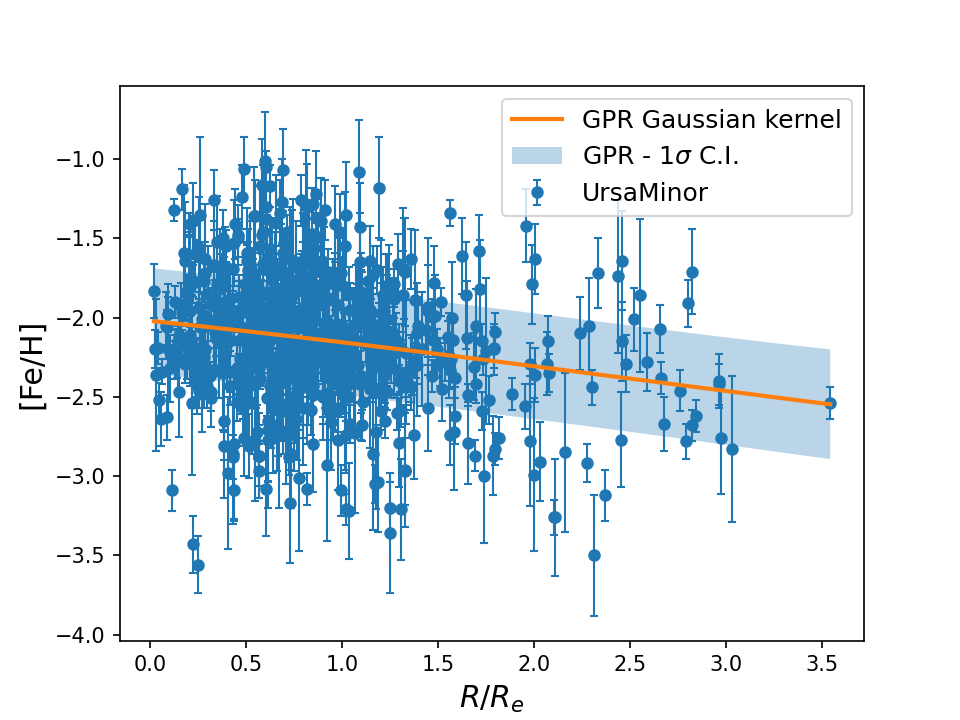}
    \includegraphics[width=0.49\textwidth]{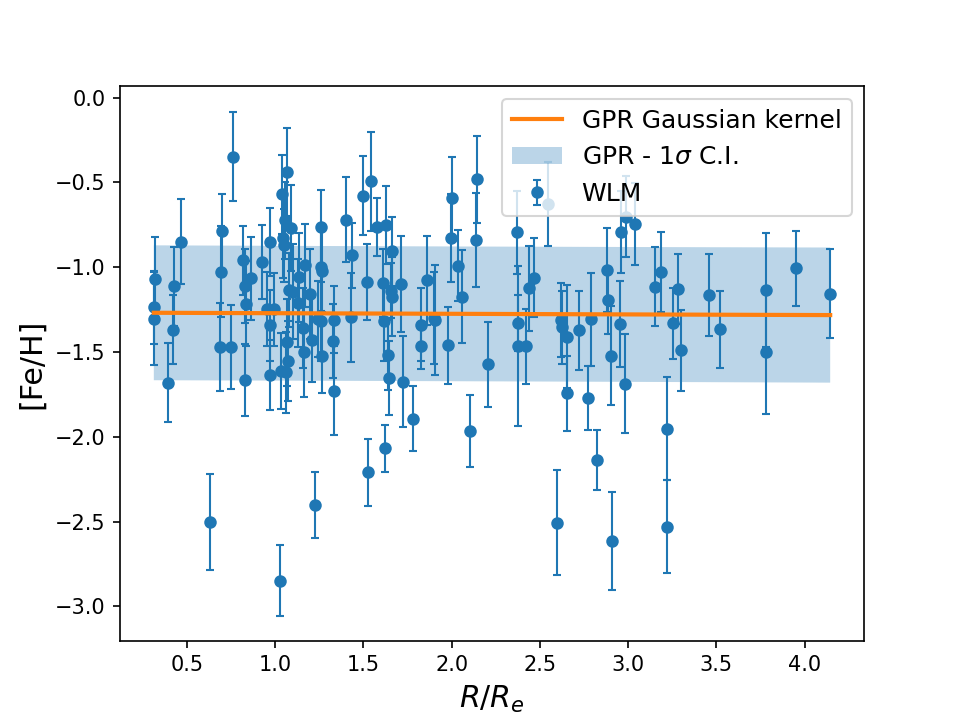}
    \caption{Follow as in Fig.~\ref{fig:mgrad_GPR_1}.
    }
    \label{fig:mgrad_GPR_5}
\end{figure*}

\section{Orbital parameters}
\label{sec:apx_orb}
We looked for possible correlations between the calculated metallicity gradients and the orbital properties of the Milky Way satellites considered in this work. We inspected the following parameters, recently made publicly available by \citet{Battaglia2022} and derived using \textit{Gaia}-eDR3 proper motions \citep{Gaia-eDR3_2021}: eccentricity, pericentric distance, orbital period, and time since the last pericentric passage. The inspected parameters were obtained assuming two Milky Way potentials (in which the MW is treated as an isolated system): a ``light-potential'', with a MW total mass of $M(<R_{\rm vir})=8.8\times10^{11}M_\odot$, and a ``heavy-potential'', with a MW total mass of $M(<R_{\rm vir})=1.6\times10^{12}M_\odot$. We refer to \citet{Battaglia2022} for further details on the MW-potentials and the orbital integration.

\begin{figure*}
    \centering
    \includegraphics[width=.9\textwidth]{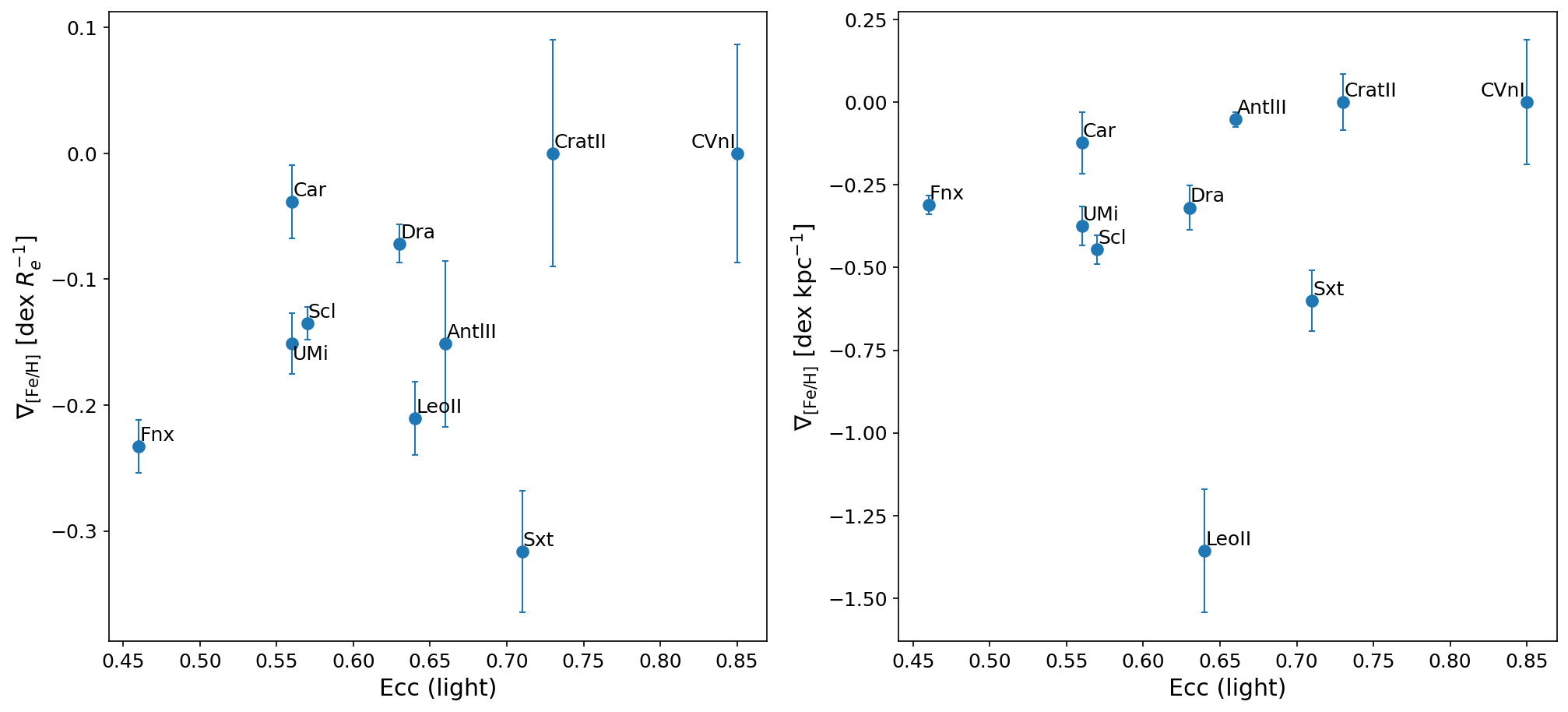}\\
    \includegraphics[width=.9\textwidth]{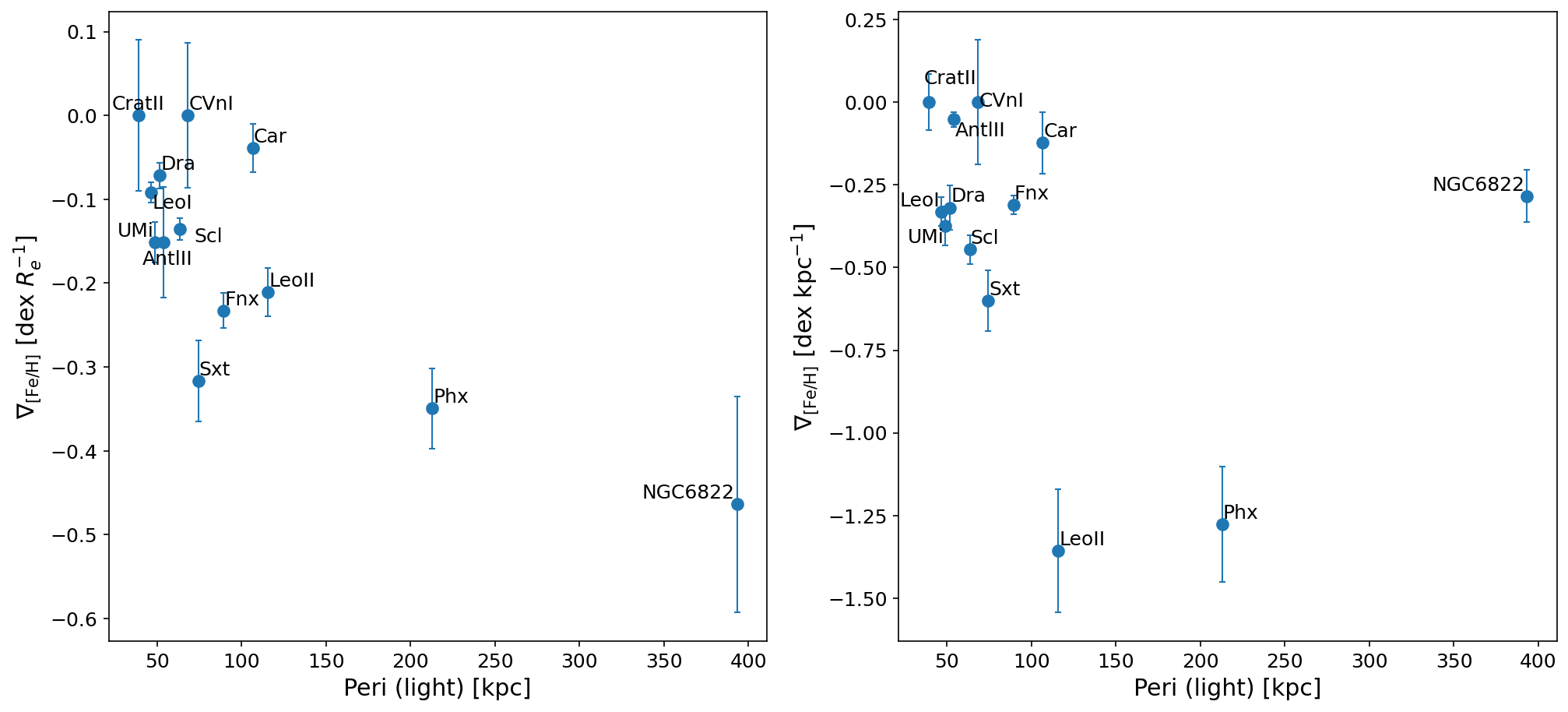}\\
    \includegraphics[width=.9\textwidth]{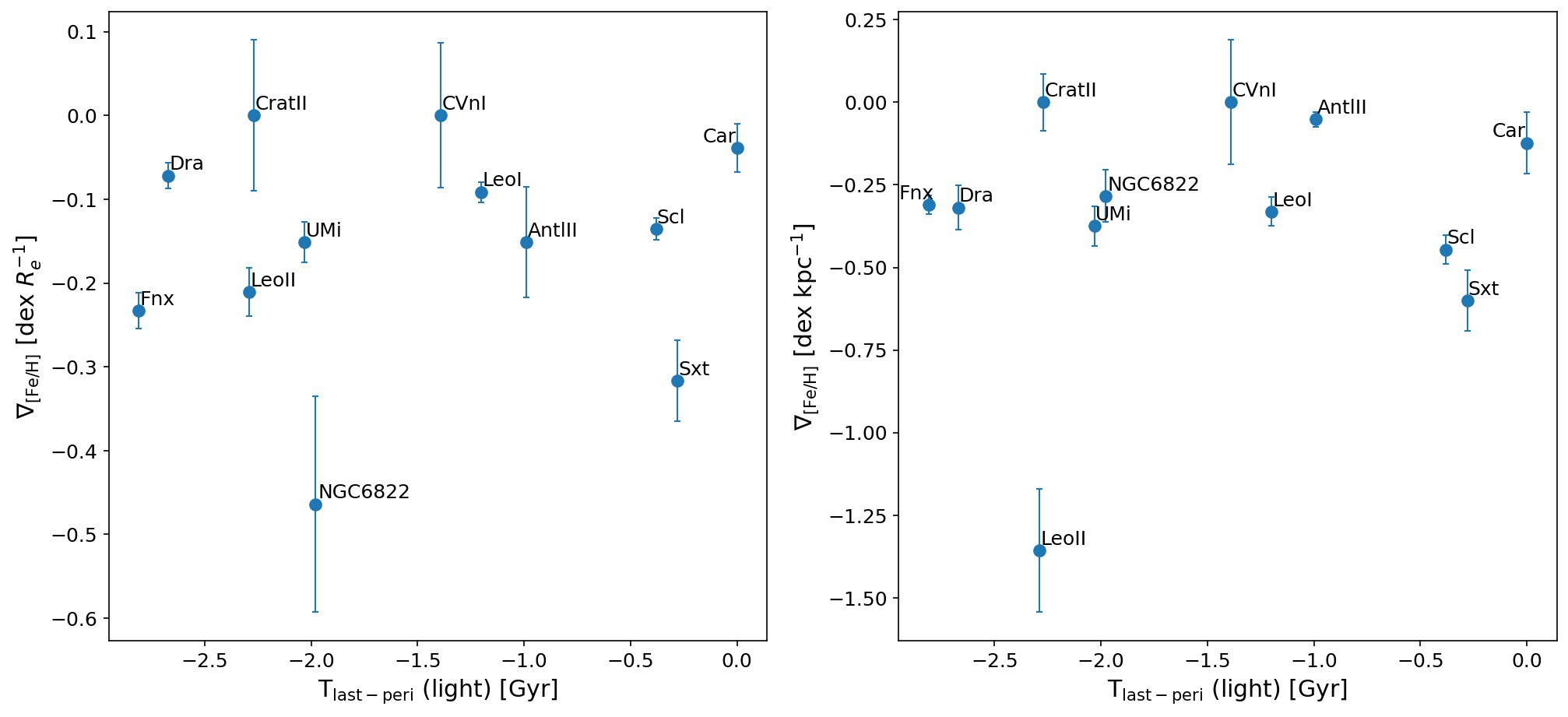}
    \caption{Distribution of metallicity gradients as a function of the orbital parameters obtained for the Milky Way satellites, assuming a ``light-potential'' for the MW \citep[see][for further details]{Battaglia2022}.
    On the x-axes are shown the eccentricities (top panels), the pericentric distances (middle panels), and the time since the last pericentric passage (bottom panels), of the considered sample.
    On the y-axes, the metallicity gradients in units of the 2D SMA half-light radius (left panels), and in units of the physical radius (right panels). 
    }
    \label{fig:mgrad_orb_params_light}
\end{figure*}

\begin{figure*}
    \centering
    \includegraphics[width=.9\textwidth]{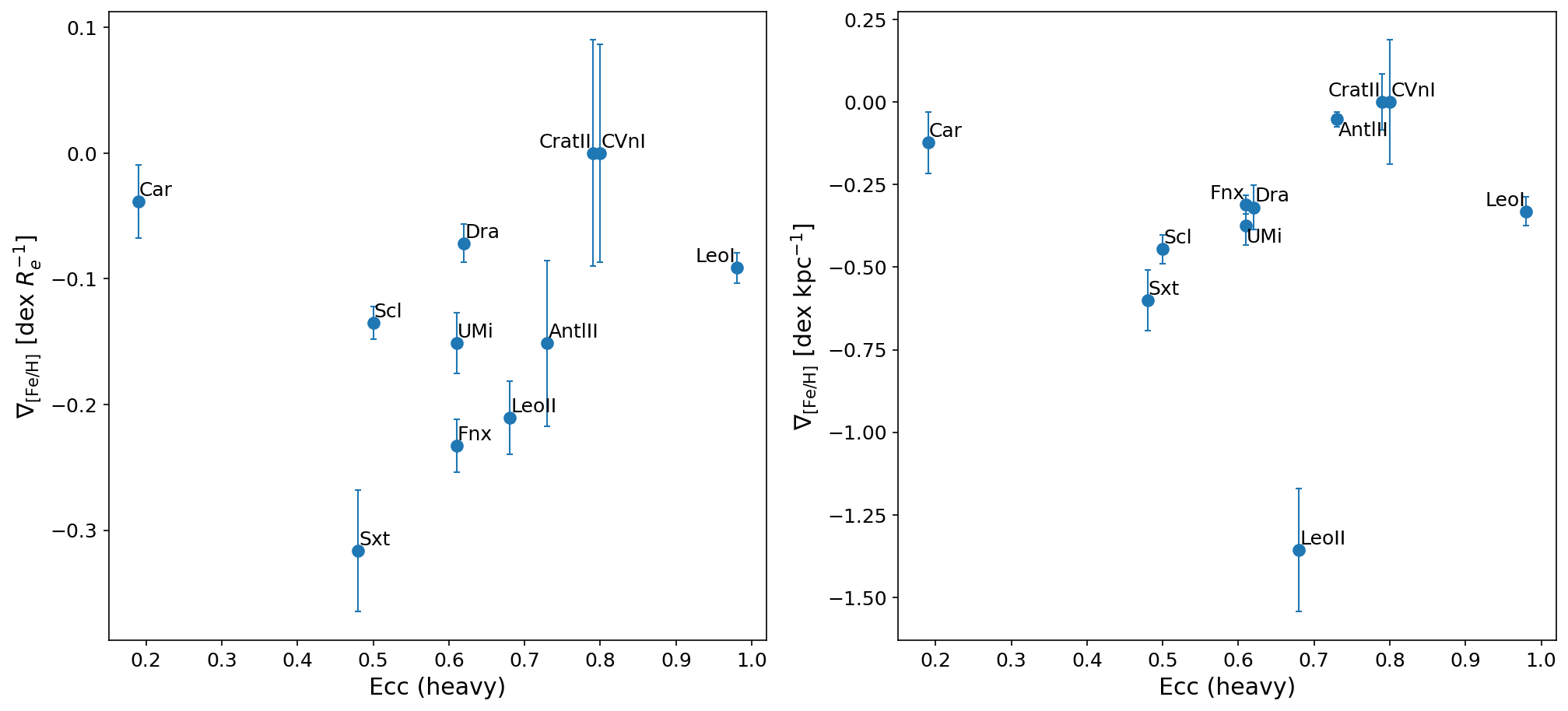}\\
    \includegraphics[width=.9\textwidth]{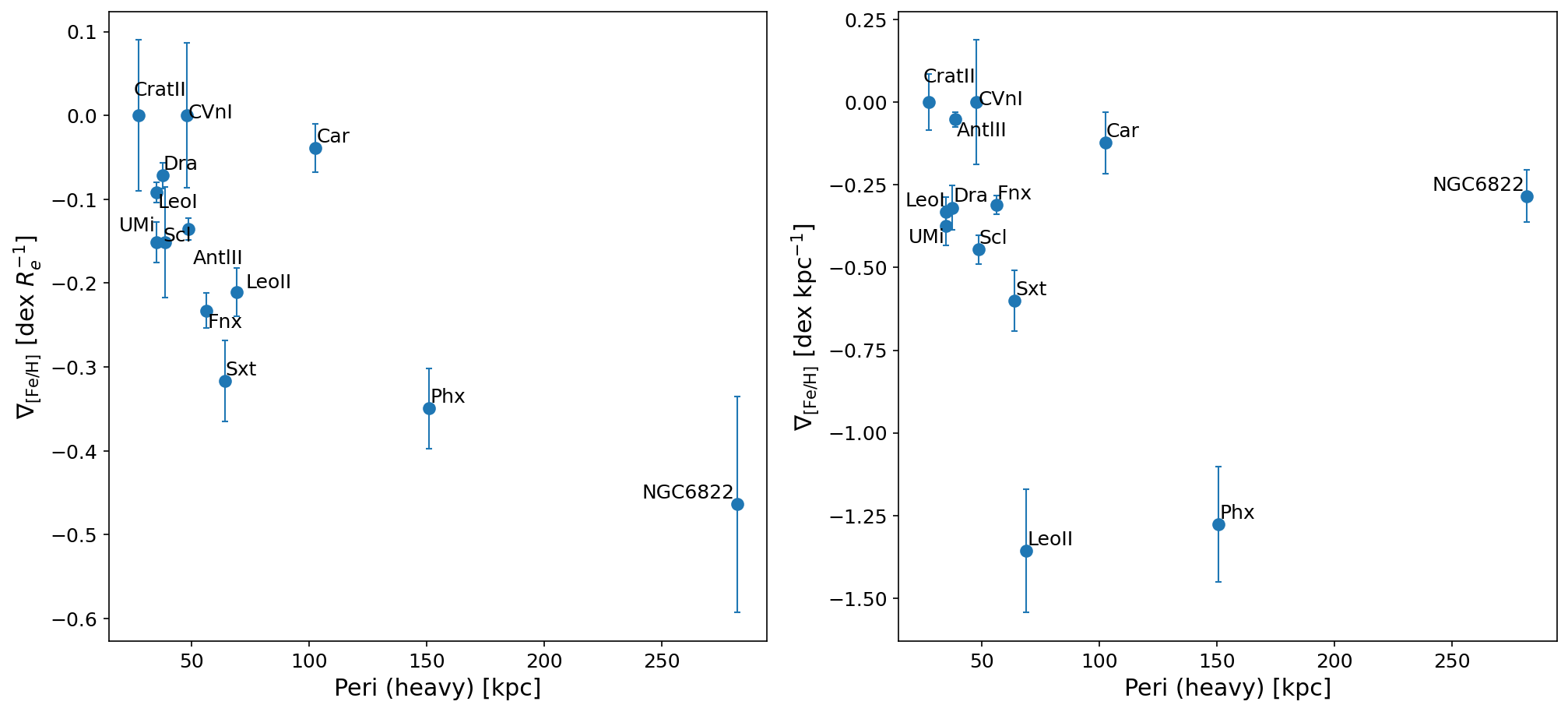}\\
    \includegraphics[width=.9\textwidth]{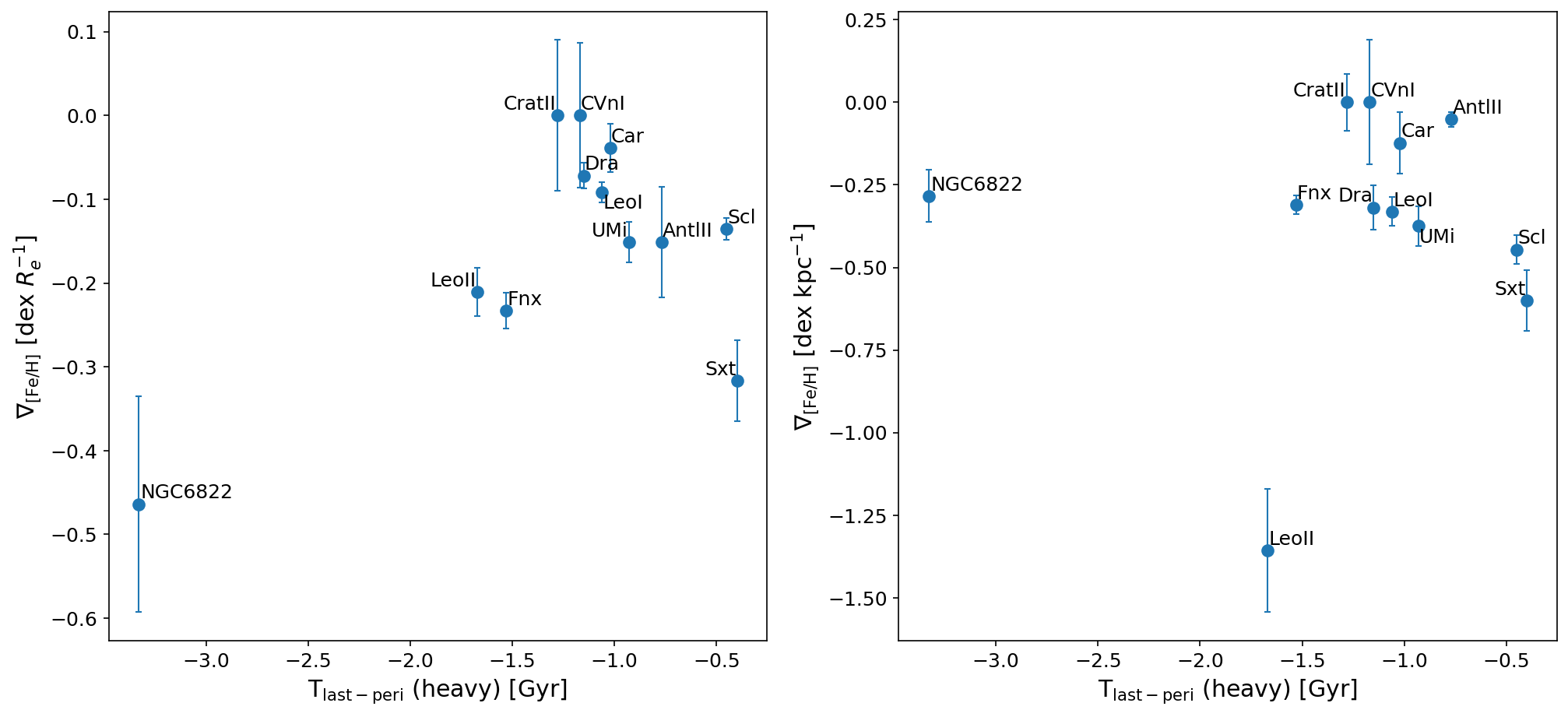}
    \caption{As in Fig.~\ref{fig:mgrad_orb_params_light}, but with the orbital parameters derived assuming a ``heavy-potential'' for the Milky Way \citep[see][for further details]{Battaglia2022}.
    }
    \label{fig:mgrad_orb_params_heavy}
\end{figure*}

As reported in the main text, and as shown in Figs.~\ref{fig:mgrad_orb_params_light} and \ref{fig:mgrad_orb_params_heavy}, the considered sample is too small to derive definitive conclusions on any possible correlation between the metallicity gradients and the orbital parameters.
For example, one can observe a correlation between $\nabla_{\rm [Fe/H]} (R/R_e)$ and the pericentric distance for both MW-potentials. Calculating the Spearman's correlation coefficient, we find indeed a high value of $\sim$\,$-0.7$, in both cases, indicative of a strong anti-correlation.
While this result may be related to the potential role played by tidal interactions in shaping the radial metallicity profile of satellite galaxies, it is also true that this correlation is mainly driven by the two outermost points, Phoenix and NGC~6822, which also show signs of a past merger event, as we have seen in the main text. In addition, Phoenix is one of the systems in \citet{Battaglia2022} with an uncertain 3D velocity, considered not particularly reliable in terms of orbital parameters. By removing them, the Spearman's correlation coefficient sensibly reduces to a value of $\sim$\,$-0.4$, for both MW-potentials. Therefore, it remains difficult to understand the eventual role of tidal interactions in influencing metallicity gradients.

\section{Comparison with simulations}
\label{sec:apx_MARVEL}

\begin{table*}
\caption{Calculated metallicity averages (i.e., arithmetic mean, median and mode; Cols.~2-4) and radial metallicity gradients (Cols.~6-7) for the samples of galaxies analyzed in this work (column 1) calculated using ``circular'' radii. For reference in column 5 are the 2D circular half-light radii; values from \citet{Battaglia2022} and the references listed in Table~\ref{tab:data_ref}.}
\label{tab:met_grad_circ}
\centering
\begin{tabular}{cccc|crr}
\hline
\hline
  \multicolumn{1}{c}{Galaxy} &
  \multicolumn{1}{c}{Mean} &
  \multicolumn{1}{c}{Median} &
  \multicolumn{1}{c}{Mode} &
  \multicolumn{1}{c}{$R_e^{\rm circ}$} &
  \multicolumn{1}{c}{$\nabla_{\rm [Fe/H]} (R/R_e)_{\rm circ}$} &
  \multicolumn{1}{c}{$\nabla_{\rm [Fe/H]} (R)_{\rm circ}$} \\
  \multicolumn{1}{c}{} &
  \multicolumn{3}{c}{[dex]} &
  \multicolumn{1}{c}{[\arcmin]} &
  \multicolumn{1}{c}{[dex $R_e^{-1}$] } &
  \multicolumn{1}{c}{[dex kpc$^{-1}$]} \\
\hline 
SMC     &  $-1.03$  & $-0.99$  & $-0.99$  &  66.3  & $-0.128 \pm 0.006$  & $-0.103 \pm 0.00$ \\
Fnx     &  $-1.16$  & $-1.03$  & $-1.02$  &  15.5  & $-0.24  \pm 0.02 $  & $-0.38  \pm 0.03$ \\
LeoI    &  $-1.45$  & $-1.44$  & $-1.41$  &  2.93  & $-0.12  \pm 0.01 $  & $-0.51  \pm 0.05$ \\
Scl     &  $-1.79$  & $-1.98$  & $-2.11$  &  9.9   & $-0.21  \pm 0.01 $  & $-0.88  \pm 0.05$ \\
LeoII   &  $-1.63$  & $-1.62$  & $-1.43$  &  2.37  & $-0.14  \pm 0.03 $  & $-0.91  \pm 0.19$ \\
Car     &  $-1.75$  & $-1.68$  & $-1.68$  &  8.1   & $-0.03  \pm 0.03 $  & $-0.13  \pm 0.11$ \\
Sxt     &  $-2.10$  & $-2.28$  & $-2.34$  &  18.3  & $-0.28  \pm 0.05 $  & $-0.62  \pm 0.12$ \\
AntlII  &  $-1.08$  & $-1.36$  & $-1.16$  &  60.0  & $-0.10  \pm 0.05 $  & $-0.05  \pm 0.02$ \\
UMi     &  $-2.13$  & $-2.15$  & $-2.13$  &  12.2  & $-0.09  \pm 0.02 $  & $-0.31  \pm 0.09$ \\
Dra     &  $-2.01$  & $-1.94$  & $-1.71$  &  8.04  & $-0.12  \pm 0.01 $  & $-0.61  \pm 0.08$ \\
CVnI    &  $-1.92$  & $-1.94$  & $-1.85$  &  5.5   & $ 0.00  \pm 0.09 $  & $ 0.00  \pm 0.26$ \\
CratII  &  $-1.91$  & $-1.92$  & $-1.75$  &  29.3  & $ 0.00  \pm 0.09 $  & $ 0.00  \pm 0.09$ \\
M32     &  $-1.27$  & $-1.15$  & $-0.84$  &  0.41  & $ 0.00  \pm 0.04 $  & $ 0.00  \pm 0.43$ \\
N205    &  $-0.86$  & $-0.78$  & $-0.90$  &  1.86  & $ 0.00  \pm 0.02 $  & $ 0.00  \pm 0.05$ \\
N185    &  $-0.96$  & $-0.93$  & $-0.87$  &  2.60  & $-0.23  \pm 0.03 $  & $-0.49  \pm 0.06$ \\
N147    &  $-0.55$  & $-0.46$  & $-0.46$  &  4.9   & $ 0.01  \pm 0.05 $  & $ 0.01  \pm 0.05$ \\
AndVII  &  $-1.26$  & $-1.20$  & $-1.41$  &  3.3   & $ 0.00  \pm 0.14 $  & $ 0.00  \pm 0.19$ \\
AndII   &  $-1.25$  & $-1.24$  & $-1.24$  &  5.5   & $-0.36  \pm 0.07 $  & $-0.35  \pm 0.07$ \\
AndV    &  $-1.84$  & $-1.88$  & $-1.90$  &  1.3   & $-0.11  \pm 0.05 $  & $-0.42  \pm 0.18$ \\
N6822   &  $-1.06$  & $-1.02$  & $-0.85$  &  10.14 & $-0.47  \pm 0.13 $  & $-0.34  \pm 0.09$ \\
IC1613  &  $-1.19$  & $-1.08$  & $-1.20$  &  6.77  & $-0.06  \pm 0.08 $  & $-0.04  \pm 0.05$ \\
WLM     &  $-1.29$  & $-1.24$  & $-1.17$  &  2.8   & $ 0.00  \pm 0.03 $  & $ 0.00  \pm 0.04$ \\
VV124   &  $-1.46$  & $-1.52$  & $-1.94$  &  0.85  & $-0.14  \pm 0.03 $  & $-0.43  \pm 0.09$ \\
LeoA    &  $-1.55$  & $-1.59$  & $-1.71$  &  1.75  & $-0.12  \pm 0.04 $  & $-0.32  \pm 0.10$ \\
PegDIG  &  $-1.39$  & $-1.31$  & $-1.02$  &  2.53  & $-0.11  \pm 0.07 $  & $-0.20  \pm 0.13$ \\
SagDIG  &  $-1.74$  & $-1.84$  & $-1.58$  &  0.95  & $ 0.00  \pm 0.06 $  & $ 0.00  \pm 0.20$ \\
Cet     &  $-1.80$  & $-1.72$  & $-1.83$  &  2.6   & $-0.14  \pm 0.05 $  & $-0.25  \pm 0.09$ \\
Aqu     &  $-1.60$  & $-1.59$  & $-1.79$  &  1.12  & $-0.18  \pm 0.06 $  & $-0.51  \pm 0.16$ \\
Phx     &  $-1.58$  & $-1.52$  & $-1.16$  &  1.9   & $-0.44  \pm 0.04 $  & $-1.93  \pm 0.18$ \\
Tuc     &  $-1.67$  & $-1.62$  & $-1.37$  &  0.8   & $-0.06  \pm 0.05 $  & $-0.30  \pm 0.22$ \\
\hline
\hline
\end{tabular}
\end{table*}

We report here Table~\ref{tab:met_grad_circ} containing the values of the metallicity gradients for our sample of observed dwarf galaxies, calculated using 2D ``circular'' radii and limiting the analysis to twice the 2D projected geometric half-light radius (also in table) for comparison with the results of \citet{Mercado2021}. Further reported on table are the calculated metallicity averages (i.e., arithmetic mean, median and mode) of our galaxies.

\end{appendix}

\end{document}